\documentclass[11pt]{article}
\usepackage{latexformat}
\pdfoutput=1
\usepackage{amsmath,bbm,array,amsfonts,graphicx,wrapfig,lscape,float,mathtools,multirow,longtable,amsthm,upgreek,mathrsfs}
\usepackage[dvipsnames]{xcolor}
\usepackage{stackrel}
\usepackage[all]{xy}
\usepackage[bottom]{footmisc}
\usepackage{tikz}
\usepackage{physics}
\usepackage{bm}
\usepackage[capitalize]{cleveref}
\usepackage{color}
\usepackage{enumerate}
\usetikzlibrary{decorations.pathreplacing}
\usepackage{diagbox}
\numberwithin{equation}{section}
\numberwithin{figure}{section}
\numberwithin{table}{section}
\usepackage{pgfplots}
\pgfplotsset{compat=1.14}
\usepackage{makecell}
\usepackage[utf8]{inputenc}
\allowdisplaybreaks

\newcommand{\blue}{\color{blue}}

\newcommand{\red}{\color{red}}

\newcommand{\BZ}{\mathbb{Z}}

\def\SO{\mathrm{SO}}

\def\SU{\mathrm{SU}}

\def\USp{\mathrm{USp}}

\def\su{\mathfrak{su}}

\def\so{\mathfrak{so}}
\def\sp{\mathfrak{sp}}
\def\usp{\mathfrak{usp}}

\newcommand{\bbZ}{\mathbb{Z}}

\newcommand{\Hom}{\mathrm{Hom}}

\newcommand{\bes}[1]{\begin{equation} \begin{split} #1\end{split} \end{equation}}

\usepackage{tikz}
\usetikzlibrary{positioning}
\usetikzlibrary{arrows}
\usetikzlibrary{decorations.pathreplacing}
\usetikzlibrary{shapes.geometric}
\usetikzlibrary{calc}
\usetikzlibrary{decorations.pathmorphing}
\usetikzlibrary{shapes.misc}

\usetikzlibrary{positioning,trees,decorations.pathmorphing,decorations.markings,decorations.pathreplacing,calc,shapes,patterns,arrows,chains,arrows.meta,fit,fadings,decorations.markings,graphs,graphs.standard,quotes,plotmarks,calligraphy,decorations.pathreplacing}
\usetikzlibrary{intersections, calc}
\usetikzlibrary{shapes.geometric}
\usetikzlibrary{matrix, arrows.meta}
\usetikzlibrary{shapes.geometric}
\usetikzlibrary{decorations.pathmorphing}
\usetikzlibrary{shapes.misc}
\usetikzlibrary{arrows.meta}
\usetikzlibrary{angles, quotes}

\tikzset{flavour/.style={draw=none,minimum size=0.3mm,fill=white, regular polygon,regular polygon sides=4,draw}}
\tikzset{gaugeBig/.style={inner sep=1mm,draw=none,fill=white,minimum size=2mm,circle, draw}}
\tikzset{bd/.style={circle, draw=black, inner sep=0pt, fill=black, minimum size=2mm}}
\tikzset{wd/.style={circle, draw=black, inner sep=0pt, fill=white, minimum size=2mm}}
\tikzset{Dynkin/.style={circle, draw=black, inner sep=0pt, fill=white, minimum size=2mm}}
\tikzstyle{ligne}=[draw, very thick] 
\tikzstyle{gridline}=[draw, gray] 
\tikzset{gauge/.style={circle, draw,inner sep=2.5pt}}
\tikzset{gaugeo/.style={circle, draw,inner sep=2.5pt,fill=orange}}
\tikzset{gaugec/.style={circle, draw,inner sep=2.5pt,fill=cyan}}
\tikzset{gauger/.style={circle, draw,inner sep=2.5pt,fill=red}}
\tikzset{gaugeb/.style={circle, draw,inner sep=2.5pt,fill=blue}}
\tikzset{gaugeg/.style={circle, draw,inner sep=2.5pt,fill=green}}
\tikzset{gaugem/.style={circle, draw,inner sep=2.5pt,fill=magenta}}
\tikzset{hasse/.style={circle, fill,inner sep=2pt}}
\tikzset{shrinky/.style={circle, fill,inner sep=1pt}}
\tikzset{sized/.style={circle, draw, inner sep=1.5pt}}
\tikzset{seven/.style={circle, draw,inner sep=3pt}}

\tikzset{dotto/.style={circle, orange, draw,inner sep=1.5pt,fill=orange}}
\tikzset{dottp/.style={circle, purple, draw,inner sep=1.5pt,fill=purple}}
\tikzset{dottc/.style={circle, cyan, draw,inner sep=1.5pt,fill=cyan}}
\tikzset{dottr/.style={circle, red, draw,inner sep=1.5pt,fill=red}}
\tikzset{dottb/.style={circle, blue, draw,inner sep=1.5pt,fill=blue}}
\tikzset{dottg/.style={circle, green, draw,inner sep=1.5pt,fill=green}}
\tikzset{dottm/.style={circle, magenta, draw,inner sep=1.5pt,fill=magenta}}

\definecolor{cE8}{HTML}{0000FF}

\tikzset{
	flavor/.style={rectangle, draw=black!100, thick, minimum size=2mm},
	gauge/.style={circle, thick, draw=black!100,fill=white!100,  minimum size=2mm, inner sep=0pt},
	bodyE8/.style={draw=cE8!100},
}

\usepackage[toc,page]{appendix}

\usepackage{tocloft}

	\title{Orbi-Instantons and Class \texorpdfstring{$\mathcal{S}$}{S} Theories of Type D}
	
	\author[a]{Jiakang Bao,}
        \author[b,c]{Noppadol Mekareeya,}
        \author[d,e]{Gabi Zafrir,}
        \author[f]{Hao Y. Zhang}
	
	\affiliation[a]{
		Graduate School of Physics, University of Tokyo, Tokyo 113-0033, Japan}
        \affiliation[b]{INFN, sezione di Milano-Bicocca, Piazza della Scienza 3, I-20126 Milano, Italy}
        \affiliation[c]{The Institute for Fundamental Study ``The Tah Poe Academia Institute'', Naresuan University, Phitsanulok 65000, Thailand}
        \affiliation[d]{Department of Physics, University of Haifa at Oranim, Kiryat Tivon 36006, Israel}
        \affiliation[e]{Haifa Research Center for Theoretical Physics and Astrophysics, University of Haifa, Haifa 3498838, Israel}
        \affiliation[f]{Kavli Institute for the Physics and Mathematics of the Universe (WPI), University of Tokyo, Kashiwa, Chiba 277-8583, Japan}
	
	\emailAdd{jiakang.bao@phys.s.u-tokyo.ac.jp}
        \emailAdd{n.mekareeya@gmail.com}
        \emailAdd{gabi.zafrir@oranim.ac.il}
        \emailAdd{hao.zhang@ipmu.jp}
	
	\preprint{
		\begin{flushright}
			
		\end{flushright}
	}

	\abstract{We investigate the landscape of 6d $\mathcal{N}=(1,0)$ D-type orbi-instanton superconformal field theories (SCFTs) and their torus compactifications to four-dimensional class $\mathcal{S}$ theories. By analysing a general class of 6d F-theory constructions via generalised quivers, we demonstrate that---in contrast to the well-characterised A-type series---the dimensional reductions that admit a 4d class $\mathcal{S}$ description on a Riemann sphere with three untwisted D-type punctures constitute only a subset of the full orbi-instanton landscape. For this subclass, we show that the punctures can be effectively characterised by two sets of integers: the $s$-labels and the $m$-labels. The $s$-labels, or ``Kac-type labels'', serve as the D-type analogues to the Kac labels used in A-type theories; we establish their correspondence with ``modified excess numbers'' in the associated 3d mirror theories (magnetic quivers). The $m$-labels are further introduced to streamline the mapping from 6d generalised quivers to their class $\mathcal{S}$ descriptions. Furthermore, we analyse physical distinctions arising from 6d $\theta$ angles and explore the hierarchy of Higgs branch flows. In doing so, we uncover instances of ``hidden Higgsings''---renormalization group flows present in the 6d parent theories that are not manifest in the puncture closures of the corresponding class $\mathcal{S}$ descriptions.
	}

\dedicated{To the memory of Yitzhak Frishman
 and Kellogg S.~Stelle.}

\begin{document}
	\maketitle

\section{Introduction and Summary}\label{intro}

The classification of six-dimensional (6d) superconformal field theories (SCFTs) via F-theory was established in \cite{Heckman:2013pva,Heckman:2015bfa}. Within this framework, a prominent family of 6d $\mathcal{N}=(1,0)$ theories---known as \textit{orbi-instanton} theories---arises as the low-energy worldvolume description of M5-branes probing an M9-brane on asymptotically locally Euclidean (ALE) spaces \cite{Heckman:2015bfa, Heckman:2018pqx}. These theories are characterised by small $E_8$ instantons on orbifolds of ADE type, which motivates their nomenclature.

Directly analysing these theories is challenging due to their strongly coupled nature. However, the tensor branch description derived from F-theory provides a robust analytical tool \cite{Heckman:2013pva,Heckman:2015bfa}; see \cite{Heckman:2018jxk} for a comprehensive review. In an elliptically fibred Calabi--Yau (CY) threefold with a non-compact base where gravity decouples, the effective string spectrum originates from D3-branes wrapping compact $\mathbb{CP}^1$ curves in the base.

This geometric construction is conveniently encoded using \textit{generalised quivers}. In this notation, curves with self-intersection number $-n$ are represented by the integer $n$, and their relative positions indicate their intersections. In the M-theory picture, the curve volumes correspond to the physical distances between M5-branes (and the $E_8$ wall). Furthermore, singular fibres may introduce non-trivial gauge algebra decorations on the curves. The resulting matter content and global symmetries can be determined following \cite{Bertolini:2015bwa}.\footnote{Throughout this paper, global symmetries in generalised quivers are denoted by $[F]$, where $F$ represents the number of full hypermultiplets.}

An alternative perspective on orbi-instanton theories involves their compactification on a two-torus ($T^2$), which is expected to yield a 4d theory. For the specific case of M5-branes probing an M9-brane in flat space, the resulting 4d theory is the strongly coupled Minahan-Nemeschansky $E_8$ SCFT. A natural question is whether replacing flat space with an ALE space similarly yields strongly coupled 4d SCFTs. In all well-understood cases, the answer appears to be affirmative, suggesting that dimensional reduction provides a mechanism for engineering 4d SCFTs akin to the class $\mathcal{S}$ construction \cite{Gaiotto:2009we}. Indeed, evidence suggests a non-trivial correspondence: the torus reduction of orbi-instanton theories often coincides with a 6d $\mathcal{N}=(2,0)$ theory on a punctured Riemann surface; see, for instance, \cite{Ohmori:2015pua,Mekareeya:2017jgc,Baume:2021qho}. In other words, the torus reduction of an orbi-instanton theory appears to result in a class $\mathcal{S}$ theory. This pattern is most transparent for A-type orbi-instanton theories, studied in \cite{Mekareeya:2017jgc}. These theories reduce to a Riemann sphere with three A-type punctures; upon further circle reduction, they yield star-shaped 3d mirror theories \cite{Benini:2010uu,Tachikawa:2015bga} (also known as magnetic quivers \cite{Cabrera:2019izd,Cabrera:2019dob}) with unitary gauge nodes.

While these observations might suggest that all torus-reduced orbi-instanton theories are contained within class $\mathcal{S}$, it remains unclear if this is a general feature or merely an artifact of the non-exhaustive list of understood cases. In this paper, we investigate this by focusing on \textbf{D-type orbi-instanton theories} and their torus compactifications. Our analysis suggests that not all such theories possess class $\mathcal{S}$ descriptions. This suggests that the 4d theories obtained from 6d $(2,0)$ theories on a Riemann sphere with three untwisted D-type punctures correspond only to a subset of the dimensional reductions of D-type orbi-instantons. Consequently, the class $\mathcal{S}$ framework may not fully capture the landscape of orbi-instanton torus reductions.\footnote{This discrepancy becomes more pronounced for E-type cases, where there are finitely many E-type punctures but infinitely many orbi-instanton theories. We demonstrate that this issue already manifests in the D-type series and defer the E-type analysis to future work.}

To refine this statement, recall that an orbi-instanton theory (defined by ADE group of type $G$ and $k$ M5-branes) is specified by a group homomorphism from $\Gamma$ to $E_8$ (where $\Gamma \subset \text{SU}(2)$ is a discrete subgroup) and a nilpotent orbit. While A-type homomorphisms are classified in \cite{kac1990infinite}, the D-type classification remains incomplete despite previous efforts \cite{frey2001conjugacy}. Recently, Frey and Rudelius \cite{Frey:2018vpw} analysed D-type and E-type discrete homomorphisms via 6d SCFTs and their F-theory constructions. In particular, the orbi-instanton theory can be obtained by \textit{affinising}\footnote{By affinisation, we mean that we attach one additional curve to the generalised quiver. This curve can be added to, say, the leftmost curve or the next-to-leftmost curve.} the generalised quiver associated to a D-type nilpotent orbit/partition $\bm{p}$ with an extra curve and a specific gauge algebra decoration. Thus, a D-type discrete homomorphism is labelled by a D-type partition and a Lie algebra. The partition can be translated into three punctures \cite{Mekareeya:2016yal}, and the 6d matter content can be determined via \cite{Bertolini:2015bwa}. Based on the partition, we categorise the theories into three cases:
\begin{itemize}
    \item $\bm{p}^\text{T}_1\geq8$,
    \item $\bm{p}^\text{T}_1<8$ and $\bm{p}^\text{T}_1+\bm{p}^\text{T}_2\geq6$,
    \item $\bm{p}^\text{T}_1+\bm{p}^\text{T}_2<6$.
\end{itemize}
Here, $\bm{p}^\text{T}_i$ denotes the $i^\text{th}$ entry of the transpose of $\bm{p}$. We find that many theories in the first two cases admit class $\mathcal{S}$ descriptions, while the third case (and certain instances in the first two) does not.\footnote{In \cref{6d4drelations}, we divide these cases into two types based on whether the leftmost end features an $\mathfrak{su}$ gauge algebra or a bifurcation. The type with an $\mathfrak{su}$ end always belongs to the first case ($\bm{p}^\text{T}_1\geq8$). For the bifurcating end type, if the leftmost end features a $-1$ curve with a $\mathfrak{usp}$ gauge algebra decoration (which can be viewed as a bifurcation), it belongs to the second case; otherwise, it belongs to the first case.}

In \cite{Frey:2018vpw}, it was noted that the classification therein might be defined up to the outer automorphisms of the D-type Lie algebras. This is because D-type partitions may be \textit{very even}, corresponding to two distinct nilpotent orbits. Such a distinction is not manifest in the 6d tensor branch description. Here, we demonstrate that this distinction between very even orbits yields no difference for the class $\mathcal{S}$ theories, either. Therefore, from a physical perspective (specifically, the characterisation of D-type homomorphisms $\Gamma \rightarrow E_8$ by 6d SCFTs), such pairs of very even orbits may not yield two distinct homomorphisms.

However, similar to the very even partitions for nilpotent orbits, we do observe certain one-to-two correspondences where a single 6d curve configuration potentially corresponds to two distinct discrete homomorphisms. Physically, this is due to the presence of 6d $\theta$ angles. Such differences would also be manifest in the 4d class $\mathcal{S}$ descriptions.\footnote{We shall discuss this in more detail in \cref{thetaangle}.}

Moreover, Higgsings provide another vital diagnostic. The Higgsing of 6d theories can be performed following the algorithms in \cite{Bao:2024wls,Bao:2025pxe} (see also \cite{Heckman:2015ola,Heckman:2016ssk,Heckman:2018pqx,Hassler:2019eso,Fazzi:2022hal,Fazzi:2022yca, Baume:2023onr}). On the 4d side, Higgsings are achieved by the partial closure of punctures. In terms of 3d mirror theories/magnetic quivers, the Higgs mechanism follows algorithms including quiver subtraction \cite{Hanany:2018uhm, Cabrera:2018ann} and quiver decay and fission \cite{Bourget:2023dkj,Bourget:2024mgn,Lawrie:2024wan} (see also \cite{Bourget:2019aer,Lawrie:2024zon,Bennett:2026xpm}).\footnote{Due to the extensive literature on this topic, we include only a representative subset of references.} It turns out that some 6d Higgsings are hidden from the class $\mathcal{S}$ descriptions, in the sense that they cannot be directly seen from the punctures.

Let us comment on the characterisation of the orbi-instanton theories. For A-type cases, the homomorphism $\mathbb{Z}_k \rightarrow E_8$ is uniquely characterised by Kac labels \cite{kac1990infinite}---non-negative integers assigned to the nodes of the affine $E_8$ Dynkin diagram. These labels allow one to derive the 6d generalised quiver, the 4d class $\mathcal{S}$ theory on a three-punctured sphere, and the 3d mirror unitary quivers \cite{Mekareeya:2017jgc}. For the latter, the Kac labels coincide with the excess numbers \cite{Gaiotto:2008ak} of the gauge groups. In contrast, D-type cases lack labels that uniquely characterise the homomorphisms from $\widehat{D}_k$ to $E_8$, complicating their classification. The purpose of this paper is therefore not to provide an exhaustive classification, but to discuss infinite subclasses of these theories. Despite the lack of precise D-type Kac labels, we find that for theories admitting a torus reduction to 4d class $\mathcal{S}$, the punctures can be characterised by specific sets of integers, which we refer to as $s$-labels and $m$-labels. The $s$-labels (or ``Kac-type labels'') relate to a notion similar to the excess number in 3d mirrors, which we term ``modified excess numbers''. The $m$-labels, meanwhile, simplify the derivation from 6d quivers to class $\mathcal{S}$ descriptions. While not being unique characterisations, these labels are highly effective for theory identifications.

Finally, we briefly comment on our methodology. In analogy with the A-type cases, one expects D-type orbi-instanton theories to reduce to D-type class $\mathcal{S}$ theories on a sphere with three punctures of a specific structure. We introduce $s$-labels to specify these punctures. Given a choice of orbifold group $\widehat{D}_k$, there is a finite number of consistent $s$-labels; for small values of $k$ (specifically $k=3,4,5$), we exhaustively construct all possible class $\mathcal{S}$ theories of the required form. Simultaneously, we use the results of \cite{Frey:2018vpw} to list all $\widehat{D}_k$-type orbi-instanton theories. We then match the two sets using various physical data: global symmetry, 't Hooft anomalies, and the spectrum of Coulomb branch operators. We find that all class $\mathcal{S}$ theories of the required form possess the right properties to be torus reductions of orbi-instanton theories, but that some orbi-instanton theories lack a corresponding class $\mathcal{S}$ candidate.

This study allows us to distinguish between theories that expect a class $\mathcal{S}$ description and those that do not. For the former, we propose rules to associate a 4d D-type class $\mathcal{S}$ theory with either a 6d quiver or a Frey-Rudelius label (a D-type partition and a Lie algebra). We emphasize that these associations are based on extensive observations and matching of properties rather than formal proofs. While the interrelation through RG flows suggests these results hold for higher $k$, some degree of caution is advised regarding potential special cases that may only emerge at larger $k$.

We close with a remark on the relative sizes of the two classes. Our exhaustive analysis at $k=4,5$ shows that roughly $68\%$ of the D-type orbi-instanton theories in this range admit a class $\mathcal{S}$ description for their 4d torus reduction. Consistent with this, most cases in the Frey-Rudelius labelling admit a D-type class $\mathcal{S}$ realisation upon torus compactification. Taken together, these findings indicate that a substantial fraction of D-type orbi-instanton theories reduce to 4d D-type class $\mathcal{S}$ theories on $T^2$. Some caution is nevertheless warranted: our exhaustive search is restricted to low values of $k$, and the precise count of theories within each Frey-Rudelius case is difficult to predict, owing to the incomplete understanding of the embeddings of $\widehat{D}_k$ into $E_8$. Finally, we note that some D-type class $\mathcal{S}$ theories arising in our analysis can also be associated with an embedding of a cyclic group into $E_8$. This observation may have interesting implications for the embedding of $\widehat{D}_k$ into $E_8$ and its relation to the embeddings of its cyclic normal subgroup.

\paragraph{Organisation of the paper} The paper is organised as follows. In \cref{Atype}, we briefly review A-type orbi-instanton theories and their class $\mathcal{S}$ descriptions. In \cref{Dtype}, we list the three untwisted D-type punctures, introduce their labels, and describe the consistency checks for dimensional reduction. In \cref{6d4drelations}, we detail the algorithms for obtaining 4d class $\mathcal{S}$ theories from 6d quivers, including the role of 6d $\theta$ angles. In \cref{PartitionDesc}, we discuss $m$-labels and the Frey-Rudelius analysis. Finally, in \cref{examples}, we illustrate our findings with $\widehat{D}_3$, $\widehat{D}_4$ and $\widehat{D}_5$ examples.

\section{Orbi-Instanton Theories of A-Type}\label{Atype}
In this section, we shall briefly review the class $\mathcal{S}$ descriptions of A-type orbi-instantons. More details can be found in \cite{Mekareeya:2017jgc}. The main punchline is that, the descendent theories in 5d, 4d and 3d can be uniquely characterised by (the Kac label of) the discrete homomorphism $\bbZ_k \rightarrow E_8$ labelling the A-type orbi-instanton theory.

Given an orbi-instanton theory specified by a discrete homomorphism from $\mathbb{Z}_k$ to $E_8$ and a nilpotent orbit in $\mathfrak{sl}(k+1)$, its torus compactification gives rise to a class $\mathcal{S}$ theory. The class $\mathcal{S}$ theory is specified by the following three A-type untwisted regular punctures:
\begin{subequations} \label{classSAtype}
    \begin{align}
        &\lambda=[\lambda_i]=\Big[n-n_6,~n-n_6-n_5,~n-n_6-n_5-n_4,~n-n_6-n_5-n_4-n_3,\nonumber\\
        &\quad\quad n-n_6-n_5-n_4-n_3-n_2,~n-n_6-n_5-n_4-n_3-n_2-n_1,~1^k\Big]\;,\\
        &\mu=[\mu_i]=[2n+2n_{4'}+n_{2'}+n_{3'},~2n+n_{4'}+n_{2'},~2n+n_{4'}+n_{3'}]\;,\\
        &\nu=[\nu_i]=[3n+2n_{4'}+n_{2'}+2n_{3'}, 3n+2n_{4'}+n_{2'}+n_{3'}]\;.
    \end{align}
\end{subequations}
Here, $n$ is a positive integer determined by the length of the 6d quiver (i.e., the number of M5-branes), and $n_i$ are non-negative integers depending on different homomorphisms. Notice that we have the part $1^k$ in $\lambda$, corresponding to the trivial nilpotent orbit. One may also change this to the other orbits.

As shown in \cite{kac1990infinite}, the discrete homomorphisms from $\mathbb{Z}_k$ to $E_8$ (up to conjugations by the automorphisms of $E_8$) can be classified by a set of non-negative integers:
\begin{align}
    \begin{split}
    &s_{3'}\\
    s_1\quad s_2\quad s_3\quad s_4\quad s_5\quad&s_6\quad s_{4'}\quad s_{2'}\;.
    \end{split}
\end{align}
We shall refer to them as the Kac labels. The Kac labels should satisfy the condition that their weighted sum (weighted by the dual Coxeter labels) is equal to $k$. More concretely, recall that the dual Coxeter labels of affine $E_8$ are given by
\begin{align}
    &3'\nonumber\\
    1\quad2\quad3\quad4\quad5\quad&6\quad4'\quad2'\;,\label{dualCoxeter}
\end{align}
where the apostrophes on the numbers appearing multiple times are used to distinguish the Kac labels. Then
\begin{equation}
    s_1+2s_2+3s_3+4s_4+5s_5+6s_6+4s_{4'}+2s_{2'}+3s_{3'}=k
\end{equation}
for $\mathbb{Z}_k$.

It turns out that for the A-type orbi-instanton theories, $n_i$ are precisely the Kac labels in the above partitions. In other words, $n_i=s_i$.

The algorithm determining the exact partitions from the 6d quiver configurations, as well as other relevant discussions such as the 5d brane web descriptions, can be found in \cite{Mekareeya:2017jgc}. Here, we shall only mention how the global symmetries can be read off from the 3d mirror quivers (a.k.a.~the magnetic quivers) of the class $\mathcal{S}$ theories.

Given the three punctures $\lambda$, $\mu$ and $\nu$, the corresponding 3d mirror theory is a star-shaped quiver with three legs and only unitary gauge nodes. The central node has rank equal to $|\lambda|=|\mu|=|\nu|$. The ranks of the remaining nodes along the three legs are then given by $|\lambda|-\lambda_1$, $|\lambda|-\lambda_1-\lambda_2$, $\dots$, and likewise for the other two legs, as each leg corresponds to a partition.

The global symmetry can be determined by examining the balance of the 3d mirror theory. The excess number/balance of a node is \cite{Gaiotto:2008ak}
\begin{equation}
    b=N_f-2N\;,
\end{equation}
where $N$ is the rank of the node. In other words, $b$ is equal to the sum of the ranks of the adjacent nodes minus twice the rank of the node. We say that a node is overbalanced (balanced, resp.~underbalanced) if $b>0$ ($b=0$, resp.~$b<0$). A theory with all the nodes satisfying $b\geq0$ is called a \textit{good} theory. Now, the balanced nodes in the 3d mirror theory would form some Dynkin diagram(s) corresponding to the non-abelian part of the global symmetry. The number of $\text{U}(1)$ in the global symmetry is given by the number of unbalanced nodes minus one.

\section{Descriptions in D-Type Class \texorpdfstring{$\mathcal{S}$}{S}}\label{Dtype}
Let us consider the torus reduction of the 6d SCFTs associated with M5-branes probing a $\mathbb{C}^2/\Gamma$ singularity in the presence of an M9-plane, where we shall mainly concentrate on the case of $\Gamma=\text{Dic}_{k-2}$. For simplicity, we shall also denote the dicyclic group $\text{Dic}_{k-2}$ as $\widehat{D}_k$.

We propose that the resulting 4d theory can be described as a class $\mathcal{S}$ theory associated with the reduction of the D-type $(2,0)$ theory on a 3-punctured sphere, at least for certain choices of the embeddings of $\widehat{D}_k \rightarrow E_8$. The form of the class $\mathcal{S}$ theories is highly reminiscent of the A-type class $\mathcal{S}$ theories describing the torus reduction in the case of $\Gamma=\mathbb{Z}_k$ \cite{Mekareeya:2017jgc}, and can be thought of as a D-type generalisation of these.

It is convenient to specify the class $\mathcal{S}$ theories using a collection of integers that determine the puncture data. In the A-type case ($\Gamma=\mathbb{Z}_k$), these integers can be identified with the Kac labels that determine the embeddings of $\mathbb{Z}_k \rightarrow E_8$. However, to our knowledge, there is no such nice picture for understanding the embeddings of $\widehat{D}_k \rightarrow E_8$. See, for instance, \cite{Frey:2018vpw}. Nevertheless, we can still introduce certain sets of integers specifying the puncture data of the class $\mathcal{S}$ theory. However, as we shall see, these do not bijectively determine the embeddings of $\widehat{D}_k \rightarrow E_8$.

\subsection{Punctures and Labels}\label{labels}
Let us introduce our proposal for the 4d theory in terms of a D-type class $\mathcal{S}$ theory. Specifically, the 4d theory is equivalent to the reduction of the $D_{3n+\delta}$ $(2,0)$ theory on a sphere with 3 punctures. The specific punctures are\footnote{In this paper, unless otherwise specified, we shall always take the trivial nilpotent orbit associated to one end of the orbi-instanton 6d quiver, which gives the $1^{2k}$ part in the puncture $\lambda$. For the other choices of nilpotent orbits, we can simply replace the corresponding part in $\lambda$.}
\begin{subequations} \label{classSform}
    \begin{align}
        &\lambda=[\lambda_i]=\Big[n-n_6,~n-n_6-n_5,~n-n_6-n_5-n_4,~n-n_6-n_5-n_4-n_3,\nonumber\\ \label{classSforma}
        &\quad\quad n-n_6-n_5-n_4-n_3-n_2,~n-n_6-n_5-n_4-n_3-n_2-n_1,~1^{2k}\Big]\;,\\ \label{classSformb}
        &\mu=[\mu_i]=[2n+l+m+x,~2n+m+x,~2n+x,~1]\;,\\ \label{classSformc}
        &\nu=[\nu_i]=[3n+\delta+y, 3n+\delta-y]\;.
    \end{align}
\end{subequations}
Here, the numbers $(n_6,n_5,\dots,n_1,x,m,l,y)$ determine the embedding, and $n$ is a number related to the number of M5-branes. The demand that the number of boxes in each puncture should be equal enforces that
\begin{subequations} \label{classSrest}
    \begin{align}
    &6n_6+5n_5+4n_4+3n_3+2n_2+n_1+3x+2m+l=2k-1\;,\\
    &3x+2m+l = 2\delta - 1\;.
\end{align}
\end{subequations}
In general, for a chosen value of $k$, the constraint leads to only finitely many solutions, corresponding to the fact that there is only a finite number of embeddings\footnote{There are several cases that require negative numbers for some of the labels. We include them below in the examples of \cref{examples} for completeness, although they do not appear to provide new cases.} of $\widehat{D}_k \rightarrow E_8$.

In the A-type cases, we have the Kac labels that specify the puncture data. For the D-type cases, finding a bijection between some integer tuple labels and the class $\mathcal{S}$ description becomes much more difficult (let alone the labels for the full set of discrete homomorphisms). Moreover, unlike the A-type cases, it is also more involved to determine the constraints on the potential labels such that they would give good theories. Here, we shall introduce two types of labels that would play certain roles in the orbi-instanton theories and class $\mathcal{S}$ theories, although they do not provide complete classifications of the discrete homomorphisms.

We should say a few words about the chosen form of the proposed class $\mathcal{S}$ theory \eqref{classSform}. In general, these classes of theories should have an $\mathfrak{so}_{2k}$ flavour symmetry associated with the $\widehat{D}_k$ singularity, which in the 6d quiver is associated to the symmetry rotating the $k$ $\mathfrak{usp}$ fundamental hypers situated at one end of the quiver (which here we shall generally take to be the right end). As such, the 4d theory resulting from its torus reduction should have an $\mathfrak{so}_{2k}$ flavour symmetry. Furthermore, the results of \cite{Ohmori:2014kda}, which we briefly review below, allows us to determine the 4d central charge of this symmetry from its mixed anomaly with gravity in the 6d SCFT. This suggests that the 4d theory associated to the torus reduction of any 6d SCFT in this class should have an $\mathfrak{so}_{2k}$ flavour symmetry with central charge $k_{\mathfrak{so}_{2k}}=4k+8$. The structure of the puncture in \eqref{classSforma} precisely ensures the presence of such symmetry. In fact, this signals out D-type class $\mathcal{S}$ as in all other cases, one would not generically have an $\mathfrak{so}_{2k}$ flavour symmetry\footnote{It is possible to have $\mathfrak{so}$ flavour symmetries also in $A_\text{odd}$ class $\mathcal{S}$ with twisted punctures, but there it would be impossible to generically reproduce the central charge since that requires $2k+6$ columns while the number of columns in that class is always odd.}.

This naturally suggests the choice of D-type class $\mathcal{S}$ and fixes the choice of one puncture to be of the form in \eqref{classSforma}. The remaining two punctures are then motivated from analogy with the A-type case given in \eqref{classSAtype}, where we want to take the other two punctures to consist of three and two columns respectively. However, as the number of columns in D-type class $\mathcal{S}$ is always even, we have added a single box to the end of the second puncture. The resulting form is indeed quite successful in describing the torus reduction of 6d D-type orbi-instanton theories. Specifically, for low values of $k$, one can perform a complete scan over all possible numbers $(n_6,n_5,\dots,n_1,x,m,l,y)$ obeying the restrictions  \eqref{classSrest}. The resulting class $\mathcal{S}$ theories can then be compared against possible 6d SCFTs in this class by matching their properties against the expectations from the 6d SCFTs. We review the specific properties we can match and how in \cref{consistencychecks}. We performed this exhaustive scan of class $\mathcal{S}$ theories for $k=3,4,5$. Most of the resulting theories were bad, in the sense of \cite{Gaiotto:2012uq}, but all cases that were good or ugly matched with the torus reduction of a certain SCFT in this class. For each of these values of $k$, we collect the list of good or ugly $\mathcal{S}$ theories, possible 6d SCFTs and the matching between them in \cref{examples}. Throughout the text, we shall occasionally refer to specific theories in these list as theory $N$, with $N$ being the numbering of the 6d SCFT used in \cref{examples}.

We also note that for all 6d SCFTs in this class where we do find a class $\mathcal{S}$ description for its torus reduction, it is of the form in \eqref{classSform}. However, there are families of 6d SCFTs in this class for which we do not find a class $\mathcal{S}$ description of this form that can match their torus reduction. The matching between 6d SCFTs in this class and D-type class $\mathcal{S}$ theories of this form, including when it exists and when it appears not to exist, would be discussed in detail in the next section. This suggests that there are certain 6d SCFTs in this class whose torus reduction is not described by a theory in class $\mathcal{S}$. Nevertheless, it is still possible for these to be described in D-type class $\mathcal{S}$, but with a different set of punctures than \eqref{classSformb} and \eqref{classSformc}. One notable possibility is to use twisted punctures instead of untwisted ones. We have explored some options in this direction, with negative results so far, but cannot claim to have exhausted all options.

\paragraph{The $s$-labels} We find that in the D-type cases, there is a set of integers given by
\begin{subequations}
    \begin{align}
        &s_1=\left\lceil\frac{\lambda_5}{2}\right\rceil-\left\lfloor\frac{\lambda_6}{2}\right\rfloor\;,\\
        &s_2=\left\lfloor\frac{\lambda_4}{2}\right\rfloor-\left\lceil\frac{\lambda_5}{2}\right\rceil\;,\\
        &s_3=\left\lceil\frac{\lambda_3}{2}\right\rceil-\left\lfloor\frac{\lambda_4}{2}\right\rfloor\;,\\
        &s_4=\left\lfloor\frac{\lambda_2}{2}\right\rfloor-\left\lceil\frac{\lambda_3}{2}\right\rceil\;,\\
        &s_5=\left\lceil\frac{\lambda_1}{2}\right\rceil-\left\lfloor\frac{\lambda_2}{2}\right\rfloor\;,\\
        &s_6=\frac{|\lambda|}{2}-\left\lceil\frac{\lambda_1}{2}\right\rceil-\left\lceil\frac{\mu_1}{2}\right\rceil-\left\lceil\frac{\nu_1}{2}\right\rceil\;,\\
        &s_{4'}=\left\lceil\frac{\mu_1}{2}\right\rceil-\left\lfloor\frac{\mu_2}{2}\right\rfloor\;,\\
        &s_{2'}=\left\lfloor\frac{\mu_2}{2}\right\rfloor-\left\lceil\frac{\mu_3}{2}\right\rceil\;,\\
        &s_{3'}=\left\lceil\frac{\nu_1}{2}\right\rceil-\left\lfloor\frac{\nu_2}{2}\right\rfloor\;,
    \end{align}
\end{subequations}
where $|\lambda|$ is the number of total boxes in any of the three partitions (i.e., $|\lambda|=|\mu|=|\nu|=6n+2\delta$), such that
\begin{equation}
    s_1+2s_2+3s_3+4s_4+5s_5+6s_6+4s_{4'}+2s_{2'}+3s_{3'}=k\label{Kacweightedsum}
\end{equation}
for $\widehat{D}_k$. We shall refer to them as the $s$-labels. Since the weighted sum of the $s_i$ equals $k$, which is similar to the Kac labels in the A-type cases, we shall also call them the Kac-type labels. As to be discussed shortly, the Kac-type labels are related to the excess numbers/balances in the 3d mirror theory descriptions. 

Since the partitions are weakly decreasing, that is, $\lambda_{i+1}\geq\lambda_i$, $\mu_{i+1}\geq\mu_i$, $\nu_{i+1}\geq\nu_i$, it is not hard to see that $s_1,s_2,s_3,s_4,s_5,s_{4'},s_{2'},s_{3'}\geq-1$. There is no lower bound for $s_6$. However, for all the examples of good theories we have, it turns out that $s_6\geq-1$ as well\footnote{If a set of punctures gives a bad (or ugly) theory, then $s_6$ can be smaller than $-1$. For instance, the case with $n_4=1$ and $l=3$ for odd $n$ is a bad theory, and it has $s_6=-2$.}. In general, unlike the Kac labels for the A-type cases which are all non-negative integers, there does not seem to have simple bounds of the $s$-labels here that determine when a theory can be good or ugly.

Moreover, we should emphasise that the $s$-labels do not bijectively correspond to the discrete homomorphisms. In \cref{examples}, we shall use superscripts to distinguish the coincident $s$-labels for different theories\footnote{In the $D_4$ case, $s=(1,0,1,-1,1,-1,1,-1,1)$ corresponds to two homomorphisms. They are cases 16 and 17 (where the numbering is from \cref{D4table} below). We may denote them as $s_\text{I}$ and $s_\text{II}$ respectively.}.

\paragraph{The $m$-labels} In principle, to get the good theories, we can say that $s$ should satisfy the physical conditions from the consistency checks in \cref{consistencychecks} below. Of course, the actual derivation would be too involved. Nevertheless, we notice that in \cite{Frey:2018vpw}, a discrete homomorphism of type D was labelled by some partition $\bm{p}$ together with a Lie algebra $\mathfrak{g}$. The constraints on $\bm{p}$ and $\mathfrak{g}$ can then be translated into the constraints on the $s$-labels. In fact, we find that it would be more convenient to consider another set of integers $m_i$, which we shall refer to as the $m$-labels. They are more directly related to both the 6d quivers and the enumerations using $\bm{p}$ and $\mathfrak{g}$.

In short, the integers $m_i$ are the positions of the hypers in the 6d quiver, counting from the end corresponding to the (trivial) nilpotent orbit. Generically, there are 8 of them coming from the 8 D8-branes in the Type IIA setup (although there are exceptions which need nine $m_i$). The details of the explanations and the rules for the $m$-labels will be given in \cref{6d4drelations}. As we shall see shortly, they are particularly useful when determining the class $\mathcal{S}$ description from the 6d quiver. How the $m$-labels are related to $\bm{p}$ and $\mathfrak{g}$ will be discussed in \cref{PartitionDesc}.

Of course, as the dicyclic groups are not abelian, unlike the A-type cases, we do not expect a set of integers (each corresponding to a node in the affine $E_8$ Dynkin diagram) to fully classify the homomorphisms from these groups to $E_8$. The classification of the discrete homomorphisms from the non-abelian groups to $E_8$ would require more sophisticated labels.

\subsection{Three-Dimensional Mirror Theories}\label{magneticquivers}
One may also consider the 3d mirror theory of the class $\mathcal{S}$ theory compactified on $S^1$. As a star-shaped quiver, each leg would correspond to a puncture. More concretely, the central node is $\mathfrak{d}_{3n+\delta}$, and each leg is a chain of alternating $\mathfrak{d}$- and $\mathfrak{c}$-type nodes. On the leg corresponding to $\lambda$, the rank of the $p^\text{th}$ node is\footnote{The numbering $p$ starts from the central node $\mathfrak{d}_{3n+\delta}$ to the end of the leg, and the central node has $p=1$.}
\begin{equation}
    r_p(\lambda)=\begin{cases}
        \left\lceil\frac{1}{2}\left(|\lambda|-\sum\limits_{i=1}^{p-1}\lambda_i\right)\right\rceil\;,&\text{odd $p$, i.e., $\mathfrak{d}$-type node}\;,\\
        \left\lfloor\frac{1}{2}\left(|\lambda|-\sum\limits_{i=1}^{p-1}\lambda_i\right)\right\rfloor\;,&\text{even $p$, i.e., $\mathfrak{c}$-type node}\;,
    \end{cases}
\end{equation}
and likewise for $\mu$ and $\nu$. The excess number/balance is \cite{Gaiotto:2008ak}
\begin{equation}
    b=\begin{cases}
        N_f-2N+1\;,&\mathfrak{d}_N\text{ node}\;,\\
        N_f-2N-1\;,&\mathfrak{c}_N\text{ node}\;.
    \end{cases}
\end{equation}
We shall also define the modified excess number
\begin{equation}
    s:=N_f-2N
\end{equation}
for any node. It turns out that in the 3d mirror theory
\begin{equation}
    \includegraphics[scale=1.3]{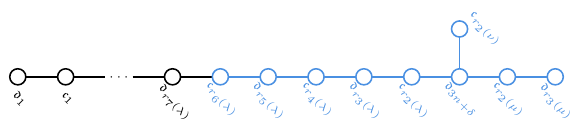}\;,\label{MQ}
\end{equation}
the blue part corresponds to the affine $E_8$ Dynkin diagram with the modified excess numbers equal to the Kac-type labels (and the black part corresponds to the part $1^{2k}$ in $\lambda$). By definition, \eqref{Kacweightedsum} can be written as
\begin{equation}
b_1+2b_2+3b_3+4b_4+5b_5+6b_6+4b_{4'}+2b_{2'}+3b_{3'}=k-2
\end{equation}
in terms of the actual excess numbers. For brevity, we will abbreviate the quiver \eqref{MQ} as
\begin{align}
    \begin{split}
        &{\blue 2r_2(\nu)}\\
        2 \quad 2 \quad \cdots \quad 2r_7(\lambda) \quad {\blue 2r_6(\lambda) \quad 2r_5(\lambda) \quad 2r_4(\lambda) \quad 2r_3(\lambda) \quad 2r_2(\lambda) \quad} &{\blue 6n+2\delta \quad} {\blue 2r_4(\mu) \quad 2r_2(\mu) }
    \end{split}\label{MQabbrev}
\end{align}
where, as indicated in \eqref{MQ}, the $2r_7(\lambda)$ node is of the $\so$-type, the leftmost blue node {\blue $2r_6(\lambda)$} is of the $\usp$-type, and the two types are alternating in the entire quiver.

\subsection{Consistency Checks}\label{consistencychecks}
To test the claims in the paper, we perform a variety of consistency checks that we summarise here. For the examples in the rest part of the paper, we have checked that the quantities listed in this section would agree between the 6d and the 4d computations.

\paragraph{Coulomb branch spectra of operators} One may compute the dimensions of the Coulomb branch operators of the class $\mathcal{S}$ theories. From the 6d generalised quiver for a very Higgsable theory, the prescription was proposed in \cite[Appendix B]{Ohmori:2018ona} (see \cite{Heckman:2022suy} for a generalisation to cases involving an automorphism twist). From the puncture data of the 4d theory, we can use the Tinkertoys. For the untwisted D-type punctures in this paper, this was studied in \cite{Chacaltana:2011ze} (which was also summarised in \cite[Section 2]{Chacaltana:2013oka}). We can then compare the two results and see if they would agree. The Coulomb branch spectrum may be expressed as a polynomial $\sum\limits_\Delta c_\Delta q^\Delta$, indicating $c_\Delta$ operators of dimension $\Delta$.

Since this only concerns the Coulomb branch, it is hard to tell whether a theory is good or ugly from this. Nevertheless, if the theory is bad, there would be negative $c_\Delta$ in the result.

\paragraph{Global symmetries} The global symmetry should be preserved when compactifying an orbi-instanton theory on a torus. In 6d, we can directly read off the global symmetry from the generalised quiver. In 4d, we can compute the Schur index of the class $\mathcal{S}$ theory \cite{Gadde:2011uv,Lemos:2012ph}. At order $\tau^2$, the coefficient is given by the character of the adjoint representation (in the orthogonal basis) of the global symmetry group. For brevity, we shall only list the unrefined index here, where this coefficient gives the dimension of the adjoint representation.

When the theory is ugly, there would be a non-trivial term at order $\tau$, and the coefficient $n_\text{free}$ counts the number of free chirals in the IR. In other words, $n_\text{free}$ is twice the number of the free hypers. We can remove their contributions from the index by dividing it by $\text{PE}\left[n_\text{free}\tau/\left(1-\tau^2\right)\right]$ (where PE is the plethystic exponential). Then the index would become $1+\chi_\text{adj}\tau^2+\dots$. For a bad theory, the index would have irregular coefficients as it would not converge.

\paragraph{Anomalies} If we write the the anomaly polynomial $I_8$ of the 6d theory as in \cite{Ohmori:2014kda}, namely,
\bes{ \label{6dAnomPol}
        I_8 \supset \alpha p_1(T)^2 + \beta p_1(T)c_2(R) + \gamma p_2(T) + \sum_i \kappa_i p_1(T) \text{Tr}\,F_i^2\;,
}
we can obtain the central charges $a$, $c$ and the flavour central charges $k_i$ of the $i^\text{th}$ flavour symmetry of the 4d theory as follows \cite[(6.4)]{Ohmori:2015pua}:
\bes{ \label{anom4d}
a = 24\alpha - 12\beta - 18\gamma\;, \qquad
c = 64\alpha - 12\beta - 8\gamma\;, \qquad
k_i = 192\kappa_i\;.
}
The (quaternionic) dimension of the Higgs branch of the 4d theory is given by\footnote{This follows by matching the $\text{U}(1)_R$ 't Hooft anomalies of the 4d SCFT with those of the low-energy theory on a generic point of the Higgs branch, which the 6d picture suggests should just be a collection of free hypers.}
\begin{equation} \label{dimHiggs4d}
    \dim_\mathbb{H}\mathcal{H}=24(c-a)=-1440\gamma\;.
\end{equation}
The anomalies for some cases are listed explicitly in \cref{sec:embeddingcyclic}, in particular for the embeddings via the cyclic groups.

From the 3d mirror theory, whose Coulomb branch (aka moduli space of dressed monopole operators of the magnetic quiver) is the same as the Higgs branch of the corresponding 4d theory as a space, we can also obtain the dimension. It is equal to the sum of the ranks of the nodes in the orthosymplectic mirror theory.

\section{Relations between 6d Quivers and 4d Class \texorpdfstring{$\mathcal{S}$}{S} Theories}\label{6d4drelations}
We have discussed the punctures of the class $\mathcal{S}$ theories in the previous section, along with some examples. As mentioned, we can perform an exhaustive scan over the class $\mathcal{S}$ theories with the proposed form for low values of $k$. These can then be matched against the torus reduction of the 6d SCFTs using the consistency checks discussed in the previous section. One can then build upon these results and uncover the general relation between the 6d SCFTs, described via the 6d low-energy quiver gauge theories, and the 4d class $\mathcal{S}$ theories resulting from their torus reductions. The resulting proposals can then be checked using myriad examples, where we construct a 6d SCFT and its associated 4d class $\mathcal{S}$ theory and check that their properties indeed match the expectations from \cref{consistencychecks}. Additionally, as the theories are related to each other by Higgsing, it is possible to leverage the obtained descriptions to acquire similar proposals even for 6d SCFTs of types appearing only at higher values of $k$ than those used in the exhaustive scan. The end results of this endeavour are explicit relations between the form of the 6d quiver, described by $n$ M5-branes probing a $\mathbb{C}^2/\widehat{D}_k$ singularity in the presence of an M9-plane, and the 4d class $\mathcal{S}$ theory resulting from its torus reduction, which is systematically discussed below. As previously mentioned, these relations are checked using the consistency checks in \cref{consistencychecks}, at least for a number of specific examples which are also used in the text below to illustrate the relations. We shall not write this for each case to avoid cluttering, so we shall stress here instead that we explicitly checked that every case appearing passes all the consistency checks in \cref{consistencychecks}.

\subsection{Two Basic Types}\label{twotypes}
The 6d quivers in the D type orbi-instanton class for which we find a D-type class $\mathcal{S}$ description can be broken into two two basic types, containing the majority of cases, and a disparate list of exceptions. The different types are distinguished by the behaviour at the leftmost end of the quiver (this edge is the one chosen here to correspond to the intersection with the M9-plane). Here we shall describe the relation between the 6d quiver and 4d class $\mathcal{S}$ theory for the two basic types, while the exceptions are discussed in a later subsection. 

\subsubsection{An \texorpdfstring{$\mathfrak{su}$}{su} End and a \texorpdfstring{$\mathfrak{usp}$}{usp} End}\label{suusp}
Let us begin with the class of 6d quivers specified by a long tail of $\mathfrak{so}$-$\mathfrak{usp}$ alternating quiver ending with an $\mathfrak{su}$ gauge algebra on a $-2$ curve on the leftmost end, and a $\mathfrak{usp}_{2k-8}$ gauge algebra with $k$ flavours on the rightmost end. Once the flavours are specified, the ranks of the groups are completely determined by the anomaly cancellation condition. In general, besides the $2k$ half-hypers on the rightmost end, we have $8$ half-hypers distributed somewhere along the quiver\footnote{There are certain cases where we have more than 8 flavours in the quiver. This happens, for instance, in cases 2 and 10 for $k=5$ as in \cref{D5table}. These cases require separate treatments and are considered later in this section.} (they become full hypers on the $\mathfrak{su}$ decoration). These arise from the M9-plane, which contains 8 D8-branes when reduced to IIA (they give half-hypers on the $\mathfrak{so}$-$\mathfrak{usp}$ due to the orientifold, but become full hypers on $\mathfrak{su}$ as they meet their image there).

We can then specify the quiver by the location of the flavours, in addition to $k$ and the number of tensors $n$ (note that $n$ is always even in this type of quivers). Specifically, we number the gauge algebras with the rightmost $\mathfrak{usp}_{2k-8}$ having number 1, the adjacent $\mathfrak{so}$ on its left having number 2, and so on until the leftmost $\mathfrak{su}$ which has number $n$. 

Going from right to left, we then use $m_1$ for the number of gauge algebras under which the first fundamental half-hyper is charged, $m_2$ is then used for the number of gauge algebras under which the second rightmost half-hyper is charged, and so forth for $m_{3,\dots,8}$. For flavours on $\mathfrak{su}$, one counts full hypers rather than half-hypers. We remark that these $m_i$'s will later enter the puncture data for our class $\mathcal{S}$ description.

Let us illustrate this with some examples.
\begin{itemize}
    \item The case
\bes{
[2]\,\, {\overset{\mathfrak{su}_7}2}  \,\,\underset{[2]}{\overset{\mathfrak{usp}_{12}}1} \,\, \overset{\mathfrak{so}_{22}}4 \,\, \underset{[1]}{\overset{\mathfrak{usp}_{16}}1} \,\, \overset{\mathfrak{so}_{24}}4 \,\,\overset{\mathfrak{usp}_{16}}1 \,\, [12]
} 
has $m_8=m_7=6$ (the leftmost $2$ hypers $[2]$ are charged under $\mathfrak{su}_7$), $m_6=m_5=m_4=m_3=5$ (the other $4$ half-hypers $[2]$ are charged under $\usp_{12}$), $m_2=m_1=3$ (the $2$ half-hypers $[1]$ are charged under the second $\usp_{16}$ which is the third gauge algebra from the right).
\item The case
\bes{
{\overset{\mathfrak{su}_5}2}  \,\,\underset{[2]}{\overset{\mathfrak{usp}_{10}}1} \,\, \underset{[1]}{\overset{\mathfrak{so}_{22}}4} \,\, \underset{[1]}{\overset{\mathfrak{usp}_{16}}1} \,\, \overset{\mathfrak{so}_{24}}4 \,\,\overset{\mathfrak{usp}_{16}}1 \,\, [12]
} 
has $m_8=m_7=m_6=m_5=5$ (the $4$ half-hypers $[2]$ are associated with $\usp_{10}$), $m_4=m_3=4$ (the left $2$ half-hypers $[1]$ below $\so_{22}$ are charged under $\so_{22}$), $m_2=m_1=3$ (the right $2$ half-hypers $[1]$ below $\usp_{16}$ are charged under the second $\usp_{16}$ from the right).
\item The case
\bes{
[6]\,\, {\overset{\mathfrak{su}_9}2}  \,\,\underset{[1]}{\overset{\mathfrak{usp}_{12}}1} \,\, \overset{\mathfrak{so}_{20}}4 \,\, \overset{\mathfrak{usp}_{12}}1 \,\, [10]
} 
has $m_8=m_7=m_6=m_5=m_4=m_3=4$ (the leftmost $6$ hypers $[6]$ are associated with $\mathfrak{su}_9$), $m_2=m_1=3$ (the $2$ half-hypers $[1]$ are associated with the second $\usp_{12}$ from the right).
\end{itemize}
Here, an E-string theory is treated as a $\mathfrak{usp}_0$, and an empty $-2$ curve is treated as $\mathfrak{su}_1$. For instance,
\bes{
[4]\,\, {\overset{\mathfrak{su}_2}2}  \,\, 1 \,\, \overset{\mathfrak{so}_9}4 \,\, \underset{\left[\frac{1}{2}\right]}{\overset{\mathfrak{usp}_2}1} \,\, \overset{\mathfrak{so}_{10}}4 \,\, \overset{\mathfrak{usp}_2}1 \,\, [5]
} 
has $m_8=m_7=m_6=m_5=6$ (the leftmost $4$ hypers $\left[4 \right]$ are associated with $\su_{2}$), $m_4=m_3=m_2=5$ (the empty $-1$ is treated as a $\mathfrak{usp}_0$, so it needs $16$ half-hypers, $13$ of which it ``receives" from the adjacent nodes, leaving $3$ hypers that are then associated with $\usp_{0}$), $m_1=3$ (the $1$ half-hyper $\left[\frac{1}{2} \right]$ is associated with the second $\usp_{2}$ from the right). Likewise,
\bes{
2  \,\, \underset{[4]}{\overset{\mathfrak{usp}_2}1} \,\, \overset{\mathfrak{so}_{10}}4 \,\, \overset{\mathfrak{usp}_2}1 \,\, \overset{\mathfrak{so}_{10}}4 \,\, \overset{\mathfrak{usp}_2}1 \,\, [5]
} 
has $m_8=m_7=m_6=m_5=m_4=m_3=m_2=m_1=5$ (the $8$ half-hypers $\left[ 4 \right]$ are associated with the leftmost $\usp_{2}$). Note that here the empty $-2$ is treated as having an $\su_1$ gauge group, so it needs two hypers which it ``receives" from the adjacent $\su_2$ gauge algebra. Hence, there are no flavours associated with this node.

\subsubsection{A Bifurcating End and a \texorpdfstring{$\mathfrak{usp}$}{usp} End}\label{bifurusp}
Another type of quivers we shall consider consists again of an $\mathfrak{so}$-$\mathfrak{usp}$ alternating quiver with a $\mathfrak{usp}_{2k-8}$ gauge algebra with $k$ flavours on the rightmost end, but with a bifurcation of two $-1$ curves on the leftmost end. As before, the quiver has $8$ half-hypers distributed somewhere along it (becoming $8$ pairs of half-hypers on the bifurcation point), arising from the M9-plane.

The quiver can then be specified by the location of the flavours, in addition to $k$ and the number of tensors, which we take to be $n+1$ (note that $n$ is always odd then). The rules for specifying the flavours are the same as before, except for the bifurcation point. Here, the number of flavours is the average number of half-hypers on both $-1$ curves (sum of the half-hypers divided by $2$). We will also denote by $\Delta n$ the difference in the ranks of the gauge algebras on the two $-1$ curves (ordered such that $\Delta n \geq 0$). This together with the $m_i$'s is sufficient to specify the quiver. Finally, we note that the flavours on the bifurcation point are awarded the number $n$, and that as before, an empty $-1$ curve will be regarded as $\mathfrak{usp}_0$ for flavour computations.

Let us illustrate this with some examples.
\begin{itemize}
    \item The case
\bes{ 
[2] \,\, \overset{\mathfrak{usp}_2}1 \,\, \overset{[2]}{\underset{[2]}{\underset{1}{\underset{\mathfrak{usp}_2}{\overset{\mathfrak{so}_{16}}4}}}} \,\, \overset{\mathfrak{usp}_8}1\,\, \overset{\mathfrak{so}_{16}}4\,\, \overset{\mathfrak{usp}_8}1 \,\, [8]
}
has $m_8=m_7=m_6=m_5=5$ (the two $\usp_2$ gauge algebras at the bifurcation have a total of $8$ half-hypers) and $m_4=m_3=m_2=m_1=4$ (the $4$ vector half-hypers are associated to the $[2]$ of the $\so_{16}$). Note that here we have 6 tensors, so $n=5$, and that $\Delta n=0$.
\item The case
\bes{ 
[4] \,\, \overset{\mathfrak{usp}_4}1 \,\, \overset{[2]}{\underset{1}{\overset{\mathfrak{so}_{16}}4}} \,\, \overset{\mathfrak{usp}_8}1\,\, \overset{\mathfrak{so}_{16}}4\,\, \overset{\mathfrak{usp}_8}1 \,\, [8]
}
has $m_8=m_7=m_6=m_5=5$ (the $\usp_4$ gauge algebra has 8 half-hypers while the empty $-1$ has none as it gets the needed 16 half-hypers from the adjacent $-4$ curve), $m_4=m_3=m_2=m_1=4$ (the $4$ vector half-hypers are associated to the $[2]$ of the $\so_{16}$). There are 6 tensors, so $n=5$. However, note that now $\Delta n=2$.
\item The case
\bes{ 
[2] \,\, \overset{\mathfrak{usp}_4}1 \,\, \underset{1}{\underset{\mathfrak{usp}_2}{\overset{\mathfrak{so}_{20}}4}} \,\, \underset{\left[\frac{3}{2}\right]}{\overset{\mathfrak{usp}_{18}}1}\,\, \underset{[1]}{\overset{\mathfrak{so}_{29}}4}\,\, \underset{\left[\frac{1}{2}\right]}{\overset{\mathfrak{usp}_{22}}1} \overset{\mathfrak{so}_{30}}4\,\, \overset{\mathfrak{usp}_{22}}1 \,\, [15]
}
has $k=15$, $n=7$, $\Delta n=1$, $m_1=3$, $m_2=m_3=4$, $m_4=m_5=m_6=5$, and $m_7=m_8=7$.
\end{itemize}

\paragraph{Two variants} It is possible to include in this class also the following cases\footnote{This is essentially as these cases that can be generated by Higgsing an empty $-1$ curve. Then it appears that formally this can be taken into account by assigning a $\mathfrak{usp}_{-2}$ gauge algebra to it.}.
\begin{itemize}
    \item The leftmost end has the configuration with $-1$, $-3$ curves. One regards this as a bifurcation at the $-3$ curve, where on the two $-1$ curves, one has a $\mathfrak{usp}_{-2}$ gauge algebra and one has none\footnote{We remark that a similar prescription of extrapolating to $\mathfrak{usp}_{-2}$ here and $\mathfrak{usp}_{-4}$ below is also used in  \cite{Hassler:2019eso} to study T-brane Higgsings of 6d SCFTs.}.
    \item The leftmost end has a $-2$ curve with an $\mathfrak{so}$ gauge algebra. It is regarded as a bifurcation, where each of the two $-1$ curves has a $\mathfrak{usp}_{-2}$ gauge algebra.
\end{itemize}

We note several details about this class.
\begin{itemize}
    \item The rest of the parameters are determined as before. Note that the number of tensors will be equal to $n$ in the $-1$, $-3$ case, and to $n-1$ in the $-2$ with an $\mathfrak{so}$ case as we have fewer $-1$ curves. The $m_i$'s are also determined as before where the $\mathfrak{usp}_{-2}$ would have $6$ hypers. Additionally, for the $\mathfrak{so}$ on the $-3$ or the $-2$ curve, we ignore the spinor matters when counting the flavours (so only the vectors are counted).
    \item It is important that all bifundamentals are in the vector representations of the $\mathfrak{so}$ gauge algebras. As such, quivers containing bifundamentals in the spinors of $\mathfrak{so}_7$, or for $\mathfrak{so}_8$ bifundamentals that cannot all be made to be in the vector (by a suitable triality transformation), do not belong to this class.
    \item It appears that one can also include cases where there is a $\mathfrak{g}_2$ on the $-3$ or the $-2$ curve. In this case, one regards one of the $-1$ curves as having a $\mathfrak{usp}_{-4}$ gauge algebra. To count the flavours on $\mathfrak{g}_2$, one counts the number of half-hypers in the $-3$ curve case, and the number of half-hypers minus 4 in the $-2$ curve case. It further appears that formally an $\mathfrak{su}_3$ on the $-3$ curve can also be described by regarding the eliminated $-1$ curve as having a $\mathfrak{usp}_{-4}$ gauge algebra. For this case, there are no flavours on this $\mathfrak{su}_3$.   
\end{itemize}

Let us illustrate this with some examples.
\begin{itemize}
    \item Consider the case
\bes{
1\,\, \underset{\left[\frac{1}{2}{\bf 32}\right][1]}{\overset{\mathfrak{so}_{12}}3}  \,\,\underset{[2]}{\overset{\mathfrak{usp}_8}1} \,\, \overset{\mathfrak{so}_{16}}4 \,\,\overset{\mathfrak{usp}_8}1 \,\, [8]\;.
}
According to the above prescription, we replace the spinor half-hyper of the $\so_{12}$ gauge algebra by an ``imaginary'' $-1$ curve as follows:
\bes{
\underset{\red [2]}{1}\,\, \overset{\overset{\red \usp_{-2}}{1}}{ \underset{[1]} {\overset{\mathfrak{so}_{12}}3} } \,\,\underset{[2]}{\overset{\mathfrak{usp}_8}1} \,\, \overset{\mathfrak{so}_{16}}4 \,\,\overset{\mathfrak{usp}_8}1 \,\, [8]\;.
}
We have $k=8$, $n=5$, and $\Delta n=1$ (one empty $-1$ and one $-1$ with $\mathfrak{usp}_{-2}$ give $\Delta n=0-(-1)=1$). We also have $m_1=m_2=m_3=m_4=3$ (four half-hypers of $\mathfrak{usp}_8$), $m_5=m_6=4$ (two vector half-hypers of $\mathfrak{so}_{12}$, where we ignore the spinor half-hyper) and $m_7=m_8=5$. Let us further remark on $m_7$ and $m_8$. Here, we have an empty $-1$ curve and one with $\mathfrak{usp}_{-2}$. The former needs $8$ hypers while the latter needs $6$. The $\mathfrak{so}_{12}$ provides $6$, satisfying the $\mathfrak{usp}_{-2}$ one, but we still need $2$ more for the empty one, which would then give $m_7$ and $m_8$. We illustrate the ``imaginary'' $-1$ curve, along with these two hypers, in {\red red} in the above diagram.
\item Consider the case
\bes{
[2{\bf 16}]\,\, \overset{\mathfrak{so}_{10}}2  \,\,\underset{[3]}{\overset{\mathfrak{usp}_8}1} \,\, \overset{\mathfrak{so}_{16}}4 \,\,\overset{\mathfrak{usp}_8}1 \,\, [8]\;.
}
Here, we have $k=8$ and $n=5$ (again we treat this as having an ``imaginary'' bifurcation at the end supporting two $-1$ curves with one $\mathfrak{usp}_{-2}$), but $\Delta n=0$ as illustrated below:
\bes{
\underset{\red [1]}{\overset{\red \usp_{-2}}{1}} \,\, \overset{\overset{\red [1]}{\overset{\red \usp_{-2}}{1}}}{\overset{\mathfrak{so}_{10}}2}  \,\,\underset{[3]}{\overset{\mathfrak{usp}_8}1} \,\, \overset{\mathfrak{so}_{16}}4 \,\,\overset{\mathfrak{usp}_8}1 \,\, [8]\;.
}
We also have $m_1=m_2=m_3=m_4=m_5=m_6=3$ (six half-hypers of $\mathfrak{usp}_8$) and $m_7=m_8=5$. Let us further remark on $m_7$ and $m_8$. Note that $m_7$ and $m_8$ essentially come from the ``imaginary'' bifurcation, depicted in {\red red} in the above diagram. Specifically, we replace the spinors with two $-1$ curves, each supporting a $\mathfrak{usp}_{-2}$. The latter requires $6$ flavours, but it only gets $5$ from the $\mathfrak{so}_{10}$ gauge algebra. As such, each one should be considered as having an additional flavour, which would give $m_7$ and $m_8$.
\item Consider the case
\bes{
[4]\,\, \overset{\mathfrak{g}_2}2  \,\, 1 \,\, \overset{\mathfrak{so}_9}4\,\, \overset{\mathfrak{usp}_2}1 \,\, \overset{\mathfrak{so}_{11}}4\,\, \underset{\left[\frac{1}{2}\right]}{\overset{\mathfrak{usp}_4}1} \,\, \overset{\mathfrak{so}_{12}}4 \,\,\overset{\mathfrak{usp}_4}1 \,\, [6]\;.
}
Here, we have $k=6$, $n=9$, and $\Delta n=1$ (one $-1$ with $\mathfrak{usp}_{-2}$ and one with $\mathfrak{usp}_{-4}$ give $\Delta n=-1-(-2)=1$). We also have $m_1=3$ (half-hypers of $\mathfrak{usp}_4$), $m_2=m_3=m_4=m_5=8$ and $m_6=m_7=m_8=9$. For the $\mathfrak{g}_2$, we count the number of half-hypers minus 4. As it has $8$ half-hypers, we assign to it 4 flavours which give $m_{2,\dots,5}$. Finally, we have the hidden $-1$ curves, where the one with $\mathfrak{usp}_{-2}$ needs 6 flavours and the one with $\mathfrak{usp}_{-4}$ needs 4, as illustrated in {\red red} below:
\bes{
\underset{\red \left[\frac{1}{2} \right]}{\overset{\red \mathfrak{\usp_{-4}}}1}\,\, \overset{\overset{\red \left[\frac{5}{2}\right]}{\overset{\red \usp_{-2}}{1}}}{\underset{[4]}{\overset{\mathfrak{g}_2}2}}  \,\, 1 \,\, \overset{\mathfrak{so}_9}4\,\, \overset{\mathfrak{usp}_2}1 \,\, \overset{\mathfrak{so}_{11}}4\,\, \underset{[\frac{1}{2}]}{\overset{\mathfrak{usp}_4}1} \,\, \overset{\mathfrak{so}_{12}}4 \,\,\overset{\mathfrak{usp}_4}1 \,\, [6]\;.
}
As both see exactly 7 half-hypers from $\mathfrak{g}_2$, we see that we are missing a total of 6 half-hypers (one from the $\mathfrak{usp}_{-4}$ one and five from the $\mathfrak{usp}_{-2}$ one), which would give $m_{6,\dots,8}$.
\end{itemize}

\subsubsection{Class \texorpdfstring{$\mathcal{S}$}{S} Theory Descriptions}\label{classSfortwobasictypes}
Now, we present the 4d class $\mathcal{S}$ theory associated with the 6d quiver. For this, we define
\begin{equation} \label{MandDm}
    M=\frac{1}{2}(m_6+m_5+m_4+m_3+m_2+m_1)\;,\quad\Delta m_{\pm} = \frac{1}{2}(m_8 \pm m_7)\;.
\end{equation}
The 4d class theory is then given by the following partitions. For $M+k+\Delta m_{+} = \text{odd}$, we have
\bes{ \label{ClassSpres1}
& \lambda=\left[m_6,m_5,m_4,m_3,m_2,m_1,1^{2k}\right]\;, \\
& \mu=\left[M+k+\Delta m_{-}-n-1, M+k-\Delta m_{-}-n-1,2n+1,1\right]\;, \\
& \nu=\left[M+k+n-\Delta m_{+},M+k-n+\Delta m_{+}\right]\;.
}
For $M+k+\Delta m_{+} = \text{even}$, we have
\bes{ \label{ClassSpres2}
& \lambda=\left[m_6,m_5,m_4,m_3,m_2,m_1,1^{2k}\right]\;, \\
& \mu=\left[M+k-\Delta m_{+}-1, M+k+\Delta m_{+}-2n-1,2n+1,1\right]\;, \\
& \nu=\left[M+k+\Delta m_{-},M+k-\Delta m_{-}\right]\;.
}
There are several notes on the above prescription:
\begin{itemize}
    \item The distinction between $M+k+\Delta m_{+}$ being even or odd is only relevant for $n$ even, where only one case yields a consistent class $\mathcal{S}$ theory. The case of $n$ odd is more interesting. Here, if $m_8=m_7=n$, then the two cases lead to the same class $\mathcal{S}$ theory, and the distinction is irrelevant. However, when $m_8,m_7<n$, we always have $M+k+\Delta m_{+}$ being even\footnote{One can show that in these models, we have $k=2r+\frac{1}{2}\sum\limits_{i=1}^8(n+1-m_i)$, where $r$ is the rank of the $\mathfrak{usp}$ at the bifurcation. As such, $M+k+\Delta m_{+} = 2r+4n+4$ and is always even in this class.}, and both cases make sense. In these cases, the same 6d quiver actually corresponds to two different 6d SCFTs, differing by the choice of the $\theta$ angles. See the discussions about this in \cref{thetaangle}.
    \item In certain cases, $M$ can be fractional. The above prescription would then yield an ill-defined theory. In these cases, one should repeat this but with $m_6 \leftrightarrow m_8$. See \cref{nonintegerM} for further details.
    \item While $\Delta n$ does not explicitly appear, it is important to note that the above theory only holds if $\Delta n \leq 1$. Cases with $\Delta n > 1$ appear to not have a 4d D-type class $\mathcal{S}$ description (there are some subtitles as will be discussed below).
    \item Cases with non-integer $M$ give either bad or ugly class $\mathcal{S}$ theories while quivers with $\mathfrak{g}_2$ appear to give ugly class $\mathcal{S}$ theories. Otherwise, the theories appear to be good.
\end{itemize}

Let us illustrate this with a few examples.
\begin{itemize}
    \item Consider the case
\bes{
[4]\,\, {\overset{\mathfrak{su}_3}2}  \,\,\underset{[2]}{\overset{\mathfrak{usp}_2}1} \,\, \overset{\mathfrak{so}_{10}}4 \,\, \overset{\mathfrak{usp}_2}1 \,\, [5]\;.
} 
Here, we have $k=5$, $n=4$, $m_8=m_7=m_6=m_5=4$, $m_4=m_3=m_2=m_1=3$. We see that $M=10$, $\Delta m_+ = 4$, $\Delta m_- = 0$, and the 4d class $\mathcal{S}$ theory is $\left[4^2,3^4,1^{10}\right]$, $\left[15^2\right]$, $\left[10^2,9,1\right]$. This indeed agrees with theory 36 in Appendix \ref{D5} as expected.\footnote{Here and subsequently, we refer to theories with flavour symmetry algebra $D_3$, $D_4$ or $D_5$ with the labels as indicated in Appendix \ref{examples}.}
\item Similarly, we see that
\bes{
[4]\,\, {\overset{\mathfrak{su}_2}2}  \,\, 1 \,\, \overset{\mathfrak{so}_9}4 \,\, \underset{\left[\frac{1}{2}\right]}{\overset{\mathfrak{usp}_2}1} \,\, \overset{\mathfrak{so}_{10}}4 \,\, \overset{\mathfrak{usp}_2}1 \,\, [5]
}
has $k=5, n=6$, $m_8=m_7=m_6=m_5=6$, $m_4=m_3=m_2=5$, $m_1=3$, and as such, $M=15$, $\Delta m_+ = 6$, $\Delta m_- = 0$. The 4d class $\mathcal{S}$ theory is $\left[6^2,5^3,3,1^{10}\right]$, $\left[20^2\right]$, $\left[13^3,1\right]$. This indeed agrees with theory 39 in Appendix \ref{D5} as expected.
\item Consider instead the case
\bes{
\overset{\mathfrak{su}_2}2  \,\, \underset{\left[\frac{1}{2}\right]}{\overset{\mathfrak{usp}_4}1} \,\, \overset{\mathfrak{so}_{19}}4 \,\, \underset{\left[\frac{1}{2}\right]}{\overset{\mathfrak{usp}_{18}}1} \,\, \underset{[3]}{\overset{\mathfrak{so}_{32}}4} \,\, \overset{\mathfrak{usp}_{24}}1 \,\, [16]
}
corresponding to $k=16, n=6$, $m_8=5$, $m_7=3$, $m_6=m_5=m_4=m_3=m_2=m_1=2$. We then have $M=6$, $\Delta m_+ = 4$, $\Delta m_- = 1$, and the 4d class $\mathcal{S}$ theory is $\left[2^6,1^{32}\right]$, $\left[23,21\right]$, $\left[17,13^2,1\right]$. One can see that the global symmetry is indeed $\USp(6)\times\text{U}(1)\times\SO(32)$ as expected.
\item Similarly, consider the case
\bes{
\overset{\mathfrak{su}_2}2  \,\, \overset{\mathfrak{usp}_4}1 \,\, \overset{\mathfrak{so}_{20}}4 \,\, \underset{\left[\frac{3}{2}\right]}{\overset{\mathfrak{usp}_{20}}1} \,\, \underset{[2]}{\overset{\mathfrak{so}_{33}}4} \,\, \underset{\left[\frac{1}{2}\right]}{\overset{\mathfrak{usp}_{26}}1} \,\, \overset{\mathfrak{so}_{34}}4 \,\, \overset{\mathfrak{usp}_{26}}1 \,\, [17]
}
corresponding to $k=17, n=8$, $m_8=m_7=m_6=5$, $m_5=m_4=m_3=m_2= 4$, $m_1=3$. We then have $M=12$, $\Delta m_+ = 5$, $\Delta m_- = 0$, and the 4d class $\mathcal{S}$ theory is $\left[5,4^4,3,1^{34}\right]$, $\left[29^2\right]$, $\left[23,17^2,1\right]$. One can see that the global symmetry is indeed $\SO(3)\times\USp(4)\times\text{U}(1)\times\SO(34)$ as expected.
\item Consider the case
\bes{ 
1 \,\, \underset{1}{\overset{\mathfrak{so}_9}4} \,\, \underset{\left[\frac{1}{2}\right]}{\overset{\mathfrak{usp}_2}1}\,\, \overset{\mathfrak{so}_{10}}4\,\, \overset{\mathfrak{usp}_2}1 \,\, [5]
}
corresponding to $k=5$, $n=5$ and $\Delta n = 0$. We also have $m_1=3$, $m_2=m_3=m_4=m_5=m_6=m_7=m_8=5$ (both empty $-1$ curves are missing $7$ half-hypers). As such, we have $M=14$, $\Delta m_+ = 5$, $\Delta m_- = 0$, and the 4d class $\mathcal{S}$ theory is $\left[5^5,3,1^{10}\right]$, $\left[19^2\right]$, $\left[13^2,11,1\right]$. This indeed agrees with theory 17 in Appendix \ref{D5} as expected.
\item Consider the case
\bes{
1\,\, \underset{[1{\bf 16}][2]}{\overset{\mathfrak{so}_{10}}3}  \,\,\overset{\mathfrak{usp}_2}1 \,\, \overset{\mathfrak{so}_{10}}4 \,\,\overset{\mathfrak{usp}_2}1 \,\, [5]
}
corresponding to $k=5$, $n=5$ and $\Delta n = 1$. We have $m_1=m_2=m_3=m_4=4$ and $m_5=m_6=m_7=m_8=5$ (the empty $-1$ is missing 3 flavours while the $\mathfrak{usp}_{-2}$ is missing 1). As such, we have $M=13$, $\Delta m_+ = 5$, $\Delta m_- = 0$, and the 4d class $\mathcal{S}$ theory is $\left[5^2,4^4,1^{10}\right]$, $\left[18^2\right]$, $\left[12^2,11,1\right]$. This agrees with theory 33 in Appendix \ref{D5} as expected.
\item Consider the case
\bes{
[2{\bf 16}][2]\,\, \overset{\mathfrak{so}_9}2  \,\, \underset{\left[\frac{1}{2}\right]}{\overset{\mathfrak{usp}_2}1} \,\, \overset{\mathfrak{so}_{10}}4\,\, \overset{\mathfrak{usp}_2}1 \,\, [5]
}
corresponding to $k=5$, $n=5$ and $\Delta n = 0$. We have $m_1=3$, $m_2=m_3=m_4=m_5=4$ and $m_6=m_7=m_8=5$ (both $-1$ curves with $\mathfrak{usp}_{-2}$ are missing 3 half-hypers). As such, we have $M=12$, $\Delta m_+ = 5$, $\Delta m_- = 0$, and the 4d class $\mathcal{S}$ theory is $\left[5,4^4,3,1^{10}\right]$, $\left[17^2\right]$, $\left[11^3,1\right]$. This agrees with theory 27 in Appendix \ref{D5} as expected.
\item Consider the case
\bes{
[3]\,\, \overset{\mathfrak{g}_2}2  \,\, \underset{\left[\frac{3}{2}\right]}{\overset{\mathfrak{usp}_2}1} \,\, \overset{\mathfrak{so}_{10}}4\,\, \overset{\mathfrak{usp}_2}1 \,\, [5]
}
corresponding to $k=5$, $n=5$ and $\Delta n = 1$. We have $m_1=m_2=m_3=3$, $m_4=m_5=4$ and $m_6=m_7=m_8=5$ (the $-1$ curve with $\mathfrak{usp}_{-2}$ is missing 5 half-hypers, and the one with $\mathfrak{usp}_{-4}$ is missing 1). As such, we have that $M=11$, $\Delta m_+ = 5$, $\Delta m_- = 0$, and the 4d class $\mathcal{S}$ theory is $\left[5,4^2,3^3,1^{10}\right]$, $\left[16^2\right]$, $\left[11,10^2,1\right]$. This agrees with theory 43 in Appendix \ref{D5} as expected. Note that this is an ugly class $\mathcal{S}$ theory.
\item Consider the case
\bes{ 
\left[\frac{1}{2}\right] \,\, \overset{\mathfrak{usp}_2}1 \,\, \underset{\underset{\left[\frac{1}{2}\right]}{\overset{\mathfrak{usp}_2}1}}{\overset{\mathfrak{so}_{19}}4} \,\, \underset{\left[\frac{5}{2}\right]}{\overset{\mathfrak{usp}_{18}}1}\,\, \underset{[1]}{\overset{\mathfrak{so}_{28}}4}\,\, \overset{\mathfrak{usp}_{20}}1 \,\, \overset{\mathfrak{so}_{28}}4\,\, \overset{\mathfrak{usp}_{20}}1 \,\, [14]
}
corresponding to $k=14$, $n=7$ and $\Delta n = 0$. We also have $m_1=m_2=4$, $m_3=m_4=m_5=m_6=m_7=5$ and $m_8=7$. As such, we have $M=14$, $\Delta m_+ = 6$, $\Delta m_- = 1$, and the 4d class $\mathcal{S}$ theory is $\left[5^4,4^2,1^{28}\right]$, $\left[29,27\right]$, $\left[21,19,15,1\right]$. One can see that the global symmetry of this theory is $\USp(4)\times\SU(2)\times\SO(28)$ as expected.
\item Consider the case
\bes{ \label{Biexampleso12}
1\,\, \underset{\left[\frac{1}{2}{\bf 32}\right]}{\overset{\mathfrak{so}_{12}}3}  \,\,\underset{[3]}{\overset{\mathfrak{usp}_{10}}1} \,\, \overset{\mathfrak{so}_{18}}4 \,\,\overset{\mathfrak{usp}_{10}}1 \,\, [9]
}
corresponding to $k=9$, $n=5$ and $\Delta n = 1$. We have $m_1=m_2=m_3=m_4=m_5=m_6=3$ and $m_7=m_8=5$ (the $-1$ curve with $\mathfrak{usp}_0$ is missing 4 half-hypers). As such, we have that $M=9$, $\Delta m_+ = 5$, $\Delta m_- = 0$, and the 4d class $\mathcal{S}$ theory is $\left[3^6,1^{18}\right]$, $\left[18^2\right]$, $\left[12^2,11,1\right]$. One can check that the global symmetry of this theory is $\SU(4)\times\SU(2)\times\SU(2)\times\SO(18)$ as expected.
\item Consider the case
\bes{
[2]\,\, \overset{\mathfrak{g}_2}2  \,\, \underset{\left[\frac{3}{2}\right]}{\overset{\mathfrak{usp}_4}1} \,\, \underset{[1]}{\overset{\mathfrak{so}_{14}}4}\,\, \overset{\mathfrak{usp}_6}1 \,\, [7]
}
corresponding to $k=7$, $n=5$ and $\Delta n = 1$. We have $m_1=m_2=2$, $m_3=m_4=m_5=3$ and $m_6=m_7=m_8=5$. As such, we have that $M=9$, $\Delta m_+ = 5$, $\Delta m_- = 0$, and the 4d class $\mathcal{S}$ theory is $\left[5,3^3,2^2,1^{14}\right]$, $\left[16^2\right]$, $\left[11,10^2,1\right]$. Note that this is an ugly class $\mathcal{S}$ theory, but still the global symmetry appears to be $\USp(4)\times\SU(2)\times\SU(2)\times\SO(14)$ as expected.
\end{itemize}

\subsubsection{Cases Differing by the 6d \texorpdfstring{$\theta$}{theta} Angle}\label{thetaangle}
A special feature of $\USp$ in 6d is that they allow a $\mathbb{Z}_2$-valued $\theta$ angle, owning to the fact that $\pi_5 (\USp) = \mathbb{Z}_2$. Here, we shall briefly summarise its effect, referring to \cite{Mekareeya:2017jgc} for more information. The $\theta$ angle only affects the theory through its non-perturbative spectrum, and as such, gauge theories differing by the value of the $\theta$ angle appear identical but are UV completed by different 6d SCFTs. Like in 4d, the $\theta$ angle can be absorbed into a fundamental flavour mass term, and as such, is physically irrelevant in the presence of massless flavours. Specifically, we can change the $\theta$ angle by changing the mass sign of a single fundamental flavour, but note that doing this on an even number of flavours will leave the $\theta$ angle unchanged.

As an example, consider the case of a $\mathfrak{usp}_{2r}$ gauge algebra with $2r+8$ fundamental hypers. This theory is known to be UV completed by a 6d SCFT, whose matter spectrum includes a Higgs branch chiral ring generator in a chiral spinor of the $\SO(4r+16)$ flavour symmetry. Note that this matter arises non-perturbatively in the gauge theory. The choice of the chirality of the spinor is determined by the $\theta$ angle of the $\mathfrak{usp}$. Note that we can change the chirality by inverting the mass sign of one of the flavours (parity of $\SO(4r+16)$), so the value of the $\theta$ angle is physically irrelevant in this case as expected. Nevertheless, if we gauge the $\SO(4r+16)$ or part of it so that no individual flavour is left, then we may no longer be able to change the $\theta$ angle. In these cases, it becomes physically relevant, and different angles lead to different 6d SCFTs. We shall next see some examples of this in the class of theories considered here. We also refer the reader to \cite{Mekareeya:2017jgc,Distler:2022yse} for examples in other classes of theories.

All quivers considered here consist of a $\mathfrak{usp}_{2k-8}$ gauge algebra with $k$ hypermultiplets at the rightmost end. The flavours at this end imply that the $\theta$ angle of that group is physically irrelevant. We then have an $\mathfrak{so}$ followed by another $\mathfrak{usp}$. Naturally, if there are fundamental flavours under that $\mathfrak{usp}$, then its $\theta$ angle would also be irrelevant. However, even if there are no flavours, we can still switch its $\theta$ angle by performing a parity transformation on the adjacent $\mathfrak{so}$. Note that this simultaneously changes the $\theta$ angles of the other connected $\mathfrak{usp}$, so we cannot change the $\theta$ angle of an individual group in this way. However, since we have flavours at the first $\mathfrak{usp}$, we can essentially use this to switch the $\theta$ angles of the other $\mathfrak{usp}$ one by one. As such, in most cases, the $\theta$ angles of the $\mathfrak{usp}$ are completely irrelevant, except for one case, where we have a bifurcation.

\paragraph{The bifurcation case} Consider the case where we have a bifurcation at the leftmost end. In this case, performing a parity automorphism on the middle $\mathfrak{so}$ will change both angles simultaneously. If at least one of the $\mathfrak{usp}$ at the bifurcation has fundamental flavours, then both $\theta$ angles are physically irrelevant, as they can be changed through the flavour and gauge parity transformations. However, in the special case when there are no flavours at the bifurcation, corresponding to $m_8<n$, the difference in the $\theta$ angles between the two $\mathfrak{usp}$ is physically relevant and would lead to different 6d SCFTs. We claim that this difference is manifest in the 4d theory by different 4d class $\mathcal{S}$ theories, where the difference corresponds to which of the two options, $M+k+\Delta m_+$ being even or odd, one chooses\footnote{Recall that in such cases, $M+k+\Delta m_+$ is always even. Here, we just use ``odd'' (resp.~``even'') to represent the choice \eqref{ClassSpres1} (resp.~\eqref{ClassSpres2}) for simplicity.}.

We should say a bit more about this difference in the 6d case before discussing the implications on the 4d theories. As mentioned above, the 6d SCFT UV completing a $\mathfrak{usp}_{2r}$ gauge theory with $2r+8$ hypermultiplets has a Higgs branch chiral ring operator in the spinor of the $\SO(4r+16)$ flavour symmetry. This operator is in the $(r+3)$-dimensional representation of $\SU(2)_R$. The choice of chirality of the spinor under the flavour $\SO$ group is then determined by the $\theta$ angle. If the $\theta$ angles of the two $\mathfrak{usp}$ at the bifurcation are identical, then one can make a gauge invariant operator from the two, while this is not possible directly if the angles differ. Thus, the case with identical angles has an extra Higgs branch chiral ring operator in the $(2r+5)$-dimensional representation of $\SU(2)_R$, so the Schur index would differ at order $\tau^{2r+4}$. This in particular distinguishes between the two different 6d SCFTs.

Note that this holds even if $r<1$. Specifically, when $r=0$, we have two rank 1 E-string SCFTs. The $E_8$ moment map then decomposes into ${\bf 120} + {\bf 128}$ under the $\SO(16)$ subgroup, with the latter playing the role of the spinor operator. Furthermore, for $r=-1$, we have $\SO(12)$ as our flavour symmetry, and the $-1$ curves play the role of spinor half-hypers. For $\SO(12)$, there are two spinor chiralities with the same contributions to the gauge anomalies, so both are possible. The $\theta$ angle then distinguishes between the chirality of the spinor. Note that for $r=-1$, the additional Higgs branch operator is in the ${\bf 3}$ of $\SU(2)_R$, which is the charge of a moment map operator. This is because the case of identical angles carries a $\text{U}(1)$ flavour symmetry, rotating the two half-hypers, which is absent in the different angles case. As such, unlike generic $r$, for $r=-1$, the difference between the two SCFTs is manifest at the level of the flavour symmetry algebra.

\paragraph{Examples} Let us illustrate this with some examples.
\begin{itemize}
    \item It is convenient to start with the case of $r=-1$ where the theories are more easily distinguished. Specifically, consider the cases
\bes{
[1 {\bf 32}][1] \,\, \overset{\mathfrak{so}_{12}}2  \,\,\underset{[3]}{\overset{\mathfrak{usp}_{10}}1} \,\, \overset{\mathfrak{so}_{18}}4 \,\,\overset{\mathfrak{usp}_{10}}1 \,\, [9]
}
and
\bes{
\left[\frac{1}{2}{\bf 32}\right]\left[\frac{1}{2}{\bf 32'}\right][1] \,\, \overset{\mathfrak{so}_{12}}2  \,\,\underset{[3]}{\overset{\mathfrak{usp}_{10}}1} \,\, \overset{\mathfrak{so}_{18}}4 \,\,\overset{\mathfrak{usp}_{10}}1 \,\, [9]\;.
}
Here, the two quivers represent different 6d SCFTs, but they both correspond to $n=5$, $k=9$, $\Delta n=0$, $m_8=m_7=4$, $m_6=m_5=m_4=m_3=m_2=m_1=3$. As mentioned above, we can treat the quiver by replacing the spinor half-hypers with $\mathfrak{usp}_{-2}$ gauge algebras, in which case the chirality of the spinor is accounted by the $\theta$ angle of the associated $\mathfrak{usp}_{-2}$ group. The claim then is that the 4d theories resulting from the torus compactification of these 6d SCFTs are given by \eqref{ClassSpres1} and \eqref{ClassSpres2}, but we use the case of even $M+k+\Delta m_+$ if $\theta_1 = \theta_2$ and the odd case if $\theta_1 \neq \theta_2$. For instance, in the above example we get the class $\mathcal{S}$ theories
\begin{equation}
    \left[3^6,1^{18}\right]\;,\quad\left[18^2\right]\;,\quad\left[13,11^2,1\right]
\end{equation}
for $\theta_1 = \theta_2$, and
\begin{equation}
    \left[3^6,1^{18}\right]\;,\quad\left[19,17\right]\;,\quad\left[12^2,11,1\right]
\end{equation}
for $\theta_1 \neq \theta_2$. We note that both theories have very similar global symmetries, specifically $\SU(4)\times\SU(2)\times\SO(18)$ matching the symmetry expected from the quiver, except for the former having an extra $\text{U}(1)$. This additional $\text{U}(1)$ precisely corresponds to the one rotating the spinor hyper of $\text{SO}(12)$ in the first quiver. For reference, we provide the 3d mirror theories for the first and second cases\footnote{Recall that we write the 3d mirror theories as in \eqref{MQabbrev}, where we omit the types of the gauge nodes (and the edges). Each number denotes twice of the rank of the corresponding gauge node.}:
\bes{
&2\,\, 2 \,\, 4 \,\, 4 \,\, \cdots \,\, 16 \,\, 16 \,\, 18 \,\, {\blue 20 \,\, 24\,\, 26 \,\, 30 \,\, 32 \,\, \overset{\scalebox{0.9}{18}}{36} \,\, 22 \,\, 12}\;, \\
&2\,\, 2 \,\, 4 \,\, 4 \,\, \cdots \,\, 16 \,\, 16 \,\, 18 \,\, {\blue 20 \,\, 24\,\, 26 \,\, 30 \,\, 32 \,\, \overset{\scalebox{0.9}{16}}{36} \,\, 24 \,\, 12}\;.
}
Also note that we can get both theories from the quiver \eqref{Biexampleso12} by Higgsing the empty $-1$ curve. This entails the Higgsing through the $\SU(2)\times\SU(2)$ flavour symmetry group. Indeed, both class $\mathcal{S}$ theories can be reached in this way, differing by which $\SU(2)$ is Higgsed.

\item Once we go to higher values of $r$, the distinction becomes harder to spot. For $r=0$, as an example, let us consider the case
\begin{equation}
    1 \,\, \overset{1}{\underset{[1]}{{\overset{\mathfrak{so}_{16}}4}}} \,\, \underset{[2]}{\overset{\mathfrak{usp}_{14}}1}\,\,\underset{[1]}{\overset{\mathfrak{so}_{24}}4}\,\,\overset{\mathfrak{usp}_{16}}{1}\,\,[12]\;.
\end{equation}
It corresponds to $n=5$, $k=12$, $\Delta n=0$, $m_8=m_7=4$, $m_6=m_5=m_4=m_3=3$, $m_2=m_1=2$. Notice that $M+k+\Delta m_+$ is even. We get
\begin{equation}
    \left[3^4,2^2,1^{24}\right]\;,\quad\left[20^2\right]\;,\quad\left[15,13,11,1\right]
\end{equation}
for the identical $\theta$ angles, and
\begin{equation}
    \left[3^4,2^2,1^{24}\right]\;,\quad\left[21,19\right]\;,\quad\left[14^2,11,1\right]
\end{equation}
for the different $\theta$ angles. Here, it is harder to distinguish between the two as both have the same flavour symmetry $\USp(2)\times\SO(4)\times\USp(2)\times\SO(24)$. However, the prescription predicts that these should actually correspond to different 4d theories, specifically having a different Higgs branch. Indeed, we find that the Schur indices are
\begin{equation}
    \mathcal{I}=1+288\tau^2+64\tau^3+42020\tau^4+\dots
\end{equation}
for the first case, and
\begin{equation}
    \mathcal{I}=1+288\tau^2+64\tau^3+42019\tau^4+\dots
\end{equation}
for the second case. They differ at order $\tau^4$, indicating the invariant made from the spinor moment map from the E-strings. The 3d mirror theories for the former and the latter cases are, respectively,
\bes{
&2\,\, 2 \,\, 4 \,\, 4 \,\, \cdots \,\, 22 \,\, 22 \,\, 24 \,\, {\blue 26 \,\, 28 \,\, 30 \,\, 34 \,\, 36 \,\, \overset{\scalebox{0.9}{20}}{40} \,\, 24 \,\, 12}\;, \\
&2\,\, 2 \,\, 4 \,\, 4 \,\, \cdots \,\, 22 \,\, 22 \,\, 24 \,\, {\blue 26 \,\, 28 \,\, 30 \,\, 34 \,\, 36 \,\, \overset{\scalebox{0.9}{18}}{40} \,\, 26 \,\, 12}\;.
}

\item Next, for $r=1$, consider the case
\bes{ \label{su2bitheta}
\overset{\mathfrak{usp}_2}1 \,\, \overset{[4]}{\underset{1}{\underset{\mathfrak{usp}_2}{\overset{\mathfrak{so}_{20}}4}}} \,\, \overset{\mathfrak{usp}_{12}}1\,\,[10]
}
corresponding to $n=3$, $k=10$, $\Delta n=0$, $m_1=m_2=m_3=m_4=m_5=m_6=m_7=m_8=2$. There are actually two different 6d SCFTs associated with the same quiver, one for the case of the identical $\theta$ angles and one for the different angles. The reductions of the two lead to two different 4d class $\mathcal{S}$ theories. In the case of the identical ones, we get the theory
\begin{equation}
    \left[2^6,1^{20}\right]\;,\quad\left[16^2\right]\;,\quad\left[13,11,7,1\right]
\end{equation}
corresponding to \eqref{ClassSpres1} in the $M+k+\Delta m_+ =$ even case. In the case of the different angles, we instead get
\begin{equation}
    \left[2^6,1^{20}\right]\;,\quad\left[17,15\right]\;,\quad\left[12^2,7,1\right]\;.
\end{equation}
Note that the two have the same flavour symmetry $\USp(8)\times\SO(20)$. Here, the difference in the Schur indices should appear only at order $\tau^6$, making it harder to spot. The 3d mirror theories for the former and the latter cases are, respectively,
\bes{
&2\,\, 2 \,\, 4 \,\, 4 \,\, \cdots \,\, 18 \,\, 18 \,\, 20 \,\, {\blue 22 \,\, 24 \,\, 26 \,\, 28 \,\, 30 \,\, \overset{\scalebox{0.9}{16}}{32} \,\, 18 \,\, 8}\;, \\
&2\,\, 2 \,\, 4 \,\, 4 \,\, \cdots \,\, 18 \,\, 18 \,\, 20 \,\, {\blue 22 \,\, 24 \,\, 26 \,\, 28 \,\, 30 \,\, \overset{\scalebox{0.9}{14}}{32} \,\, 20 \,\, 8}\;.
}
As we shall see below, there are also other ways to notice the difference in the $\theta$ angles.
\end{itemize}

\paragraph{Higgsing the rightmost end and very even partitions} There is another case where the difference in the $\theta$ angles can be important. Specifically, we can consider Higgsing down the $\mathfrak{usp}$ at the rightmost end such that it has no free flavour left, and all the flavours have been transferred to the adjacent $\mathfrak{so}$. Now, if there are no flavours on any $\mathfrak{usp}$, then we can still change the angles by the partity of the $\mathfrak{so}$ gauge algebras. However, that would change it in pairs, and it is possible that a single global angle remains physical, depending on what happens at the leftmost side of the quiver. What happens in this case is that the long partition can become very even. If in addition two out of the three partitions are very even, then there are two choices of class $\mathcal{S}$ theories depending on whether the two very even partitions have the same or different colours (i.e., the labels I, II for the very even partitions). We expect that these cases always appear when the parent 6d theory has two associated SCFTs differing by the choice of the $\theta$ angles.

Let us illustrate this with some examples.
\begin{itemize}
    \item Consider the case
\bes{
[8]\,\, \overset{\mathfrak{su}_{10}}2 \,\, \overset{\mathfrak{usp}_{12}}1 \,\, {\overset{\mathfrak{so}_{20}}4} \,\, \overset{\mathfrak{usp}_{12}}1 \,\, [10]\;.
}
This corresponds to the class $\mathcal{S}$ theory
\begin{equation}
    \left[4^6,1^{20}\right]\;,\quad\left[22^2\right]\;,\quad\left[17^2,9,1\right]\;.
\end{equation}
Now, consider Higgsing down the rightmost $\mathfrak{usp}$ such that we get the quiver
\bes{
[8]\,\, \overset{\mathfrak{su}_{10}}2 \,\, \overset{\mathfrak{usp}_{12}}1 \,\, \underset{[5]}{\overset{\mathfrak{so}_{20}}4} \,\, \overset{\mathfrak{usp}_{2}}1\;.
}
In the class $\mathcal{S}$ theory, this should correspond to the changing $\left[4^6,1^{20}\right] \rightarrow \left[4^6,2^{10}\right]$. Note that we have two very even partitions $\left[4^6,2^{10}\right]$ and $\left[22^2\right]$. Therefore, there are two class $\mathcal{S}$ theories one can associate depending on the choice of the colours. This should match with whether the $\theta$ angles of the two $\mathfrak{usp}$ in the 6d quiver are the same or not. This is because we can change the individual angle with a parity transformation of the $\mathfrak{so}$, but their difference remains fixed. Note that we cannot use the $\mathfrak{su}_{10}$ bifundamental to switch the angle of the last $\mathfrak{usp}$ as that would change an even number of fundamentals.

\item Consider the case \eqref{su2bitheta}, and say, we Higgs it so that we get the quiver
\bes{
\overset{\mathfrak{usp}_2}1 \,\, \overset{[9]}{\underset{1}{\underset{\mathfrak{usp}_2}{\overset{\mathfrak{so}_{20}}4}}} \,\, \overset{\mathfrak{usp}_2}1\;.
}
Now, we have three $\theta$ angles for the three $\mathfrak{usp}$. However, using the $\mathfrak{so}$ parity and the permutation symmetry of the quiver, there should be only two distinct 6d SCFTs, corresponding to $\theta_1=\theta_2=\theta_3$ and $\theta_1=\theta_2 \neq \theta_3$. We can then consider the class $\mathcal{S}$ theory. The above Higgsing corresponds to the changing $\left[2^{6},1^{20}\right] \rightarrow \left[2^{16}\right]$. Therefore, we get the theory
\begin{equation}
    \left[2^{16}\right]\;,\quad\left[16^2\right]\;,\quad\left[13,11,7,1\right]
\end{equation}
if the angles are equal, and
\begin{equation}
    \left[2^{16}\right]\;,\quad\left[17,15\right]\;,\quad\left[12^2,7,1\right]
\end{equation}
if the angles are different. Note that the former has two very even partitions but not the latter. This is because the former corresponds to the case of $\theta_1=\theta_2$, and we then expect to get two cases depending on whether $\theta_3=\theta_1$ or not. These two cases should correspond to whether the colours of the two very even partitions are the same or not. However, if $\theta_1 \neq \theta_2$, corresponding to the second case, then no matter the choice of $\theta_3$, we have a case equivalent to $\theta_1=\theta_2 \neq \theta_3$ up to a symmetry transformation. Note that this supports the previous identification of the $\theta$ angles for case \eqref{su2bitheta}. It also predicts that case 2 should be equivalent to case 1 with unequal colours.
\end{itemize}

\subsubsection{Cases with Non-Integer \texorpdfstring{$M$}{M}}\label{nonintegerM}
There are certain rare cases where $M$ comes out non-integer, in which case one needs to switch $m_6$ and $m_8$. These cases appear to give either bad or ugly class $\mathcal{S}$ theories, depending on the difference between $m_7$ and $m_6$. Specifically, if after the switch we have that $m_7=m_6-1$, then the theory is usually ugly, but if instead $m_7<m_6-1$, it is bad. Note that when $M$ is an integer, we always have $m_8\geq m_7\geq m_6$, which for most cases is sufficient to guarantee that the theory is good (there are some exceptions, notably cases with $\mathfrak{g}_2$).

Before we briefly discuss the bad cases, let us first give some examples for the ugly cases.
\begin{itemize}
    \item Consider the case
\bes{
\overset{\mathfrak{su}_2}2  \,\, \underset{\left[\frac{1}{2}\right]}{\overset{\mathfrak{usp}_4}1} \,\, \underset{[3]}{\overset{\mathfrak{so}_{19}}4} \,\,\overset{\mathfrak{usp}_{12}}1 \,\, \left[\frac{21}{2}\right]
}
corresponding to $k=10, n=4$, $m_8=3$, $m_7=m_6=m_5=m_4=m_3=m_2=2$, $m_1=1$. Here, we get $M=\frac{11}{2}$, and the class $\mathcal{S}$ theory would not be well-defined (the partition $\left[2^5,1^{21}\right]$ is not a valid D-type partition). This can be remedied by exchanging $m_8$ and $m_6$. In this case, we get the class $\mathcal{S}$ theory with punctures
\begin{equation}
    \left[3,2^4,1^{21}\right]\;,\quad\left[16^2\right]\;,\quad\left[13,9^2,1\right]\;.
\end{equation}
One can note that this is an ugly class $\mathcal{S}$ theory, which also contains a free hyper. Yet, the global symmetry appears to be $\text{U}(1)\times\USp(6)\times\SO(21)$ as expected.

It is interesting to consider what happens if we Higgs this theory to the quiver
\bes{
\overset{\mathfrak{su}_2}2  \,\, \underset{[1]}{\overset{\mathfrak{usp}_4}1} \,\, \underset{[2]}{\overset{\mathfrak{so}_{18}}4} \,\,\overset{\mathfrak{usp}_{12}}1 \,\, [11]\;.
}
This can be implemented in the class $\mathcal{S}$ theory by Higgsing down the $\USp(6)$ global symmetry. We can do this by changing the partition $\left[16^2\right] \rightarrow \left[17,15\right]$, but this leads to a bad class $\mathcal{S}$ theory\footnote{We can also implement this Higgsing by changing the partition $\left[3,2^4,1^{21}\right] \rightarrow \left[3^2,2^2,1^{22}\right]$, which still gives an ugly class $\mathcal{S}$ theory. However, this is as we have 3 vector hypers under the same gauge group in the 6d quiver. There are more complicated examples, where we have 3 vector hypers under different gauge groups, in which the two changes will lead to different theories.}. Note, however, that if we apply \eqref{ClassSpres2} on this case, we get the 4d class $\mathcal{S}$ theory
\begin{equation}
    \left[2^4,1^{22}\right]\;,\quad\left[15^2\right]\;,\quad\left[11,9^2,1\right]\;,
\end{equation}
which is actually a good theory. The two theories can be related as follows. We can use the results in \cite{Zafrir:2016jpu} to lift a class $\mathcal{S}$ theory to a 5d SCFT engineered by a brane web with an O5-plane. If we lift the bad class $\mathcal{S}$ theory, one can see that it is possible to perform Hanany-Witten moves such that we get the good class $\mathcal{S}$ theory.

\item As another interesting example, consider the case
\bes{ 
\left[\frac{1}{2}\right] \,\, \overset{\mathfrak{usp}_2}1 \,\, \overset{[3]}{\underset{\left[\frac{1}{2}\right]}{\underset{1}{\underset{\mathfrak{usp}_2}{\overset{\mathfrak{so}_{19}}4}}}} \,\, \underset{\left[\frac{1}{2}\right]}{\overset{\mathfrak{usp}_{12}}1}\,\, \overset{\mathfrak{so}_{20}}4\,\, \overset{\mathfrak{usp}_{12}}1 \,\, [10]
}
corresponding to $k=10$, $n=5$, $\Delta n = 0$, $m_1=3$, $m_2=m_3=m_4=m_5=m_6=m_7=4$ and $m_8=5$. Here, we have $M=\frac{23}{2}$, so we should replace $m_8 \leftrightarrow m_6$, after which we get that $M=12$, $\Delta m_+ =4$ and $\Delta m_- =0$. What is interesting in this case is that we have two different options for the class $\mathcal{S}$ theory, depending on whether we choose to take $M+k+\Delta m_+$ as even or odd. Specifically, in the $M+k+\Delta m_+ =$ even case, we get the class $\mathcal{S}$ theory with punctures
\begin{equation}
    \left[5,4^4,3,1^{20}\right]\;,\quad\left[22^2\right]\;,\quad\left[17,15,11,1\right]\;.
\end{equation}
If we instead treat it as the $M+k+\Delta m_+ =$ odd case, we get the class $\mathcal{S}$ theory with punctures
\begin{equation}
    \left[5,4^4,3,1^{20}\right]\;,\quad\left[23,21\right]\;,\quad\left[16^2,11,1\right]\;.
\end{equation}
Both theories are ugly class $\mathcal{S}$ theories with a single free hyper and appear to have the same flavour symmetry $\USp(6)\times\SO(20)$. Previously, we have attributed this distinction to the difference in the $\theta$ angles for the two $\mathfrak{usp}_2$. However, in this case, the distinction should be physically irrelevant due to the fundamental flavours\footnote{One can also see this as $\mathfrak{so}_{19}$ has only one single spinor representation. As such, one can always build a gauge invariant from the spinor Higgs branch chiral ring operator of the SCFT associated with the two $\mathfrak{usp}_2$, regardless of the $\theta$ angle.}. This leads us to suspect that these two 4d theories are secretly equivalent (though it is possible that they are different and only one correctly matches the reduction of the 6d SCFT).

This can be further supported by noting that we can get the above quiver from the 6d theory
\bes{ 
\left[\frac{5}{2}\right] \,\, \overset{\mathfrak{usp}_4}1 \,\, \overset{[2]}{\underset{\left[\frac{1}{2}\right]}{\underset{1}{\underset{\mathfrak{usp}_2}{\overset{\mathfrak{so}_{19}}4}}}} \,\, \underset{\left[\frac{1}{2}\right]}{\overset{\mathfrak{usp}_{12}}1}\,\, \overset{\mathfrak{so}_{20}}4\,\, \overset{\mathfrak{usp}_{12}}1 \,\, [10]
}
by Higgsing the $\mathfrak{usp}_4$ node. Using \eqref{ClassSpres1} or \eqref{ClassSpres2} ($m_8=m_7=5$), we associate with this quiver the class $\mathcal{S}$ theory with partitions
\begin{equation}
    \left[5,4^4,3,1^{20}\right]\;,\quad\left[22^2\right]\;,\quad\left[16^2,11,1\right]\;,
\end{equation}
whose flavour symmetry is $\SO(5)\times\USp(4)\times\SO(20)$, where the $\SU(2)^2$ symmetry associated with the latter two punctures gets enhanced to $\USp(4)$. The two theories are then given by Higgsing through either the $\left[22^2\right]$ partition or the $\left[16^2,11,1\right]$ one. However, as the two $\SU(2)$ groups are enhanced to $\USp(4)$, both Higgsings should give the same result. Here, the exchange of the two $\SU(2)$ groups is part of the Weyl group of $\USp(4)$.
\end{itemize}

\paragraph{Bad theories} Finally, we turn to the cases where we end up with bad theories. For instance, consider the case
\bes{
\overset{\mathfrak{su}_2}2  \,\, \underset{\left[\frac{1}{2}\right]}{\overset{\mathfrak{usp}_4}1} \,\, {\overset{\mathfrak{so}_{19}}4} \,\, \overset{\mathfrak{usp}_{18}}1 \,\, \underset{[3]}{\overset{\mathfrak{so}_{33}}4} \,\,\overset{\mathfrak{usp}_{26}}1 \,\, \left[\frac{35}{2}\right]\;.
}
We have $k=17$, $n=6$, $m_1=1$, $m_2=m_3=m_4=m_5=m_6=m_7=2$ and $m_8=5$. Here, $M=\frac{11}{2}$, so we should replace $m_8 \leftrightarrow m_6$, after which we get $M=7$, $\Delta m_+ =2$ and $\Delta m_- =0$. We then get the class $\mathcal{S}$ theory with punctures
\begin{equation}
    \left[5,2^4,1^{35}\right]\;,\quad\left[24^2\right]\;,\quad\left[21,13^2,1\right]\;.
\end{equation}
We find that this is a bad class $\mathcal{S}$ theory. Note that we can also make $M$ integer by exchanging $m_7\leftrightarrow m_1$, in which case we get the class $\mathcal{S}$ theory
\begin{equation}
    \left[2^6,1^{34}\right]\;,\quad\left[25,21\right]\;,\quad\left[19,13^2,1\right]\;.
\end{equation}
This class $\mathcal{S}$ theory is ugly, as now $m_7 = m_6 - 1$.

Generally, if the theory is bad, it might be possible to make a different shuffling of the $m_i$'s such that it would be ugly instead\footnote{There can also be cases where two shufflings give ugly class $\mathcal{S}$ theories, in which case the two should be dual.}, but there are cases where all of the resulting descriptions are bad. It is not clear whether there exists yet another description that is good or ugly, or whether these quivers have no description in terms of a good/ugly $\mathcal{S}$ theory.

\subsection{Exceptions}\label{exceptions}
While the above list covers most cases for which we find a D-type class $\mathcal{S}$ description, there are several additional cases that require separate treatment. Here, we shall explicitly write the class $\mathcal{S}$ theory associated with each case, thus completing the relation between the 6d quiver and 4d class $\mathcal{S}$ theory.

\subsubsection{Cases Using the \texorpdfstring{$\SU(4)\times\SO(10)$}{SU(4)xSO(10)} Subgroup of \texorpdfstring{$E_8$}{E8}}\label{SU4xSO8}
In the above prescription, we have treated an empty $-1$ curve as a $\mathfrak{usp}_0$ theory and as such, associated $8$ flavours with it, which are essentially associated with the $\SO(16)$ subgroup of $E_8$. When part of that group is gauged, the remaining symmetry is usually a smaller $\SO$ group similar to the case of a non-empty $-1$ curve.

An interesting case is when the empty $-1$ curve is adjacent to a $-2$ curve with an $\mathfrak{su}_l$ on one side and a $-4$ curve with an $\mathfrak{so}_m$ on the other side. Note that if the $-1$ curve was not empty, rather supporting a $\mathfrak{usp}_{2r}$, the embedding would be $\mathfrak{u}_1\times\mathfrak{su}_l\times\mathfrak{so}_m\subset\mathfrak{so}_{4r+16}$, with the $\mathfrak{u}_1$ associated with the bifundamental. Now, if we take $r=0$ and $m=8$, which is the minimal $\mathfrak{so}$ that can be supported on a $-4$ curve, then the maximal value of $l$ is $4$. However, note that the commutant of $\SU(4)$ in $\SO(16)$ is $\SO(10)$, which is also the commutant inside $E_8$. As such, a quiver where we have an $\mathfrak{su}_4$ on the $-2$ curve and an $\mathfrak{so}_{10}$ on the $-4$ curve is possible, such as case 2 in $D_5$. While formally quite similar to the first discussed case, note that the $-1$ effectively sees $9$ flavours, and as such, these cases are not covered by the previous prescription and require special treatment.

Specifically, consider the following quivers
\bes{ \label{su4Xso101}
[8]\,\, \overset{\mathfrak{su}_{4}}2 \,\, 1 \,\, {\overset{\mathfrak{so}_{10}}4} \,\, \overset{\mathfrak{usp}_{4}}1 \,\, ... \,\, \underset{[1]}{\overset{\mathfrak{so}_{2k}}4} \,\, \overset{\mathfrak{usp}_{2k-8}}1 ... \,\, {\overset{\mathfrak{so}_{2k}}4} \,\, \overset{\mathfrak{usp}_{2k-8}}1 \,\, [k]\;, }
\bes{ \label{su4Xso102}
[8]\,\, \overset{\mathfrak{su}_{4}}2 \,\, 1 \,\, {\overset{\mathfrak{so}_{10}}4} \,\, \overset{\mathfrak{usp}_{4}}1 \,\, ... \,\, \underset{[1]}{\overset{\mathfrak{usp}_{2k-8}}4} \,\, \overset{\mathfrak{so}_{2k}}1 ... \,\, {\overset{\mathfrak{so}_{2k}}4} \,\, \overset{\mathfrak{usp}_{2k-8}}1 \,\, [k]\;, }
where the first one is for odd $k$ and the second for even $k$. We have denoted the number of tensors by $n_T = n-1$. Note that here, $n$ is odd then. We claim that the 4d theory resulting from the torus reduction of the associated SCFT is given by the D-type class $\mathcal{S}$ theory with partitions
\begin{equation} \label{su4Xso10part}
    \left[n^4,(n-k+2)^2,1^{2k}\right]\;,\quad\left[(3n+2)^2\right]\;,\quad\left[(2n+1)^3,1\right]\;.
\end{equation}
This is motivated as it gives the right flavour symmetry and matches with case 2 for $k=5$. Note that unlike the previous cases, there is little control over the flavours here. Specifically, the only freedom we have is to separate the full hyper above into two half-hypers. For example, consider the quiver
\bes{ \label{su4Xso103}
[8]\,\, \overset{\mathfrak{su}_{4}}2 \,\, 1 \,\, {\overset{\mathfrak{so}_{10}}4} \,\, \overset{\mathfrak{usp}_{4}}1 \,\, ... \,\,  \underset{\left[\frac{1}{2}\right]}{\overset{\mathfrak{usp}_{2k-10}}1} \,\,{\overset{\mathfrak{so}_{2k-1}}4} \,\, \underset{\left[\frac{1}{2}\right]}{\overset{\mathfrak{usp}_{2k-8}}1} \,\, ... \,\, {\overset{\mathfrak{so}_{2k}}4} \,\, \overset{\mathfrak{usp}_{2k-8}}1 \,\, [k]\;, }
which corresponds to Higgsing the $\mathfrak{so}_{2k}$ in the $k$ odd case. It is simple to write the 4d class $\mathcal{S}$ theory associated with this case by doing the same Higgsing, leading to the partitions
\begin{equation}
    \left[n^4,n-k+3,n-k+1,1^{2k}\right]\;,\quad\left[(3n+2)^2\right]\;,\quad\left[(2n+1)^3,1\right]\;,
\end{equation}
where here $k$ and $n$ are both odd. This case can be further generalized to the following family of quivers:
\bes{
[8]\,\, \overset{\mathfrak{su}_{4}}2 \,\, 1 \,\, {\overset{\mathfrak{so}_{10}}4} \,\, \overset{\mathfrak{usp}_{4}}1 \,\, ... \,\,  \underset{\left[\frac{1}{2}\right]}{\overset{\mathfrak{usp}_{2l-8}}1} \,\,{\overset{\mathfrak{so}_{2l+1}}4} \,\, {\overset{\mathfrak{usp}_{2l-6}}1} \,\, ... \,\,  \underset{\left[\frac{1}{2}\right]}{\overset{\mathfrak{usp}_{2k-8}}1} \,\,{\overset{\mathfrak{so}_{2k}}4} \,\, ... \,\, {\overset{\mathfrak{so}_{2k}}4} \,\, \overset{\mathfrak{usp}_{2k-8}}1 \,\, [k]\;, }
where $l$ is an even number obeying $4 \leq l \leq k$. The associated 4d class $\mathcal{S}$ theory appears to be
\begin{equation} \label{su4Xso10partl}
    \left[n^4,n-l+2,n-2k+l+2,1^{2k}\right]\;,\quad\left[(3n+2)^2\right]\;,\quad\left[(2n+1)^3,1\right]\;,
\end{equation}
where $n$ is odd. Note that the cases \eqref{su4Xso102} and \eqref{su4Xso103} can be recovered if we set $l=k$ (for $k$ even) and $l=k-1$ (for $k$ odd).

For instance, for $k=8$, we have three options, $l=4,6,8$, which are described by the following quivers and class $\mathcal{S}$ theories:
\begin{itemize}
    \item For the case of $l=k=8$, we have
    \bes{
[8]\,\, \overset{\mathfrak{su}_{4}}2 \,\, 1 \,\, {\overset{\mathfrak{so}_{10}}4} \,\, \overset{\mathfrak{usp}_{4}}1 \,\, {\overset{\mathfrak{so}_{14}}4} \,\,  \underset{[1]}{\overset{\mathfrak{usp}_{8}}1} \,\,{\overset{\mathfrak{so}_{16}}4} \,\, ... \,\,  {\overset{\mathfrak{so}_{16}}4} \,\, \overset{\mathfrak{usp}_{8}}1 \,\, [8] }
and
\begin{equation}
    \left[n^4,(n-6)^2,1^{16}\right]\;,\quad\left[(3n+2)^2\right]\;,\quad\left[(2n+1)^3,1\right]\;.
\end{equation}
\item For the case of $l=6$, we have
\bes{
[8]\,\, \overset{\mathfrak{su}_{4}}2 \,\, 1 \,\, {\overset{\mathfrak{so}_{10}}4} \,\, \underset{\left[\frac{1}{2}\right]}{\overset{\mathfrak{usp}_{4}}1} \,\,{\overset{\mathfrak{so}_{13}}4} \,\, {\overset{\mathfrak{usp}_{6}}1} \,\,{\overset{\mathfrak{so}_{15}}4} \,\, \underset{\left[\frac{1}{2}\right]}{\overset{\mathfrak{usp}_{8}}1} \,\,{\overset{\mathfrak{so}_{16}}4} \,\, ... \,\, {\overset{\mathfrak{so}_{16}}4} \,\, \overset{\mathfrak{usp}_{8}}1 \,\, [8] }
and
\begin{equation}
    \left[n^4,n-4,n-8,1^{16}\right]\;,\quad\left[(3n+2)^2\right]\;,\quad\left[(2n+1)^3,1\right]\;.
\end{equation}
\item For the case of $l=4$, we have
\bes{
[8]\,\, \overset{\mathfrak{su}_{4}}2 \,\, 1 \,\, {\overset{\mathfrak{so}_{9}}4} \,\, {\overset{\mathfrak{usp}_{2}}1} \,\,{\overset{\mathfrak{so}_{11}}4} \,\, {\overset{\mathfrak{usp}_{4}}1} \,\,{\overset{\mathfrak{so}_{13}}4} \,\, {\overset{\mathfrak{usp}_{6}}1} \,\, {\overset{\mathfrak{so}_{15}}4} \,\, \underset{\left[\frac{1}{2}\right]}{\overset{\mathfrak{usp}_{8}}1} \,\,{\overset{\mathfrak{so}_{16}}4} \,\, ... \,\, {\overset{\mathfrak{so}_{16}}4} \,\, \overset{\mathfrak{usp}_{8}}1 \,\, [8] }
and
\begin{equation}
    \left[n^4,n-2,n-10,1^{16}\right]\;,\quad\left[(3n+2)^2\right]\;,\quad\left[(2n+1)^3,1\right]\;.
\end{equation}
\end{itemize}

There do not appear to be any more quivers in this class. Specifically, Higgsing the $\mathfrak{su}_4$ will lead us to a case already described previously, so all possible Higgsings are already captured. We also note that this case is only possible with an empty $-1$ curve. Realizing this with a non-empty $-1$ curve requires a bifundamental in the ${\bf 6}$ of $\mathfrak{su}_4$, but that is not possible on a $-2$ curve. As such, the cases outlined above completely exhaust the cases in this class.

\subsubsection{Cases Involving Spinor Bifundamentals}\label{spinbifund}
A second class of quivers not included in the previous discussions are those with a bifundamental in the spinor of an $\mathfrak{so}$. Due to the restriction of anomaly cancellation, this is only possible if the gauge algebra is $\mathfrak{so}_8$ or $\mathfrak{so}_7$. Let us discuss these in turn.

\paragraph{The $\mathfrak{so}_8$ case} We shall begin with the case involving an $\mathfrak{so}_8$ gauge factor with a spinor bifundamental. Because of the triality property of $\mathfrak{so}_8$, it is possible to redefine it so that the spinor becomes the vector. As such, if all bifundamentals are in the same 8-dimensional representation, then we can choose them to be in the vector, and this case reduces to one of the previous cases. However, if there are at least two bifundamentals with different 8-dimensional representations, then it is impossible to make all of them be in the vector. This case seems to only occur if the quiver ends with a $-1$ $-3$ $-1$ curve, which then take the form
\bes{
[6]\,\, \overset{\mathfrak{usp}_{2}}1 \,\, \underset{[1]}{\overset{\mathfrak{so}_{8}}3} \,\, \overset{\mathfrak{usp}_{2}}1 \,\, \overset{\mathfrak{so}_{12}}4 \,\, ... \,\, \underset{[1]}{\overset{\mathfrak{so}_{2k}}4} \,\, \overset{\mathfrak{usp}_{2k-8}}1 ... \,\, {\overset{\mathfrak{so}_{2k}}4} \,\, \overset{\mathfrak{usp}_{2k-8}}1 \,\, [k] }
for $k$ even, and
\bes{ \label{so8S1}
[6]\,\, \overset{\mathfrak{usp}_{2}}1 \,\, \underset{[1]}{\overset{\mathfrak{so}_{8}}3} \,\, \overset{\mathfrak{usp}_{2}}1 \,\, \overset{\mathfrak{so}_{12}}4 \,\, ... \,\, \underset{[1]}{\overset{\mathfrak{usp}_{2k-8}}4} \,\, \overset{\mathfrak{so}_{2k}}1 ... \,\, {\overset{\mathfrak{so}_{2k}}4} \,\, \overset{\mathfrak{usp}_{2k-8}}1 \,\, [k] }
for $k$ odd. In these cases, one notes that the $\mathfrak{so}_8$ sees one hyper in each of its three 8-dimensional representations, with two coming from the half-hyper bifundamental. As such, we cannot switch all of them to be in the vector, and this case is distinct from the previously covered ones. As before, we shall denote the number of tensors by $n_T=n-1$. Note that here $n$ is always even.

We claim that the 4d theory resulting from the torus reduction of the associated SCFT is given by the D-type class $\mathcal{S}$ theory with partitions
\begin{equation} \label{so8S1classS}
    \left[n^4,(n-k+2)^2,1^{2k}\right]\;,\quad\left[(3n+2)^2\right]\;,\quad\left[(2n+1)^3,1\right]\;.
\end{equation}
Note that this is equivalent to \eqref{su4Xso101} and \eqref{su4Xso102}, but with $n$ being even rather than odd. This is motivated as it gives the right flavour symmetry and matches with theory 12 for $k=5$. Like in the previous case, the only other possible flavour placement is to seperate the full hyper into two half-hypers of different $\mathfrak{usp}$ groups.  For instance, Higgsing down the first $\mathfrak{so}_{2k}$ in the $k$ even case gives the quiver
\bes{ \label{so8S2}
[6]\,\, \overset{\mathfrak{usp}_{2}}1 \,\, \underset{[1]}{\overset{\mathfrak{so}_{8}}3} \,\, \overset{\mathfrak{usp}_{2}}1 \,\, \overset{\mathfrak{so}_{12}}4 \,\, ... \,\,  \underset{\left[\frac{1}{2}\right]}{\overset{\mathfrak{usp}_{2k-10}}1} \,\,{\overset{\mathfrak{so}_{2k-1}}4} \,\, \underset{\left[\frac{1}{2}\right]}{\overset{\mathfrak{usp}_{2k-8}}1} \,\, ... \,\, {\overset{\mathfrak{so}_{2k}}4} \,\, \overset{\mathfrak{usp}_{2k-8}}1 \,\, [k]\;. }
As before, the associated class $\mathcal{S}$ theory will be the same one:
\begin{equation}
    \left[n^4,n-k+3,n-k+1,1^{2k}\right]\;,\quad\left[(3n+2)^2\right]\;,\quad\left[(2n+1)^3,1\right]\;,
\end{equation}
but now $k$ and $n$ are both even. Again, this case can be further generalised to the family of quivers
\bes{
[6]\,\, \overset{\mathfrak{usp}_{2}}1 \,\, \underset{[1]}{\overset{\mathfrak{so}_{8}}3} \,\, \overset{\mathfrak{usp}_{2}}1 \,\, \overset{\mathfrak{so}_{12}}4 \,\, ... \,\,  \underset{\left[\frac{1}{2}\right]}{\overset{\mathfrak{usp}_{2l-8}}1} \,\,{\overset{\mathfrak{so}_{2l+1}}4} \,\, {\overset{\mathfrak{usp}_{2l-6}}1} \,\, ... \,\,  \underset{\left[\frac{1}{2}\right]}{\overset{\mathfrak{usp}_{2k-8}}1} \,\,{\overset{\mathfrak{so}_{2k}}4} \,\, ... \,\, {\overset{\mathfrak{so}_{2k}}4} \,\, \overset{\mathfrak{usp}_{2k-8}}1 \,\, [k]\;, }
where $l$ is an odd number obeying $5 \leq l \leq k$. The associated 4d class $\mathcal{S}$ theory appears to again be
\begin{equation} \label{so8SlclassS}
    \left[n^4,n-l+2,n-2k+l+2,1^{2k}\right]\;,\quad\left[(3n+2)^2\right]\;,\quad\left[(2n+1)^3,1\right]\;,
\end{equation}
but now with $n$ even and $l$ odd. Note that, as before, the cases \eqref{so8S1} and \eqref{so8S2} can be recovered if we set $l=k$ (for $k$ odd) and $l=k-1$ (for $k$ even).

We also note that something interesting happens for $k=4$. In this case, the quiver essentially becomes
\bes{
[6]\,\, \overset{\mathfrak{usp}_{2}}1 \,\, \underset{[1][1]}{\overset{\mathfrak{so}_{8}}3} \,\, 1 \,\, \overset{\mathfrak{so}_{8}}4 \,\, ... \,\,  {\overset{\mathfrak{so}_{8}}4} \,\, 1 \,\, [4]\;, }
which is theory 10 in $D_4$. Indeed, one notes that the resulting class $\mathcal{S}$ theory matches with our findings in this case. What is interesting here is that there is only one bifundamental for $\mathfrak{so}_8$, so we can treat this case also as part of the bifurcation family. However, here, $\Delta n= 2$, so there is no class $\mathcal{S}$ description in that family. This is then a special case where we seem to find a class $\mathcal{S}$ description if we regard the bifundamental as a spinor in this family here instead.

As before, the quivers listed above seem to exhaust the quivers in this class. Specifically, we can try to Higgs down the $\mathfrak{su}_2$ at the leftmost end, in which we get to a quiver in the previous class. For instance, for $k=5$, we get theory 21. We can also Higgs down the $\mathfrak{so}_8$, in which case we get a new quiver with $\mathfrak{so}_7$ and a spinor bifundamental:
\bes{
[6]\,\, \overset{\mathfrak{usp}_{2}}1 \,\, {\overset{\mathfrak{so}_{7}}3} \,\, \overset{\mathfrak{usp}_{2}}1 \,\, \overset{\mathfrak{so}_{12}}4 \,\, ... \,\, \underset{[1]}{\overset{\mathfrak{so}_{2k}}4} \,\, \overset{\mathfrak{usp}_{2k-8}}1 ... \,\, {\overset{\mathfrak{so}_{2k}}4} \,\, \overset{\mathfrak{usp}_{2k-8}}1 \,\, [k] }
for $k$ even, and a similar one for $k$ odd. This quiver appears to have no class $\mathcal{S}$ description. This is motivated as the $\SU(2)$ global symmetry we need to Higgs is not manifest in the class $\mathcal{S}$ description, rather arising in the index through the sum over the representations. As such, it is not clear how to perform this Higgsing in terms of the punctures. For example, in $k=5$, attempting to do this from theory 12 seems to just lead us to theory 21, that is we Higgs through the global $\SO(12)$ rather than the $\SU(2)$. A similar statement seems to apply also to the case
\bes{
[4] \,\, 1 \,\, \underset{[1]}{\overset{\mathfrak{so}_{7}}3} \,\, \overset{\mathfrak{usp}_{2}}1 \,\, \overset{\mathfrak{so}_{12}}4 \,\, ... \,\, \underset{[1]}{\overset{\mathfrak{so}_{2k}}4} \,\, \overset{\mathfrak{usp}_{2k-8}}1 ... \,\, {\overset{\mathfrak{so}_{2k}}4} \,\, \overset{\mathfrak{usp}_{2k-8}}1 \,\, [k]\;. }
Again for $k$ even, with a similar one for $k$ odd. This is motivated by the fact that we find no class $\mathcal{S}$ theories associated with these cases for $k=5$, where these are theories 20 and 25, respectively.

\paragraph{The \texorpdfstring{$\mathfrak{so}_7$}{so7} on \texorpdfstring{$-3$}{-3} case} The remaining cases involve an $\mathfrak{so}_7$ gauge group with bifundamental matter in the spinor representation. These appear in two types of quivers. One are the quivers ending with a $-1$ $-3$ $-1$ sequence of tensors, with the $\mathfrak{so}_7$ on the $-3$. The second type are the quivers ending with a $-2$ $-1$ sequence of tensors, with the $\mathfrak{so}_7$ on the $-2$. Let us first discuss the case of the $\mathfrak{so}_7$ on the $-3$ curve, which is quite limited, admitting only a handful number of quivers.

We shall begin by considering the following quiver
\bes{
[8]\,\, \overset{\mathfrak{usp}_{4}}1 \,\, {\overset{\mathfrak{so}_{7}}3} \,\, 1 \,\, \overset{\mathfrak{so}_{9}}4 \,\, ... \,\, \underset{\left[\frac{1}{2}\right]}{\overset{\mathfrak{usp}_{2k-8}}1} \,\, \overset{\mathfrak{so}_{2k}}4 ... \,\, {\overset{\mathfrak{so}_{2k}}4} \,\, \overset{\mathfrak{usp}_{2k-8}}1 \,\, [k]\;. }
Here, the $\mathfrak{usp}_4$ and the $\mathfrak{so}_7$ are connected by a bifundamental in the $({\bf 4},{\bf 8})$ representation. We also note that the flavour structure is completely fixed by anomaly cancellation. We claim that the 4d theory resulting from the torus reduction of this 6d SCFT is given in terms of the D-type class $\mathcal{S}$ theory associated with a sphere with the 3 punctures
\begin{equation} \label{so16case}
    \left[n^5,n-2k+4,1^{2k}\right]\;,\quad\left[(3n+2)^2\right]\;,\quad\left[(2n+1)^3,1\right]\;,
\end{equation}
where $n$ is related to the number of tensors by $n_T=n-2$ (note that $n$ is always odd). This is motivated by the fact that we get the right global symmetry and by matching with theory 18 for $k=4$ and theory 6 for $k=5$.

We can get additional quivers of this form by going on the Higgs branch. Here, the main Higgsing we can consider is to Higgs the $\mathfrak{usp}_4$ gauge factor at the end. This leads to the following 6d quiver:
\bes{
[6]\,\, \overset{\mathfrak{usp}_{2}}1 \,\, \underset{[1]}{\overset{\mathfrak{so}_{7}}3} \,\, 1 \,\, \overset{\mathfrak{so}_{9}}4 \,\, ... \,\, \underset{\left[\frac{1}{2}\right]}{\overset{\mathfrak{usp}_{2k-8}}1} \,\, \overset{\mathfrak{so}_{2k}}4 ... \,\, {\overset{\mathfrak{so}_{2k}}4} \,\, \overset{\mathfrak{usp}_{2k-8}}1 \,\, [k]\;. }
We claim that the associated 4d class $\mathcal{S}$ theory is given by the punctures
\begin{equation} \label{so12su2case}
    \left[n^4,n-1,n-2k+5,1^{2k}\right]\;,\quad\left[(3n+2)^2\right]\;,\quad\left[(2n+1)^3,1\right]\;,
\end{equation}
where we now have $n_T=n-1$. This is again motivated by matching the global symmetry and by more detailed checks in the cases of $k=4,~5$. Note that we can get to this theory from the previous one by Higgsing, but the relation is more complicated. Specifically, we need to remove 6 boxes, while the minimal Higgsing remove only 2 boxes. This appears to give the same theory but with additional free hypers. The removal of 4 additional boxes appears to remove these hypers and gives a good class $\mathcal{S}$ description of this theory.

We can continue to explore other quivers via Higgsings. First, we can further Higgs the $\mathfrak{usp}_2$ to an empty $-1$ curve. This leads to a quiver without spinor bifundamentals, so it can already be accommodated in the bifurcation family of quivers. We can also Higgs the $\mathfrak{so}_7$ down to $\mathfrak{g}_2$, leading to the quiver
\bes{
\left[\frac{13}{2}\right]\,\, \overset{\mathfrak{usp}_{2}}1 \,\, {\overset{\mathfrak{g}_{2}}3} \,\, 1 \,\, \overset{\mathfrak{so}_{9}}4 \,\, ... \,\, \underset{\left[\frac{1}{2}\right]}{\overset{\mathfrak{usp}_{2k-8}}1} \,\, \overset{\mathfrak{so}_{2k}}4 ... \,\, {\overset{\mathfrak{so}_{2k}}4} \,\, \overset{\mathfrak{usp}_{2k-8}}1 \,\, [k]\;. }
This quiver is also part of the bifurcation family, though it has {\it no class $\mathcal{S}$ theory description} there since $\Delta n = 3$. Unfortunately, the situation is not better from this perspective, and this quiver appears to have no class $\mathcal{S}$ theory description. Specifically, the {\it $\SU(2)$ flavour symmetry we need to Higgs is not manifest in the partition, and as such it is not clear how to implement this Higgsing in the class $\mathcal{S}$ theory}. This quiver gives theory 2 in $D_4$ and theory 19 in $D_5$ (see Appendix \ref{examples}). In both cases, we have not found a class $\mathcal{S}$ description of this theory, which further supports our claim.

\paragraph{The \texorpdfstring{$\mathfrak{so}_7$}{so7} on \texorpdfstring{$-2$}{-2} case} The last family of cases involves a quiver ending with a $-2$ curve supporting an $\mathfrak{so}_7$ with a spinor bifundamental. This class does allow significant freedom in the placement of flavours, so it is more similar to the families studied at the beginning of this section. Specifically, ignoring the vector hyper (subsequently denoted by $[1\cdot {\bf 7}]$) on the $\mathfrak{so}_7$ and the $k$ fundamental hypers at the rightmost edge of the quivers, there are overall 8 half-hypers. These are either distributed along the quiver as before or are situated at the leftmost edge as spinor half-hypers (note that here we count spinor half-hypers on the $\mathfrak{so}_7$ rather than vector ones). We can again specify the locations of the flavours using the numbers $m_{1,\dots,8}$. Additionally, we shall denote the number of tensors by $n$, which is always even here. Then we claim that the 4d reduction of this family of theories is given in terms of the D-type class $\mathcal{S}$ theory with punctures
\bes{ \label{classS:so7on2}
& \left[m_6,m_5,m_4,m_3,m_2,m_1,1^{2k}\right]\;, \\ 
& \left[4+4n-m_7,4+4n-m_8\right]\;, \\  
& \left[4+4n-m_8-m_7-1,(2n+2)^2,1\right]\;.
}
The logic behind this is as follows. In this class, we always have a vector hyper of the $\mathfrak{so}_7$. We can use this vector to Higgs the $\mathfrak{so}_7$ to $\mathfrak{su}_4$, in which case the quiver reduces to the family discussed at the beginning of this section. Note that the $m_i$'s remain the same after the Higgsing as the spinor half-hyper of $\mathfrak{so}_7$ reduces to the fundamental hyper of $\mathfrak{su}_4$. It is also straightforward to see that in this class of models we have $k=4+\sum\limits_{i=1}^8\frac{n-m_i}{2}$. As such, we see that we can associate with this theory the class $\mathcal{S}$ theory with punctures
\begin{align}
    \begin{split}
        &\left[m_6,m_5,m_4,m_3,m_2,m_1,1^{2k}\right]\;,\\
        &\left[4+4n-m_7,4+4n-m_8\right]\;,\\
        &\left[4+4n-m_8-m_7-1,2n+3,2n+1,1\right]\;.
    \end{split}
\end{align}
The class $\mathcal{S}$ theory we seek, corresponding to the theory with the $\mathfrak{so}_7$ gauge factor, should have an $\SU(2)$ symmetry that upon Higgsing will give the theory with the $\mathfrak{su}_4$. Furthermore, as the $\SU(2)$ is associated with the end of the quiver, it is natural to expect it to appear at the second and the third columns of the four-column puncture, which is usually where special symmetries associated with the leftmost end appear (examples include the $\text{U}(1)$ where we have an $\mathfrak{su}_2$ on the $-2$ or two spinors of $\mathfrak{so}_{12}$ with the same chirality). In particular, \eqref{classS:so7on2} obeys these requirements.

Next, we shall present some examples of this prescription.
\begin{itemize}
    \item Consider the case
    \bes{
[1\cdot {\bf 7}][3]\,\, \overset{\mathfrak{so}_{7}}2 \,\, \underset{[1]}{\overset{\mathfrak{usp}_{2}}1} \,\,\overset{\mathfrak{so}_{10}}4 \,\, \overset{\mathfrak{usp}_{2}}1 \,\, {\overset{\mathfrak{so}_{10}}4} \,\, \overset{\mathfrak{usp}_{2}}1 \,\, [5] }
corresponding to theory 42 in $D_5$. In this case, we have $k=5$, $n=6$, $m_1 = m_2 = 5$ (two fundamental half-hypers of the leftmost $\mathfrak{usp}_2$), $m_3=m_4=m_5=m_6=m_7=m_8=6$ (six spinor half-hypers of the $\mathfrak{so}_7$). We then associate to it the class $\mathcal{S}$ theory
\begin{equation}
    \left[6^4,5^2,1^{10}\right]\;,\quad\left[22^2\right]\;,\quad\left[15,14^2,1\right]\;.
\end{equation}
This indeed agrees with the class $\mathcal{S}$ theory found for theory 42.
\item Consider the case
\bes{
[1\cdot {\bf 7}][2]\,\, \overset{\mathfrak{so}_{7}}2 \,\,{\overset{\mathfrak{usp}_{4}}1} \,\, \underset{[1]}{\overset{\mathfrak{so}_{16}}4} \,\,  \underset{[1]}{\overset{\mathfrak{usp}_{10}}1} \,\, {\overset{\mathfrak{so}_{18}}4} \,\, \overset{\mathfrak{usp}_{10}}1 \,\, [9]\;. }
Here, we have $k=9$, $n=6$, $m_1 = m_2 = 3$ (two fundamental half-hypers of the leftmost $\mathfrak{usp}_{10}$), $m_3=m_4=4$ (two vector half-hypers of $\mathfrak{so}_{16}$), $m_5=m_6=m_7=m_8=6$ (four spinor half-hypers of the $\mathfrak{so}_7$). We then associate to it the class $\mathcal{S}$ theory
\begin{equation}
    \left[6^2,4^2,3^2,1^{18}\right]\;,\quad\left[22^2\right]\;,\quad\left[15,14^2,1\right]\;.
\end{equation}
We note that the global symmetry of this theory is indeed $\USp(4)\times\SU(2)^2\times\text{U}(1)\times\SO(18)$ as expected.
\item Consider the case
\bes{
[1\cdot {\bf 7}]\,\, \overset{\mathfrak{so}_{7}}2 \,\,\underset{\left[\frac{1}{2}\right]}{\overset{\mathfrak{usp}_{8}}1} \,\, {\overset{\mathfrak{so}_{23}}4} \,\,  \underset{\left[\frac{7}{2}\right]}{\overset{\mathfrak{usp}_{22}}1} \,\, {\overset{\mathfrak{so}_{30}}4} \,\, \overset{\mathfrak{usp}_{22}}1 \,\, [15]\;. }
Here, we have $k=15$, $n=6$, $m_1 = m_2 = m_3=m_4=m_5=m_6=m_7=3$ (seven fundamental half-hypers of the leftmost $\mathfrak{usp}_{22}$), $m_8=5$ (one fundamental half-hyper of $\mathfrak{usp}_{8}$). We then associate to it the class $\mathcal{S}$ theory
\begin{equation}
    \left[3^6,1^{30}\right]\;,\quad\left[25,23\right]\;,\quad\left[19,14^2,1\right]\;.
\end{equation}
We note that the global symmetry of this theory is indeed $\SO(7)\times\SU(2)\times\SO(30)$ as expected.
\end{itemize}

Note that this class is only valid for cases where there is a bifundamental in the spinor of the $\mathfrak{so}_7$. Cases with no bifundamentals (like theory 30 in $D_5$, see Appendix \ref{D5}) or cases with a vector bifundamental (like theory 28 in $D_5$, see Appendix \ref{D5}) should be treated as part of the bifurcation family.

\subsubsection{Cases with \texorpdfstring{$\mathfrak{g}_2$}{g2}}\label{g2}
We have previously seen that quivers ending on a $-2$ curve with a $\mathfrak{g}_2$ gauge factor can be described as part of the bifurcation family. However, in that description, the $\mathfrak{g}_2$ must have at least two fundamental hypers. Nevertheless, there are 6d quivers of this type where the $\mathfrak{g}_2$ has fewer flavours, which then require special treatment. As before, we shall associate to the 6d quiver a collection of $m_i$'s specifying the locations of the half-hypers, including the $\mathfrak{g}_2$ fundamentals. A special feature of this class is that there are actually 9 such half-hypers in the quiver, so we also introduce $m_9$. Additionally, we shall denote the number of tensors by $n$, which is always even here. We claim that the associated class $\mathcal{S}$ theory is given by the punctures
\begin{align} \label{classSg2}
    \begin{split}
        &\left[m_6,m_5,m_4,m_3,m_2,m_1,1^{2k}\right]\;,\\
        &\left[3n-m_7+2, 3n-m_8+2, 3n-m_9+2,1\right]\;,\\
        &\left[6n+4-m_9-m_8-m_7,3n+3\right]\;,
    \end{split}
\end{align}
where as before, the $m_i$'s are ordered so that $m_{i+1}\geq m_i$.

Let us illustrate this with some examples.
\begin{itemize}
    \item Consider the quiver
\bes{
\overset{\mathfrak{g}_{2}}2 \,\,\underset{\left[\frac{9}{2}\right]}{\overset{\mathfrak{usp}_{8}}1} \,\, {\overset{\mathfrak{so}_{16}}4} \,\,  \overset{\mathfrak{usp}_{8}}1 \,\, [8]\;. }
Here, we have $k=8$, $n=4$, $m_1=m_2=m_3=m_4=m_5=m_6=m_7=m_8=m_9=3$. The associated class $\mathcal{S}$ theory is then given by the punctures
\begin{equation}
    \left[3^6,1^{16}\right]\;,\quad\left[11^3,1\right]\;,\quad\left[19,15\right]\;.
\end{equation}
One can note that the global symmetry is indeed $\SO(9)\times\SO(16)$ as expected.
\item As another example, consider the case
\bes{
[1]\,\, \overset{\mathfrak{g}_{2}}2 \,\,\underset{\left[\frac{5}{2}\right]}{\overset{\mathfrak{usp}_{6}}1} \,\, \underset{[1]}{\overset{\mathfrak{so}_{16}}4} \,\,  \overset{\mathfrak{usp}_{8}}1 \,\, [8]\;. }
Here, we have $k=8$, $n=4$, $m_1=m_2=2$, $m_3=m_4=m_5=m_6=m_7=3$, $m_8=m_9=4$. The associated class $\mathcal{S}$ theory is then given by the punctures
\begin{equation}
    \left[3^4,2^2,1^{16}\right]\;,\quad\left[11,10^2,1\right]\;,\quad\left[17,15\right]\;.
\end{equation}
One can note that the global symmetry is indeed $\SO(5)\times\SU(2)^2\times\SO(16)$ as expected.
\end{itemize}

Note that the class $\mathcal{S}$ theory will only make sense if $m_{7,8,9}$ are all odd, or if one is odd and the other two are even and equal. However, there are cases where all are even or where only one of them is even. These cases then behave similarly to the non-integer $M$ cases discussed previously. Specifically, it is possible to shuffle the $m_i$'s such that the resulting class $\mathcal{S}$ theory is sensible, but then the theory will be either ugly or bad. A common example of this is the cases where the $\mathfrak{g}_2$ factor has more than one fundamental hyper.

For instance, consider the case
\bes{
[2]\,\, \overset{\mathfrak{g}_2}2  \,\, \underset{\left[\frac{3}{2}\right]}{\overset{\mathfrak{usp}_4}1} \,\, \underset{[1]}{\overset{\mathfrak{so}_{14}}4}\,\, \overset{\mathfrak{usp}_6}1 \,\, [7]
}
corresponding to $k=7$, $n=4$, $m_1=m_2=2$, $m_3=m_4=m_5=3$ and $m_6=m_7=m_8=m_9=4$. A naive application of the above prescription assigns the following class $\mathcal{S}$ theory to this quiver: $\left[4,3^3,2^2,1^{16}\right]$, $\left[10^3,1\right]$, $\left[16,15\right]$, which obviously does not make sense. Nevertheless, we can get a sensible class $\mathcal{S}$ theory by making the substitution $m_7\leftrightarrow m_5$, in which case we get the theory
\begin{equation}
    \left[4^2,3^2,2^2,1^{14}\right]\;,\quad\left[11,10^2,1\right]\;,\quad\left[17,15\right]\;.
\end{equation}
This is an ugly class $\mathcal{S}$ theory whose global symmetry appears to be $\USp(4)\times\SU(2)\times\SU(2)\times\SO(14)$ as expected. This quiver was also considered previoulsly as part of the bifurcation family, where we associated with it the class $\mathcal{S}$ theory
\begin{equation}
    \left[5,3^3,2^2,1^{14}\right]\;,\quad\left[16^2\right]\;,\quad\left[11,10^2,1\right]\;.
\end{equation}
The above result suggests a duality between the two descriptions.

There can be cases where all shufflings that give sensible class $\mathcal{S}$ theories result in bad theories. For cases in which the $\mathfrak{g}_2$ sees more than one fundamental hyper, we know that there is an ugly description given by treating it in the bifurcation family (this is usually preferable because of this). For other cases, however, it is not clear whether a non-bad class $\mathcal{S}$ description exists or not.

As examples, consider the following quivers:
\bes{
[1]\,\, \overset{\mathfrak{g}_{2}}2 \,\, {\overset{\mathfrak{usp}_{6}}1} \,\, \underset{[1]}{\overset{\mathfrak{so}_{21}}4} \,\,  \overset{\mathfrak{usp}_{18}}1 \,\, {\overset{\mathfrak{so}_{31}}4} \,\,  \underset{\left[\frac{5}{2}\right]}{\overset{\mathfrak{usp}_{28}}1}  \,\, {\overset{\mathfrak{so}_{36}}4} \,\,  \overset{\mathfrak{usp}_{28}}1 \,\, [18] }
and
\bes{
[2]\,\, \overset{\mathfrak{g}_{2}}2 \,\, {\overset{\mathfrak{usp}_{4}}1} \,\, {\overset{\mathfrak{so}_{17}}4} \,\,  \overset{\mathfrak{usp}_{14}}1 \,\, {\overset{\mathfrak{so}_{27}}4} \,\,  \underset{\left[\frac{5}{2}\right]}{\overset{\mathfrak{usp}_{24}}1}  \,\, {\overset{\mathfrak{so}_{32}}4} \,\,  \overset{\mathfrak{usp}_{24}}1 \,\, [16]\;. }
In both cases, all sensible class $\mathcal{S}$ theories end up being bad. For the second one, we can find an ugly description in the bifurcation family, but for the first case, it is not clear whether such a description exists.

\subsubsection{Cases with an Empty \texorpdfstring{$-2$}{-2} Curve}\label{SU1}
There are a couple of exceptions that appear in the presence of a $-2$ curve supporting an $\mathfrak{su}_{1}$ gauge symmetry, which we shall next discuss here. 

\paragraph{Cases based on the sequence \texorpdfstring{$\mathfrak{su}_{1}-\mathfrak{usp}_{2}-\mathfrak{so}_{19}$}{su1-usp2-so19}}
In the beginning of this section, we introduced the family of 6d SCFTs ending with a $-2$ curve supporting an $\mathfrak{su}$ type gauge symmetry. In particular, this family includes cases where the $-2$ curve is empty and is connected by a $\mathfrak{usp}_{2}$ symmetry on a $-1$ curve. We have said that the empty $-2$ curve can be regarded as having an $\mathfrak{su}_{1}$ symmetry for the purpose of determining the value of the $m_i$'s. Furthermore, as a $\mathfrak{usp}_{2}$ on a $-1$ curve must see $20$ fundamental half-hypers, the connecting $\mathfrak{so}$ symmetry on the adjacent $-1$ curve can be at most $\mathfrak{so}_{18}$. However, the 6d SCFT made from an empty $-2$ curve and a $-1$ curve supporting a $\mathfrak{usp}_{2}$ symmetry is known to have $\SO(19)$ global symmetry. Thus, one can build 6d SCFTs starting from a sequence of an empty $-2$,  a $-1$ curve with $\mathfrak{usp}_{2}$ gauge symmetry and a $-4$ curve with $\mathfrak{so}_{19}$ gauge symmetry. While structurally similar, this family of quivers is not covered by the class of quivers discussed in \cref{suusp}, and instead should be considered separately.

One distinguishing feature of this family of quivers, compared to the cases in \cref{suusp}, is that they contain $9$ fundamental half-hypers scattered along the quiver. As such, to specify the quiver, we again use the labels $m_i$ for the locations of the flavours, where now $i=1,2,...,9$. As before, we also use $k$ and $n$, which here is equal to the number of tensors (and as such is always even). We claim then that the associated class $\mathcal{S}$ theory is given by the punctures
\begin{align} \label{IISU2SO19classS}
    \begin{split}
        &\lambda=\left[m_6,m_5,m_4,m_3,m_2,m_1,1^{2k}\right]\;,\\
        &\mu=\left[3n-m_7, 3n-m_8, 3n-m_9,1\right]\;,\\
        &\nu=\left[6n-m_9-m_8-m_7,3n+1\right]\;.
    \end{split}
\end{align}

Let us illustrate this with some examples.
\begin{itemize}
    \item Consider the quiver
\bes{
2 \,\,\overset{\mathfrak{usp}_{2}}1 \,\, \underset{\left[1\right]}{\overset{\mathfrak{so}_{19}}4} \,\, \underset{\left[\frac{5}{2}\right]}{\overset{\mathfrak{usp}_{18}}1} \,\, \underset{[1]}{\overset{\mathfrak{so}_{28}}4} \,\, \overset{\mathfrak{usp}_{20}}1 \,\, [14]\;. }
Here, we have $k=14$, $n=6$, $m_9=m_8=4$, $m_7=m_6=m_5=m_4=m_3=3$, $m_2=m_1=2$. The associated class $\mathcal{S}$ theory is then given by the punctures
\begin{equation}
    \left[3^4,2^2,1^{28}\right]\;,\quad\left[15,14^2,1\right]\;,\quad\left[25,19\right]\;.
\end{equation}
One can note that the global symmetry is indeed $\SU(2)\times \SU(2)\times \SO(5)\times\SO(28)$ as expected.
\item As another example, consider the case
\bes{
2 \,\,\overset{\mathfrak{usp}_{2}}1 \,\, {\overset{\mathfrak{so}_{19}}4} \,\, \underset{\left[\frac{9}{2}\right]}{\overset{\mathfrak{usp}_{20}}1} \,\, {\overset{\mathfrak{so}_{28}}4} \,\, \overset{\mathfrak{usp}_{20}}1 \,\, [14]\;. }
Here, we have $k=14$, $n=6$, $m_1=m_2=m_3=m_4=m_5=m_6=m_7=m_8=m_9=3$. The associated class $\mathcal{S}$ theory is then given by the punctures
\begin{equation}
    \left[3^6,1^{28}\right]\;,\quad\left[15^3,1\right]\;,\quad\left[27,19\right]\;.
\end{equation}
One can note that the global symmetry is indeed $\SO(9)\times\SO(28)$ as expected.
\item Finally, consider the case
\bes{
2 \,\,\overset{\mathfrak{usp}_{2}}1 \,\, \underset{[1]}{\overset{\mathfrak{so}_{19}}4} \,\, \underset{\left[3\right]}{\overset{\mathfrak{usp}_{18}}1} \,\, {\overset{\mathfrak{so}_{27}}4} \,\, \underset{\left[\frac{1}{2}\right]}{\overset{\mathfrak{usp}_{20}}1} \,\, {\overset{\mathfrak{so}_{28}}4} \,\, \overset{\mathfrak{usp}_{20}}1 \,\, [14]\;. }
Here, we have $k=14$, $n=8$, $m_9=m_8=6$, $m_7=m_6=m_5=m_4=m_3=m_2=5$, $m_1=3$. The associated class $\mathcal{S}$ theory is then given by the punctures
\begin{equation}
    \left[5^5,3,1^{28}\right]\;,\quad\left[19,18^2,1\right]\;,\quad\left[31,25\right]\;.
\end{equation}
One can note that the global symmetry is indeed $\SU(4)\times \SU(2)\times\SO(28)$ as expected.
\end{itemize}

Note that we can also treat cases ending with the sequence $\mathfrak{su}_{1}-\mathfrak{usp}_{2}-\mathfrak{so}_{18}$ with this family, where now $m_9=n-1$. Similarly, it is possible to consider cases where the leftmost $\mathfrak{so}_M$ group has $M<18$, in which case more $m_i$'s would be equal to $n-1$. One can show that the resulting class $\mathcal{S}$ theory is the same as the one we get if we treat this case as part of the class of theories introduced in \cref{suusp}, specifically using \eqref{ClassSpres1}. As such, these cases can be treated by both approaches.

Finally, we comment that like the previous cases involving $\mathfrak{g}_{2}$, if the resulting class $\mathcal{S}$ theory is sensible then it would also be good. However, if it is not sensible, we can usually get a sensible one by shuffling the $m_i$'s, but this will result in an ugly or bad class $\mathcal{S}$ theory.

\paragraph{Cases involving a rank 2 E-string}
The sequence involving an empty $-2$ curve and a $-1$ curve gives rise to a rank 2 E-string theory. There are cases of 6d SCFTs in this class that end with a rank 2 E-string. These can usually be dealt with using the prescription in \cref{suusp}, where we regard the $-2$ curve as supporting an $\mathfrak{su}_{1}$ gauge symmetry and the $-1$ curve as supporting a $\mathfrak{usp}_{0}$ gauge symmetry for the purpose of flavour considerations. This works well as long as the $\mathfrak{so}_{N}$ on the $-4$ curve adjacent to the empty $-1$ curve obeys $N\leq 14$. This is as according to the rules in \cref{suusp}: an empty $-1$ curve sees $8$ flavours, where in this case, $1$ is always contributed by the empty $-2$ curve, which supports an $\mathfrak{su}_{1}$ for the purpose of flavour calculations. However, there are cases with $N=15$ or $N=16$ that use the $\SO(16)$ subgroup of $E_8$. These cases should then be treated separately, which we shall do here.

As in the rest of the cases so far, the quiver can be described by the parameter $k$, the number of tensors $n$ (which is always even here), and the location of the flavours along the quiver denoted by the $m_i$'s. In the class of quivers, we in general have $8$ flavours along the quiver, so $i=1,2,...,8$. Given a quiver in this class, its reduction to 4d is described by a D-type class $\mathcal{S}$ theory associated with a three-punctured sphere with the punctures
\begin{align} \label{Estringr2classS}
    \begin{split}
        &\lambda=\left[m_6,m_5,m_4,m_3,m_2,m_1,1^{2k}\right]\;,\\
        &\mu=\left[k+M-\Delta m_+ -1, \left(\frac{k+M+\Delta m_+}{2}\right)^2,1\right]\;,\\
        &\nu=\left[k+M+\Delta m_-,k+M-\Delta m_-\right]\;.
    \end{split}
\end{align}

We note the followings:
\begin{itemize}
    \item Here, we treat the empty $-2$ curve as having no flavours (so no $m_i$ is equal to $n$) and also as contributing no flavours (so all the eight flavours of the empty $-1$ curve come from the adjacent $-4$ curve or from $m_i$'s associated to it). In addition, $M$, $\Delta m_+$ and $\Delta m_-$ are as defined in \eqref{MandDm}.
    \item Here, $k+M+\Delta m_+$ is always even, so the class $\mathcal{S}$ theory is well defined. This can be seen as one can then Higgs down the $\SU(2)$ flavour symmetry of the rank 2 E-string, which turns the theory into one in the bifurcation class for which $k+M+\Delta m_+$ is always even. We also note that if $N=16$, then the bifurcation theory is the one with two variants depending on the $\theta$ angle, though here we always get the variant with identical $\theta$ angles.
    \item The class $\mathcal{S}$ theory is not explicitly dependent on $n$, though it does enter through $k$ and the $m_i$'s. Still, $n$ is useful in the relation with the enumeration of the theories using partitions in \cref{PartitionDesc}.
\end{itemize}

Next, we shall illustrate this with a few examples. 
\begin{itemize}
    \item Consider the quiver
\bes{
2 \,\, 1 \,\, \underset{\left[4\right]}{\overset{\mathfrak{so}_{16}}4} \,\, {\overset{\mathfrak{usp}_{8}}1} \,\, [8]\;. }
Here, we have $k=8$, $n=4$, $m_8=m_7=m_6=m_5=m_4=m_3=m_2=m_1=2$. The associated class $\mathcal{S}$ theory is then given by the punctures
\begin{equation}
    \left[2^6,1^{16}\right]\;,\quad\left[11,8^2,1\right]\;,\quad\left[14^2\right]\;.
\end{equation}
One can note that the global symmetry is indeed $\USp(8)\times \SU(2)\times \SO(16)$ as expected.
\item Consider the quiver
\bes{
2 \,\, 1 \,\, \underset{\left[1\right]}{\overset{\mathfrak{so}_{16}}4} \,\, \underset{\left[2\right]}{\overset{\mathfrak{usp}_{14}}1} \,\, \underset{\left[1\right]}{\overset{\mathfrak{so}_{24}}4} \,\, {\overset{\mathfrak{usp}_{16}}1} \,\, [12]\;. }
Here, we have $k=12$, $n=6$, $m_8=m_7=4$, $m_6=m_5=m_4=m_3=3$, $m_2=m_1=2$. The associated class $\mathcal{S}$ theory is then given by the punctures
\begin{equation}
    \left[3^4,2^2,1^{24}\right]\;,\quad\left[15,12^2,1\right]\;,\quad\left[20^2\right]\;.
\end{equation}
One can note that the global symmetry is indeed $\SO(4)\times \SU(2)^3\times \SO(24)$ as expected.
\item Consider the quiver
\bes{
2 \,\, 1 \,\, {\overset{\mathfrak{so}_{15}}4} \,\, \underset{\left[\frac{7}{2}\right]}{\overset{\mathfrak{usp}_{14}}1} \,\, {\overset{\mathfrak{so}_{22}}4} \,\, {\overset{\mathfrak{usp}_{14}}1} \,\, [11]\;. }
Here, we have $k=11$, $n=6$, $m_8=5$, $m_7=m_6=m_5=m_4=m_3=m_2=m_1=3$. The associated class $\mathcal{S}$ theory is then given by the punctures
\begin{equation}
    \left[3^6,1^{22}\right]\;,\quad\left[15,12^2,1\right]\;,\quad\left[21,19\right]\;.
\end{equation}
One can note that the global symmetry is indeed $\SO(7)\times \SU(2)\times \SO(22)$ as expected.
\item Finally, we note that this class can also be used for cases with $N\leq 14$, which can be treated with the prescription in \cref{suusp}, and the resulting class $\mathcal{S}$ theory is the same. For example, consider the quiver
\bes{
2 \,\, 1 \,\, \underset{\left[2\right]}{\overset{\mathfrak{so}_{14}}4} \,\, \underset{\left[1\right]}{\overset{\mathfrak{usp}_{8}}1} \,\, {\overset{\mathfrak{so}_{16}}4} \,\, {\overset{\mathfrak{usp}_{8}}1} \,\, [8]\;. }
Here, we have $k=8$, $n=6$, $m_8=m_7=5$, $m_6=m_5=m_4=m_3=4$, $m_2=m_1=3$. The associated class $\mathcal{S}$ theory is then given by the punctures
\begin{equation}
    \left[4^4,3^2,1^{16}\right]\;,\quad\left[13,12^2,1\right]\;,\quad\left[19^2\right]\;.
\end{equation}
One can note that the global symmetry is indeed $\USp(4)\times \SU(2)\times U(1)^2\times \SO(16)$ as expected.

Note that we can also treat this using the prescription in \cref{suusp}. There, we have that $k=8$, $n=6$, $m_8=m_7=6$, $m_6=m_5=m_4=m_3=4$, $m_2=m_1=3$. Using \eqref{ClassSpres1}, we indeed get the same result as above.
\end{itemize}

\subsubsection{Cases Involving an \texorpdfstring{$\mathfrak{so}_{13}$}{so13} with a Spinor}\label{SO13wS}

Another exception is quivers ending with an $\mathfrak{so}_{13}$ gauge symmetry with a half-hyper in its spinor representation. Since the dimension of the spinor representation of $\mathfrak{so}_{13}$ is ${\bf 64}$, such cases cannot be recovered from the bifurcation family and instead comprise a special case. This is further apparent in that these cases involve nine fundamental half-hypers scattered along the quiver. As before, the quiver can be described by the parameter $k$, the number of tensors $n$ (which is always even here) and the location of the flavours, $m_i$ for $i=1,2,...,9$. We claim then that the 4d reduction of this class of theories is described by a D-type class $\mathcal{S}$ theory associated with a three-punctured sphere with the following punctures:
\begin{align} \label{SO13wSclassS}
    \begin{split}
        &\lambda=\left[m_6,m_5,m_4,m_3,m_2,m_1,1^{2k}\right]\;,\\
        &\mu=\left[3n-m_7+4,3n-m_8+4,3n-m_9+4,1\right]\;,\\
        &\nu=\left[6n+8-m_7-m_8-m_9,3n+5\right]\;.
    \end{split}
\end{align}

Next, we shall illustrate this with a few examples.
\begin{itemize}
    \item Consider the quiver
\bes{
\left[\frac{1}{2}{\bf 64}\right]\,\, \underset{[1]}{\overset{\mathfrak{so}_{13}}2} \,\,\underset{\left[\frac{7}{2}\right]}{\overset{\mathfrak{usp}_{12}}1} \,\, {\overset{\mathfrak{so}_{20}}4} \,\,  {\overset{\mathfrak{usp}_{12}}1} \,\, [10]\;. }
Here, we have $k=10$, $n=4$, $m_9=m_8=4$, $m_7=m_6=m_5=m_4=m_3=m_2=m_1=3$. The associated class $\mathcal{S}$ theory is then given by the punctures
\begin{equation}
    \left[3^6,1^{20}\right]\;,\quad\left[13,12^2,1\right]\;,\quad\left[21,17\right]\;.
\end{equation}
One can note that the global symmetry is indeed $\SO(7)\times \SU(2)\times \SO(20)$ as expected.
\item Consider the quiver
\bes{
\left[\frac{1}{2}{\bf 64}\right]\,\, {\overset{\mathfrak{so}_{13}}2} \,\,\underset{\left[4\right]}{\overset{\mathfrak{usp}_{14}}1} \,\, {\overset{\mathfrak{so}_{23}}4} \,\,  \underset{\left[\frac{1}{2}\right]}{\overset{\mathfrak{usp}_{16}}1} \,\, {\overset{\mathfrak{so}_{24}}4} \,\, {\overset{\mathfrak{usp}_{16}}1} \,\, [12]\;. }
Here, we have $k=12$, $n=6$, $m_9=m_8=m_7=m_6=m_5=m_4=m_3=m_2=5$, $m_1=3$. The associated class $\mathcal{S}$ theory is then given by the punctures
\begin{equation}
    \left[5^5,3,1^{24}\right]\;,\quad\left[17^3,1\right]\;,\quad\left[29,23\right]\;.
\end{equation}
One can note that the global symmetry is indeed $\SO(8)\times \SO(24)$ as expected.
\end{itemize}

Finally we note that like the previous cases involving $9$ $m_i$'s, there can be cases in which the class $\mathcal{S}$ theory given by \eqref{SO13wSclassS} does not make sense, notably if only one or all of $m_7$, $m_8$, $m_9$ are even. In this case, we can get a sensible class $\mathcal{S}$ theory by shuffling the $m_i$'s, though the resulting theory may end up being ugly or bad.

\subsection{Comparisons with \texorpdfstring{$D_4$}{D4} and \texorpdfstring{$D_5$}{D5}}\label{comparisonsD4D5}
It is interesting to compare the above prescriptions with the results we found for the $D_4$ and the $D_5$ cases as listed in \cref{examples}. We find that the above prescriptions allow us to reproduce all the class $\mathcal{S}$ theories found in both cases.

Specifically, in the $D_4$ case, there are 19 theories, for which we find class $\mathcal{S}$ descriptions for 13 cases (we again refer the reader to \cref{examples} for the 6d SCFT associated with each number): 4, 6, 8, 9, 10, 11, 12, 13, 14, 15, 16, 17, 18. Out of these,
\begin{itemize}
    \item cases 4, 11, 13, 14, 15 and 16 are described by the bifurcation family;
    \item cases 6, 8, 9 and 17 are described by the $\mathfrak{su}$ on $-2$ family;
    \item cases 10, 12 and 18 are part of the exceptions.
\end{itemize}
For the remaining cases where we have not found class $\mathcal{S}$ descriptions, we note that theories 0, 1 and 7 have a unique form involving a sequence of $-1\,\,-2\,\,(-2)\,\,-3$ curves, while theories 2, 3 and 5 have $\Delta n>1$.

Similarly, in $D_5$, out of 45 cases, we find class $\mathcal{S}$ theory descriptions in 31 cases, all of which can be reproduced from the above prescription. Out of these cases,
\begin{itemize}
    \item cases 7, 11, 17, 18, 21, 26, 27, 28, 29, 30, 33, 38, 41 and 43 are described by the bifurcation family;
    \item cases 9, 14, 31, 32, 34, 35, 36, 37, 39, 40 and 44 are described by the $\mathfrak{su}$ on $-2$ family;
    \item cases 2, 6, 10, 12, 13 and 42 are part of the exceptions.
\end{itemize}
For the remaining cases, we have not found class $\mathcal{S}$ descriptions. We note that theories 0, 1, 3, 4 have a unique form involving a sequence of $-1\,\,-2\,\,(-2)\,\,-3$ curves, while theories 5, 8, 15, 16, 19, 22, 23 and 24 have $\Delta n>1$. Cases 20 and 25 fall into the class of exceptions with no class $\mathcal{S}$ descriptions.

Overall, it appears that the above prescriptions allow us to reproduce the class $\mathcal{S}$ description given a 6d quiver in this family, if there exists a class $\mathcal{S}$ description. Furthermore, it appears to suggest that a given quiver in this class will not have a 4d D-type class $\mathcal{S}$ description if
\begin{itemize}
    \item it ends with a sequence of curves $-1\,\,-2\,\,-3$ or $-1\,\,-2\,\,-2\,\,-3$,
    \item or it is in the bifurcation family with $\Delta n>1$ (and is not part of the exceptions),
    \item or it is one of the handful of exceptions with no class $\mathcal{S}$ descriptions.
\end{itemize}

Finally, we remind the reader that these are expectations from the exhaustive search for $k=4,5$, and we do not at present have a proof of this statement. As such, it is possible that there are additional cases, appearing only for $k>5$ that were missed. We also note that the search assumes a special form that has the right property to describe the 4d torus reduction of generic 6d SCFTs in this class. It is possible that certain sporadic cases can be realized in class $\mathcal{S}$ but with different puncture structures or choices of ADE groups. This could also include cases for which we have not found a D-type class $\mathcal{S}$ description in the proposed form.      

\section{Relations with the Description in Terms of Partitions} \label{PartitionDesc}

As previously mentioned, the space of 6d SCFTs associated with the orbi-instanton theories on D-type singularities was analysed in \cite{Frey:2018vpw}, where the authors proposed an enumeration of the possible theories in terms of a D-type partition $\bm{p}$, and a choice of Lie algebra $\mathfrak{g}$. The above rules for the association of a class $\mathcal{S}$ theory with a 6d quiver can also be recast as a rule for the association of a class $\mathcal{S}$ theory with the choice of the partition and the gauge algebra, which is presented next. The constraints on the variables $m_i$ can then be determined in terms of the partition $\bm{p}$ (and the gauge algebra $\mathfrak{g}$). The results are organized in terms of the three main cases, which are then further split into subcases as done in \cite[Section 3.3.3]{Frey:2018vpw}. We refer the reader to the reference for further details on each subcase.

We begin by listing the cases for which we find no class $\mathcal{S}$ description. The results presented above suggest that the following cases have no description in D-type class $\mathcal{S}$:
\begin{itemize}
    \item Case (iii) ($\bm{p}^\text{T}_1 + \bm{p}^\text{T}_2<6$).
    \item Case (ii) ($\bm{p}^\text{T}_1<8$, $\bm{p}^\text{T}_1+\bm{p}^\text{T}_2\geq6$): subcases (a), (b), (c) and subcase (f) for $P>0$, where $\mathfrak{g}=\mathfrak{usp}_{2P}$.
    \item Case (i) ($\bm{p}^\text{T}_1 \geq 8$): subcase (h) when $\left|P-\frac{\bm{p}^\text{T}_1}{2}+4\right|> 1$, where for this subcase we have $\mathfrak{g}=\mathfrak{usp}_{2P}$.
\end{itemize}

For the remaining cases, we do find a description in terms of D-type class $\mathcal{S}$ as presented above. This description is determined from $k$, $n$, and the $m_i$'s. Here, $k$ is determined from the choice of the binary dihedral group while $n$ is related to the number of tensors and is not determined from the partition data. The $m_i$'s are determined from the partition data and the choice of the gauge algebra. Finally, we need to specify one of the above cases, which then determines how this data is translated into the three punctures specifying the 4d class $\mathcal{S}$ theory. Below we list the various cases in \cite{Frey:2018vpw} for which we do find a class $\mathcal{S}$ description. For each case, we provide the rules to determine the $m_i$'s from $\bm{p}$ and $\mathfrak{g}$ ($k$ and $n$ are taken to be known), and to which case it is associated to in the above list.

\begin{itemize}
    \item Case (i): $\bm{p}^\text{T}_1 \geq 8$
    \begin{itemize}
    \item Subcases (a) and (b), except (a) for $N=4, \bm{p}^\text{T}_1 = 8, \bm{p}^\text{T}_2 \leq 2$, and (b) for $\bm{p}^\text{T}_2 = 9$ or $\bm{p}^\text{T}_1 = 8$, $ \bm{p}^\text{T}_2 \geq 7$: Here, we have $\mathfrak{g}=\mathfrak{su}_{N}$, and we take $\bm{p} = \left[n_i,1^{2N+l}\right]$, where $i$ goes over all columns whose height is bigger than $1$, so $n_i>1$ (note that the number of such columns is at most eight). We have $m_i = n-n_i$. Then $l$ of the remaining $m_i$'s are equal to $n-1$, and the rest are equal to $n$. The resulting quiver is of the type discussed in \cref{suusp}, and the associated class $\mathcal{S}$ theory is as given in \eqref{ClassSpres1} and \eqref{ClassSpres2}.
    \item Subcase (a) for $N=4, \bm{p}^\text{T}_1 = 8, \bm{p}^\text{T}_2 \leq 2$: This subcase describes the family of exceptions in \ref{SU4xSO8}. The class $\mathcal{S}$ theory is determined by $k$, $n$ and $\bm{p}_1, \bm{p}_2$. Specifically, if $\bm{p}_1 = \bm{p}_2 = $even, then we get the case of \eqref{su4Xso10part}. If instead $\bm{p}_1, \bm{p}_2$ are odd, then we get the case \eqref{su4Xso10partl} with $l=2k-\bm{p}_1-3$.
    \item Subcase (b) for $\bm{p}^\text{T}_2 = 9$: This subcase describes the family of exceptions in \cref{SU1} associated with the $\mathfrak{su}_{1}-\mathfrak{usp}_{2}-\mathfrak{so}_{19}$ sequence. The partition in this case takes the form $[n_i,1]$, for $i=1,2,...,9$ and $n_i>1$. The class $\mathcal{S}$ theory is then given by \eqref{IISU2SO19classS}, with $m_i = n-n_i$.
    \item Subcase (b) for $\bm{p}^\text{T}_1 = 8$, $ \bm{p}^\text{T}_2 \geq 7$: This subcase describes the family of exceptions in \cref{SU1} associated with the rank 2 E-string. The partition in this case takes the form $[n_i]$, for $i=1,2,...,8$, where here $n_i$ can also be equal to one. The class $\mathcal{S}$ theory is then given by \eqref{Estringr2classS}, with $m_i = n-n_i$.
    \item Subcase (c) for $\bm{p}^\text{T}_1 \neq 8$: Here, we have that $\mathfrak{g}=\mathfrak{so}_{7}$, and we take $\bm{p} = \left[n_i,1^{8+l}\right]$, where as before $n_i>1$. We have $m_i = n-n_i$. Then $l$ of the remaining $m_i$'s are equal to $n-1$, and the rest are equal to $n$. The resulting quiver belongs to the family of exceptions ending with an $\mathfrak{so}_{7}$ on a $-2$ curve, discussed in \cref{spinbifund}. The associated class $\mathcal{S}$ theory is then given by \eqref{classS:so7on2}.
    \item Subcase (c) for $\bm{p}^\text{T}_1 = 8$: Here, we have that $\mathfrak{g}=\mathfrak{so}_{7}$, and we take $\bm{p} = \left[2k-7,1^{7}\right]$. This case belongs to the bifurcation family discussed in \cref{bifurusp}, and the associated class $\mathcal{S}$ theory is then given by \eqref{ClassSpres2} with $m_8=m_7=m_6=m_5=m_4=n$, $m_3=m_2=n-1$, $m_1=n-2k+6$.
    \item Subcase (d): Here, we have that $\mathfrak{g}=\mathfrak{so}_{7}$, and we take $\bm{p} = \left[n_i,1^{7+l}\right]$, where as before $n_i>1$. We always have that $m_8 = m_7 = m_6 = m_5 = m_4 =n$. The remaining three $m_i$'s are then determined from $n_i$ and $l$, where $l$ of the remaining $m_i$'s are equal to $n-2$ and the rest are given by $m_i = n-n_i-1$. The resulting quiver belongs to the bifurcation family discussed in \cref{bifurusp}, and the associated class $\mathcal{S}$ theory is then given by \eqref{ClassSpres2}.
    \item Subcase (e): Here, we have that $\mathfrak{g}=\mathfrak{so}_{8}$, and we take $\bm{p} = \left[n_i,1^{8+l}\right]$, where as before $n_i>1$. We always have that $m_8 = m_7 = m_6 = m_5 =n$. The remaining four $m_i$'s are then determined as follows. First, for all $n_i$, we have that $m_i = n-n_i-1$. Next, $l$ of the remaining $m_i$'s are equal to $n-2$, and if there are still $m_i$'s left undetermined, then these are equal to $n-1$. The resulting quiver belongs to the bifurcation family discussed in \cref{bifurusp}, and the associated class $\mathcal{S}$ theory is then given by \eqref{ClassSpres2}.
    \item Subcase (f) for $12 \geq M\geq 9$: Here, we have that $\mathfrak{g}=\mathfrak{so}_{M}$. Given a choice of $M$, we take $\bm{p} = \left[n_i,1^{M+l}\right]$, where as before $n_i>1$. Starting with $m_8$, $12-M$ of the $m_i$'s are equal to $n$. The remaining $m_i$'s are then determined as follows. First, for all $n_i$, we have that $m_i = n-n_i-1$. Next, $l$ of the remaining $m_i$'s are equal to $n-2$, and if there are still $m_i$'s left undetermined, then these are equal to $n-1$. The resulting quiver belongs to the bifurcation family discussed in \cref{bifurusp}, and the associated class $\mathcal{S}$ theory is then given by \eqref{ClassSpres1} and \eqref{ClassSpres2}. We note that for $M=12$, there are two distinct 6d quivers and 4d class $\mathcal{S}$ theories differing by the choice of the $\theta$ angle.
    \item Subcase (f) for $M=13$: Here, we have that $\mathfrak{g}=\mathfrak{so}_{13}$, and we take $\bm{p} = \left[n_i,1^{13+l}\right]$, where as before $n_i>1$. This case then belongs to the exceptions ending with an $\mathfrak{so}_{13}$ gauge algebra with spinor half-hyper, discussed in \cref{SO13wS}, for which the class $\mathcal{S}$ theory is given in \eqref{SO13wSclassS}. The nine $m_i$'s are determined from the $n_i$'s and $l$, where we first have that $m_i = n-n_i$ for all $i$, and $l$ of the remaining $m_i$'s are equal to $n-1$. The remaining undetermined $m_i$'s, if any, are equal to $n$.
    \item Subcase (g): Here, we have that $\mathfrak{g}=\mathfrak{g}_{2}$, and we take $\bm{p} = \left[n_i,1^{7+l}\right]$, where as before $n_i>1$. We denote the number of $n_i$'s by $p$ and define $x=9-l-p$. Then if $x>2$, the theory belongs to the bifurcation family discussed in \cref{bifurusp}, and the associated class $\mathcal{S}$ theory is then given by \eqref{ClassSpres2}, with the $m_i$'s given as follows. First, we have that $m_8 = m_7 = m_6 =n$. Second, $p$ of the remaining $m_i$'s are determined from the $n_i$'s to be $m_i = n-n_i-1$. Finally, $l$ of the remaining $m_i$'s are equal to $n-2$, and if there are still $m_i$'s left undetermined, then these are equal to $n-1$. Alternatively, if $x\leq 2$, then the resulting theory is described by the exceptions associated with $\mathfrak{g}_{2}$, which were discussed in \cref{g2}. The resulting class $\mathcal{S}$ theory is then given by \eqref{classSg2}, with the $m_i$'s determined as follows. First $p$ of the $m_i$'s are determined from the $n_i$'s to be $m_i = n-n_i$. Out of the remaining $m_i$'s, $l$ are equal to $n-1$, and the rest are equal to $n$.
    \item Subcase (h) for $\left|P-\frac{\bm{p}^\text{T}_1}{2}+4\right| \leq 1$: Here, we have that $\mathfrak{g}=\mathfrak{usp}_{2P}$, and we take $\bm{p} = \left[n_i,2^{2P+l},1^y\right]$, where now $n_i>2$ and we denote the number of $n_i$'s by $p$. The $m_i$'s are then determined as follows. First, $p$ of the $m_i$'s are determined by $m_i = n-n_i+1$. Next, $l$ of the remaining $m_i$'s are equal to $n-1$, and finally, the remaining $m_i$'s are equal to $n$. The resulting quiver belongs to the bifurcation family discussed in \cref{bifurusp}, and the associated class $\mathcal{S}$ theory is then given by \eqref{ClassSpres1} and \eqref{ClassSpres2}. We note that when $l+p=8$, there are two distinct 6d quivers and 4d class $\mathcal{S}$ theories differing by the choice of the $\theta$ angle.
    \end{itemize}
    \item Case (ii): $\bm{p}^\text{T}_1<8$, $\bm{p}^\text{T}_1+\bm{p}^\text{T}_2\geq6$
    \begin{itemize}
    \item Subcase (d): Here, $\mathfrak{g}=\mathfrak{usp}_{2P}$ and we note that the partition is always of the form $\bm{p} = \left[2k-5,1^5\right]$. As such, the quiver is fully specified by $0\leq P \leq 2$. For $P=2$, the resulting 4d class $\mathcal{S}$ theory is given by \eqref{so16case}, while for $P=1$, it is given by \eqref{so12su2case}. Finally, the $P=0$ case is part of the bifurcation family and is given by \eqref{ClassSpres2} with $m_8=...=m_2=n$, $m_1=n-2k+6$.
    \item Subcase (e): Here, $\mathfrak{g}=\mathfrak{usp}_{2P}$, and we note that the partition is always of the form $\bm{p} = \left[2k-q-4,q,1^4\right]$. As such, the quiver is determined by $0\leq P \leq 1$ and $q$. There are then two options. For $P=1$, we get the 4d class $\mathcal{S}$ theory given by \eqref{so8S1classS} for $q=k-2$ and \eqref{so8SlclassS} with $l=q+2$ for $q<k-2$. Like the previous subcase, the $P=0$ case is described by the bifurcation family, \eqref{ClassSpres2}. The associated $m_i$'s are given by $m_8=...=m_3=n$, $m_2=n-q+1$, $m_1=n-2k+q+5$.
    \item Subcase (f) for $P=0$: Here, $\mathfrak{g}=\mathfrak{usp}_{0}$, that is an empty $-1$ curve. We shall write the partition as $\bm{p} = [n_i]$, for $i=1,2,...,6$. Note that unlike the previous cases, here $n_i$ can also be equal to one. The associated $m_i$'s are then given by $m_8=m_7=n$, $m_i = n-n_i+1$. The resulting class $\mathcal{S}$ theory is then part of the bifurcation family, and can be determined from \eqref{ClassSpres2}.
    \end{itemize}
\end{itemize}

\section{Embeddings via Cyclic Groups}\label{sec:embeddingcyclic}
In this section, we study the orbi-instanton theories associated with embeddings of $\widehat{D}_k$ into $E_8$ via cyclic groups, as classified in \cite[Section 7.3.3]{Heckman:2015bfa}. We focus on cases where the flavour symmetry algebras are classical. We report the 6d anomaly coefficients, the data characterising the associated 4d class $\mathcal{S}$ theories (specifically, the punctures), and the corresponding 3d mirror theories. We demonstrate that the anomalies, including the central charges, of the resulting 4d class $\mathcal{S}$ theory are in perfect agreement with \eqref{anom4d}, and the Higgs branch dimension satisfies \eqref{dimHiggs4d}.

Subsequently, we shall further abbreviate the 3d mirror theory
\bes{
\scalebox{0.85}{$
\begin{array}{ccc @{} >{\color{cE8}}c >{\color{cE8}}c >{\color{cE8}}c >{\color{cE8}}c >{\color{cE8}}c >{\color{cE8}}c >{\color{cE8}}c >{\color{cE8}}c @{}}
		 & & & & & & & & 2(3h+r_{3'}) & & \\
  2\,\, \ldots & 2k & & 2(h+r_1) & 2(2h+r_2) & 2(3h+r_3) & 2(4h+r_4) & 2(5h+r_5) & 2(6h+r_5) & 2(4h+r_{4'}) & 2(2h+r_{2'}) \\
\end{array}$}
}
simply as
\bes{
\left[ \begin{array}{ccc @{} >{\color{cE8}}c >{\color{cE8}}c >{\color{cE8}}c >{\color{cE8}}c >{\color{cE8}}c >{\color{cE8}}c >{\color{cE8}}c >{\color{cE8}}c @{}}
		 & & & & & & & & 2r_{3'} & & \\
  \ldots & 2k & & 2r_1 & 2r_2 & 2r_3 & 2r_4 & 2r_5 & 2r_6 & 2r_{4'} & 2r_{2'} \\
\end{array} \right]_h\;.
}

\subsection{The Embeddings \texorpdfstring{$\widehat{D}_k \rightarrow \BZ_2 \rightarrow E_8$}{Dk -> Z2 -> E8}}\label{Z2}
For the embedding via the cyclic group $\mathbb{Z}_2$, there is one case. The global symmetry is $\SO(16)$.

\paragraph{The \texorpdfstring{$\SO(16)$}{SO(16)} global symmetry} Let us consider
\bes{ \label{Fthquivso16}
[8]\,\, \underset{\left[\frac{1}{2}\right]}{\overset{\mathfrak{sp}_{2}}1} \,\,{\overset{\mathfrak{so}_{7}}3} \,\, 1 \,\, \overset{\mathfrak{so}_{9}}4 \,\, {\overset{\mathfrak{sp}_{1}}1} \,\, \overset{\mathfrak{so}_{11}}4 \,\,\overset{\mathfrak{sp}_{2}}1 \,\, \overset{\mathfrak{so}_{13}}4 \,\, \overset{\mathfrak{sp}_{3}}1 \,\, \overset{\mathfrak{so}_{15}}4 \,\, \ldots
\,\, \overset{\mathfrak{sp}_{k-5}}1 \,\, \overset{\mathfrak{so}_{2k-1}}4 \,\, \underset{\left[\frac{1}{2}\right]}{\overset{\mathfrak{sp}_{k-4}}1} \,\,\overset{\mathfrak{so}_{2k}}4 \,\, \ldots {\overset{\mathfrak{sp}_{k-4}}1}\,\, [k]
}
with $h-1$ $\so_{2k}$ gauge factors and $h$ $\sp_{k-4}$ gauge factors.
We obtain the following anomaly coefficients for the 6d theory (we remind the reader that we use the anomaly polynomial form in \eqref{6dAnomPol}):
\bes{
\alpha &=\frac{7 h}{192}+\left(\frac{7 k^2}{5760}+\frac{7 k}{192}-\frac{217}{2880}\right)\;, \\
\beta &= h^2 \left(1-\frac{k}{2}\right)+h \left(-\frac{7 k^2}{12}+\frac{31 k}{12}-\frac{73}{24}\right)+\left(-\frac{7 k^3}{36}+\frac{65 k^2}{48}-\frac{229 k}{72}+\frac{65}{24}\right)\;,\\
\gamma &=-\frac{h}{48}+\left(-\frac{k^2}{1440}-\frac{k}{48}+\frac{31}{720}\right)\;.
}
For convenience, we define $\mathfrak{h} = h+(k-3)$. The 4d class $\mathcal{S}$ theory contains the punctures
\bes{
\left[(2\mathfrak{h}-1)^5, 2\mathfrak{h}-(2k-3), 1^{2k}\right]~, \quad \left[(4\mathfrak{h}-1)^3, 1\right]~, \quad \left[(6\mathfrak{h}-1)^2\right]
}
and has the central charges
\bes{
&a = h^2 (6 k-12)+h \left(7 k^2-31 k+\frac{151}{4}\right)+\left(\frac{7 k^3}{3}-\frac{389 k^2}{24}+\frac{473 k}{12}-\frac{421}{12}\right)\;, \\
&c = h^2 (6 k-12)+h \left(7 k^2-31 k+39\right)+\left(\frac{7 k^3}{3}-\frac{97 k^2}{6}+\frac{122 k}{3}-\frac{113}{3}\right)\;,\\
&k_{\mathfrak{so}_{16}} = 12 h+(12 k-28)~, \qquad k_{\mathfrak{so}_{2k}} =4k+8\;,
}
in agreement with \eqref{anom4d}.

The corresponding 3d mirror theory is
\bes{ \label{magquivso16}
\left[ \begin{array}{ccc @{} >{\color{cE8}}c >{\color{cE8}}c >{\color{cE8}}c >{\color{cE8}}c >{\color{cE8}}c >{\color{cE8}}c >{\color{cE8}}c >{\color{cE8}}c @{}}
	 & & & & & & & & 4 & & \\
	\ldots & 2k & & 4 & 6 & 6 & 8 & 8 & 10 & 6 & 4 \\
	\end{array} \right]_h\;.
}
The quantity $24(c-a) = -1440 \gamma= 30 h+\left(k^2+30 k-62\right)$ is equal to the Coulomb branch dimension of the corresponding 3d mirror theory \eqref{magquivso16}. The Kac-type labels are
\begin{equation}
    (s_1,s_2,s_3,s_4,s_5,s_6,s_{4'},s_{2'},s_{3'})=(k-1,-1,1,-1,1,-1,1,-1,1)\;.
\end{equation}

\subsection{The Embeddings \texorpdfstring{$\widehat{D}_k \rightarrow \BZ_2 \times \BZ_2 \rightarrow E_8$}{Dk -> Z2xZ2 -> E8} (with \texorpdfstring{$k$}{k} Even)}\label{Z2xZ2}
For the embeddings via the cyclic group $\mathbb{Z}_2\times\mathbb{Z}_2$, there are three different cases. The global symmetries are $\SU(8)$, $\SO(8)\times\SO(8)$ and $\SO(12)\times\SU(2)^2$ respectively.

\paragraph{The \texorpdfstring{$\SU(8)$}{SU(8)} global symmetry} Let us consider
\bes{ \label{FthquivSU8SO2k}
[8]\,\, \overset{\mathfrak{su}_{k}}2 \,\, \overset{\mathfrak{sp}_{k-4}}1 \,\, {\overset{\mathfrak{so}_{2k}}4} \,\, \overset{\mathfrak{sp}_{k-4}}1\,\,...\overset{\mathfrak{sp}_{k-4}}1 \,\, [k] }
with $h$ $\sp_{k-4}$ gauge factors and $(h-1)$ $\so_{2k}$ gauge factors. We obtain the following anomaly coefficients for the 6d theory:
\bes{
\alpha &=\frac{7 h}{192}+\frac{7 \left(k^2+7 k+1\right)}{5760}\;, \\
\beta &= h^2 \left(1-\frac{k}{2}\right)+h \left(-\frac{k^2}{12}+\frac{k}{12}-\frac{1}{24}\right)+\left(\frac{k^2}{48}-\frac{k}{48}+\frac{1}{48}\right)\;,\\
\gamma &= -\frac{h}{48} -\frac{k^2+7 k+1}{1440}\;.
}
The 4d class $\mathcal{S}$ theory contains the punctures
\bes{ \label{classSSU8SO2k}
\left[(2h)^6, 1^{2k}\right]~, \quad \left[(4h+k-1)^2, 4h+1, 1\right]~, \quad \left[(6h+k)^2\right]
}
and has the central charges
\bes{
&a = h^2(6k-12)+h(k^2 -k +\frac74)+\frac{-5k^2+13k-5}{24}\;, \\
&c = h^2(6k-12) + h(k^2 -k +3) + \frac{-k^2+5k-1}6\;,\\
&k_{\mathfrak{su}_{8}} = 12h + 2k~, \qquad k_{\mathfrak{so}_{2k}} =4k+8\;,
}
in agreement with \eqref{anom4d}.
The quantity $24(c-a) = 30 h+\left(k^2+7 k+1\right)$ is equal to the Coulomb branch dimension of the corresponding 3d mirror theory. This holds for both $k$ even and $k$ odd. The corresponding 3d mirror theory is
\bes{
\left[\begin{array}{ccc @{} >{\color{cE8}}c >{\color{cE8}}c >{\color{cE8}}c >{\color{cE8}}c >{\color{cE8}}c >{\color{cE8}}c >{\color{cE8}}c >{\color{cE8}}c @{}}
	 & & & & & & & & k & & \\
	\ldots & 2k & & 2k & 2k & 2k & 2k & 2k & 2k & k & 2 \\
	\end{array} \right]_h\;.
}
The Kac-type labels are
\begin{equation}
    (s_1,s_2,s_3,s_4,s_5,s_6,s_{4'},s_{2'},s_{3'})=(0,0,0,0,0,0,1,(k-4)/2,0)
\end{equation}
for even $k$.

A closely related example is theory number 10 in Table \ref{D5table}, namely
\bes{
{[8 ]} \,\, \overset{\mathfrak{su}_{4}}2 \,\, 1 \,\, \overset{\mathfrak{so}_{9}}4 \,\,\underset{\left[\frac{1}{2}\right]}{\overset{\mathfrak{sp}_{1}}1} \,\,\overset{\mathfrak{so}_{10}}4 \,\, ... \overset{\mathfrak{sp}_{1}}1 \,\, [5]\;,
}
where there are $(h+1)$ $\sp_1$ gauge factors and $h$ $\so_{10}$ gauge factors. Then, we obtain
\bes{
\begin{array}{lll}
\alpha =\frac{7}{960} (5 h+19)~, &\quad \beta = -\frac{1}{24} \left(36 h^2+161 h+163\right)~, &\quad \gamma = -\frac{1}{240} (5 h+19) \\
\kappa_{\su(8)} = \frac{1}{48} (3 h+8)~, &\quad \kappa_{\so(10)} = \frac{7}{48}~. &
\end{array}
}
The 4d class $\mathcal{S}$ theory contains the punctures
\bes{
\left[(2h+5)^4, (2h+3), (3h+1), 1^{10}\right]~,\quad \left[(4h+11)^3, 1\right]~, \quad \left[(6h+17)^2\right]
}
and has the central charges
\bes{ \begin{array}{ll}
a = \frac{3}{4} (3 h+5) (8 h+23)~, &\quad c= 18 h^2+83 h+91~, \\
k_{\su(8)} = 4 (3 h+8)~, &\quad k_{\so(10)} = 28~,
\end{array}
}
in agreement with \eqref{anom4d}. The 3d mirror theory is
\bes{
\left[\begin{array}{ccc @{} >{\color{cE8}}c >{\color{cE8}}c >{\color{cE8}}c >{\color{cE8}}c >{\color{cE8}}c >{\color{cE8}}c >{\color{cE8}}c >{\color{cE8}}c @{}}
	 & & & & & & & & 16 & &\\
	\ldots & 10 & & 10 & 14 & 18 & 24 & 28& 34 & 22 & 12 \\
	\end{array} \right]_h\;.
}
The total rank of this quiver is $30h+114$, which is in agreement with $24(c-a) = -1440 \gamma$, as expected. The Kac-type labels are
\begin{equation}
    (s_1,s_2,s_3,s_4,s_5,s_6,s_{4'},s_{2'},s_{3'})=(2,0,1,-1,1,-1,1,-1,1)^\text{II}\;.
\end{equation}

\paragraph{The \texorpdfstring{$\SO(8) \times \SO(8)$}{SO(8)xSO(8)} global symmetry} Let us consider
\bes{ \label{FthSO8SO8SO2k}
[4] \,\, \overset{\mathfrak{sp}_{\frac{k-4}{2} }}1 \,\, \underset{[4]}{\underset{1}{\underset{\mathfrak{sp}_{ \frac{k-4}{2} }}{\overset{\mathfrak{so}_{2k}}4}}} \,\, \overset{\mathfrak{sp}_{k-4}}1\,\,\ldots\overset{\mathfrak{sp}_{k-4}}1 \,\, [k]
}
with $h$ $\sp_{k-4}$ gauge factors and $h$ $\so_{2k}$ gauge factors. We obtain the following anomaly coefficients for the 6d theory:
\bes{
\alpha &=\frac{7 h}{192}+\frac{7 \left(k^2+7 k+14\right)}{5760}\;, \\
\beta &= h^2 \left(1-\frac{k}{2}\right)+h \left(-\frac{k^2}{12}-\frac{5 k}{12}+\frac{23}{24}\right)+\left(-\frac{k^2}{48}-\frac{5 k}{48}+\frac{7}{24}\right)\;,\\
\gamma &= -\frac{h}{48} - \frac{k^2+7 k+14}{1440}\;.
}
The 4d class $\mathcal{S}$ theory contains the punctures
\bes{
\left[(2h+1)^6, 1^{2k}\right]~, \quad \left[(4h+k+1)^2, 4h+3, 1\right]~, \quad \left[(6h+k+3)^2\right]
}
and has the central charges
\bes{
&a = h^2 (6 k-12)+h \left(k^2+5 k-\frac{41}{4}\right)+\left(\frac{7 k^2}{24}+\frac{37 k}{24}-\frac{35}{12}\right)\;, \\
&c = h^2 (6 k-12)+h \left(k^2+5 k-9\right)+\left(\frac{k^2}{3}+\frac{11 k}{6}-\frac{7}{3}\right)\;,\\
&k_{\mathfrak{so}_{8},1,2} = 12 h+(2 k+4)~, \qquad k_{\mathfrak{so}_{2k}} =4k+8\;,
}
in agreement with \eqref{anom4d}. The corresponding 3d mirror theory is
\bes{
\left[ \begin{array}{ccc @{} >{\color{cE8}}c >{\color{cE8}}c >{\color{cE8}}c >{\color{cE8}}c >{\color{cE8}}c >{\color{cE8}}c >{\color{cE8}}c >{\color{cE8}}c @{}}
	 & & & & & & & & k+2 & & \\
	\ldots & 2k & & 2k & 2k+2 & 2k+2 & 2k+4 & 2k+4& 2k+6 & k+4 & 4 \\
	\end{array} \right]_h\;.
}
The quantity $24(c-a) =-1440\gamma= 30 h+\left(k^2+7 k+14\right)$ is equal to the Coulomb branch dimension of the corresponding 3d mirror theory. The Kac-type labels are
\begin{equation}
    (s_1,s_2,s_3,s_4,s_5,s_6,s_{4'},s_{2'},s_{3'})=(1,-1,1,-1,1,-1,1,(k-4)/2,1)
\end{equation}
for even $k$. Incidentally, for odd $k$, we have
\begin{equation}
    (s_1,s_2,s_3,s_4,s_5,s_6,s_{4'},s_{2'},s_{3'})=(1,-1,1,-1,1,0,0,(k-3)/2,0)\;.
\end{equation}

\paragraph{The \texorpdfstring{$\SO(12) \times \SU(2)^2$}{SO(12)xSU(2)2} global symmetry} Let us consider
\bes{ \label{FthSO12SU2SU2}
[6]\,\, \overset{\mathfrak{sp}_{1}}1 \,\, \underset{[1]}{\overset{\mathfrak{so}_{8}}3} \,\, \overset{\mathfrak{sp}_{1}}1 \,\, \overset{\mathfrak{so}_{12}}4 \,\, {\overset{\mathfrak{sp}_{3}}1} \,\, \overset{\mathfrak{so}_{16}}4 \,\,... \overset{\mathfrak{so}_{2k-4}}4 \,\, \overset{\mathfrak{sp}_{k-5}}1 \,\,\underset{[1]}{\overset{\mathfrak{so}_{2k}}4} \,\, \overset{\mathfrak{sp}_{k-4}}1\,\,...\overset{\mathfrak{sp}_{k-4}}1 \,\, [k]
}
with $h$ $\so_{2k}$ gauge factors and $h$ $\sp_{k-4}$ gauge factors. We obtain the following anomaly coefficients for the 6d theory:
\bes{
\alpha &=\frac{7 h}{192}+\left(\frac{7 k^2}{5760}+\frac{217 k}{11520}-\frac{7}{360}\right);, \\
\beta &= h^2 \left(1-\frac{k}{2}\right)+h \left(-\frac{k^2}{3}+\frac{13 k}{12}-\frac{25}{24}\right)+\left(-\frac{k^3}{18}+\frac{7 k^2}{24}-\frac{143 k}{288}+\frac{1}{3}\right)\;,\\
\gamma &= -\frac{h}{48}+\left(-\frac{k^2}{1440}-\frac{31 k}{2880}+\frac{1}{90}\right)\;.
}
For convenience, we define $\mathfrak{h} = h+(k-2)/2$, where we recall that $k$ is even. The 4d class $\mathcal{S}$ theory contains the punctures
\bes{
\left[(2\mathfrak{h})^4, (2\mathfrak{h}-(k-2))^2, 1^{2k}\right]~, \quad \left[(4\mathfrak{h}+1)^3, 1\right]~, \quad \left[(6\mathfrak{h}+2)^2\right]
}
and has the central charges
\bes{
&a = h^2 (6 k-12)+h \left(4 k^2-13 k+\frac{55}{4}\right)+\left(\frac{2 k^3}{3}-\frac{83 k^2}{24}+\frac{317 k}{48}-\frac{14}{3}\right)\;, \\
&c = h^2 (6 k-12)+h \left(4 k^2-13 k+15\right)+\left(\frac{2 k^3}{3}-\frac{41 k^2}{12}+\frac{29 k}{4}-\frac{16}{3}\right)\;,\\
&k_{\mathfrak{so}_{12}} = 12 h+(6 k-8)~, \,\, k_{\mathfrak{su}_{2},1} = 12 h+(6 k-4)~, \,\,
k_{\mathfrak{su}_{2},2} = 12 h+2k~, \,\,
k_{\mathfrak{so}_{2k}} =4k+8~,
}
in agreement with \eqref{anom4d}. The corresponding 3d mirror theory is
\bes{ \label{magquivSO12SU2SU2}
\left[
\begin{array}{ccc @{} >{\color{cE8}}c >{\color{cE8}}c >{\color{cE8}}c >{\color{cE8}}c >{\color{cE8}}c >{\color{cE8}}c >{\color{cE8}}c >{\color{cE8}}c @{}}
	 & & & & & & & & 2 & & \\
	\ldots & 2k & & k+2 & 4 & 4 & 4 & 4& 4 & 2 & 2 \\
	\end{array} \right]_\mathfrak{h}\;.
}
The quantity $24(c-a) = -1440 \gamma = 30 h+\left(k^2+\frac{31 k}{2}-16\right)$ is equal to the Coulomb branch dimension of the corresponding 3d mirror theory \eqref{magquivSO12SU2SU2}. The Kac-type labels are
\begin{equation}
    (s_1,s_2,s_3,s_4,s_5,s_6,s_{4'},s_{2'},s_{3'})=(0,(k-4)/2,0,0,0,0,1,-1,0)
\end{equation}
for even $k$. Incidentally, for odd $k$, we have
\begin{equation}
    (s_1,s_2,s_3,s_4,s_5,s_6,s_{4'},s_{2'},s_{3'})=(1,(k-3)/2,0,0,0,0,1,-1,0)\;.
\end{equation}

\subsection{The Embeddings \texorpdfstring{$\widehat{D}_k \rightarrow \BZ_4\rightarrow E_8$}{Dk -> Z4 -> E8} (with \texorpdfstring{$k$}{k} Odd)}\label{Z4}
For the embeddings via the cyclic group $\mathbb{Z}_4$, there are three different cases. The global symmetries are $\SU(8)$, $\SU(8)\times\SU(2)$ and $\SO(12)\times\SU(2)$ respectively.

\paragraph{The \texorpdfstring{$\SU(8)$}{SU(8)} global symmetry} The F-theory quiver at a generic point on the tensor branch is given by \eqref{FthquivSU8SO2k}, and the class $\mathcal{S}$ punctures are given by \eqref{classSSU8SO2k}. The corresponding 3d mirror theory is
\bes{
\left[ \begin{array}{ccc @{} >{\color{cE8}}c >{\color{cE8}}c >{\color{cE8}}c >{\color{cE8}}c >{\color{cE8}}c >{\color{cE8}}c >{\color{cE8}}c >{\color{cE8}}c @{}}
	 & & & & & & & & k-1 & & \\
	\ldots & 2k & & 2k & 2k & 2k & 2k & 2k & 2k & k+1 & 2 \\
	\end{array} \right]_h\;.
}
The Kac-type labels are
\begin{equation}
    (s_1,s_2,s_3,s_4,s_5,s_6,s_{4'},s_{2'},s_{3'})=(0,0,0,0,0,0,0,(k-3)/2,1)
\end{equation}
for odd $k$.

\paragraph{The \texorpdfstring{$\SU(8) \times \SU(2)$}{SU(8)xSU(8)} global symmetry} Let us consider
\bes{ \label{FthSU8SU2SO2k}
[8]\,\, \overset{\mathfrak{su}_{4}}2 \,\,
\,\, 1 \,\,
\overset{\mathfrak{so}_{10}}4 \,\, {\overset{\mathfrak{sp}_{2}}1} \,\,
\,\,
\overset{\mathfrak{so}_{14}}4 \,\, \ldots \,\,
\overset{\mathfrak{so}_{2k-4}}4\,\,
\overset{\mathfrak{sp}_{k-5}}1 \,\,
\underset{[1]}{ \overset{\mathfrak{so}_{2k}}4}\,\, \overset{\mathfrak{sp}_{k-4}}1 \,\,
...\overset{\mathfrak{sp}_{k-4}}1 \,\, [k] }
with $h$ $\so_{2k}$ gauge factors and $h$ $\sp_{k-4}$ gauge factors. We obtain the following anomaly coefficients for the 6d theory:
\bes{
\alpha &=\frac{7 h}{192}+\frac{7 \left(2 k^2+31 k-35\right)}{11520}\;, \\
\beta &= h^2 \left(1-\frac{k}{2}\right)+\frac{1}{24} h \left(-8 k^2+26 k-25\right)+\frac{1}{288} \left(-16 k^3+84 k^2-143 k+105\right)\;,\\
\gamma &= -\frac{h}{48} -\frac{2 k^2+31 k-35}{2880}\;.
}
For convenience, we define $\mathfrak{h} = h+(k-3)/2$, where we recall that $k$ is odd. The 4d class $\mathcal{S}$ theory has the punctures
\bes{
\left[(2\mathfrak{h}-1)^4, (2\mathfrak{h}-(k-1))^2, 1^{2k}\right]~, \quad \left[(4\mathfrak{h}-1)^3, 1\right]~, \quad \left[(6\mathfrak{h}-1)^2\right]
}
and has the central charges
\bes{
&a = h^2 (6 k-12)+h \left(4 k^2-13 k+\frac{55}{4}\right)+\left(\frac{2 k^3}{3}-\frac{83 k^2}{24}+\frac{317 k}{48}-\frac{245}{48}\right)\;, \\
&c = h^2 (6 k-12)+h \left(4 k^2-13 k+15\right)+\left(\frac{2 k^3}{3}-\frac{41 k^2}{12}+\frac{29 k}{4}-\frac{35}{6}\right)\;,\\
&k_{\mathfrak{su}_{8}} = 12 h+(6 k-10)~, \qquad k_{\mathfrak{su}_{2}} = 12 h+2 k~, \qquad k_{\mathfrak{so}_{2k}} =4k+8\;,
}
in agreement with \eqref{anom4d}. The corresponding 3d mirror theory is
\bes{ \label{magquivSU8SU2SO2k}
\left[
\begin{array}{ccc @{} >{\color{cE8}}c >{\color{cE8}}c >{\color{cE8}}c >{\color{cE8}}c >{\color{cE8}}c >{\color{cE8}}c >{\color{cE8}}c >{\color{cE8}}c @{}}
	& & & & & & & & 4 & & \\
	\ldots & 2k & & k+3 & 6 & 6 & 8 & 8& 10 & 6 & 4 \\
	\end{array}\right]_\mathfrak{h}\;.
}
The quantity $24(c-a) = 30 h+\left(4 m^2+27 m-32\right)$, with $k=2m-1$, is equal to the Coulomb branch dimension of the corresponding 3d mirror theory \eqref{magquivSU8SU2SO2k}. The Kac-type labels are
\begin{equation}
    (s_1,s_2,s_3,s_4,s_5,s_6,s_{4'},s_{2'},s_{3'})=(0,(k-3)/2,1,-1,1,0,1,-1,0)
\end{equation}
for odd $k$. Incidentally, for even $k$, we have
\begin{equation}
    (s_1,s_2,s_3,s_4,s_5,s_6,s_{4'},s_{2'},s_{3'})=(1,(k-4)/2,1,-1,1,0,1,-1,0)\;.
\end{equation}

\paragraph{The \texorpdfstring{$\SO(12) \times \SU(2)$}{SO(12)xSU(2)} global symmetry} Let us consider
\bes{ \label{Fthquivso12su2kodd}
[6]\,\, \overset{\mathfrak{sp}_{1}}1 \,\, \underset{[1]}{\overset{\mathfrak{so}_{7}}3} \,\, 1 \,\, \overset{\mathfrak{so}_{9}}4 \,\, {\overset{\mathfrak{sp}_{1}}1} \,\, \overset{\mathfrak{so}_{11}}4 \,\,\overset{\mathfrak{sp}_{2}}1 \,\, \overset{\mathfrak{so}_{13}}4 \,\, \overset{\mathfrak{sp}_{3}}1 \,\, \overset{\mathfrak{so}_{15}}4 \,\,... \overset{\mathfrak{so}_{2k-1}}4 \,\, \underset{\left[\frac{1}{2} \right]}{\overset{\mathfrak{sp}_{k-4}}1} \,\,\overset{\mathfrak{so}_{2k}}4 \,\, \ldots {\overset{\mathfrak{sp}_{k-4}}1}\,\, [k]
}
with $h-1$ $\so_{2k}$ gauge factors and $h$ $\sp_{k-4}$ gauge factors. We obtain the following anomaly coefficients for the 6d theory:
\bes{
\alpha &=\frac{7 h}{192}+\left(\frac{7 k^2}{5760}+\frac{7 k}{192}-\frac{35}{384}\right)\;, \\
\beta &= h^2 \left(1-\frac{k}{2}\right)+h \left(-\frac{7 k^2}{12}+\frac{31 k}{12}-\frac{67}{24}\right)+\left(-\frac{7 k^3}{36}+\frac{65 k^2}{48}-\frac{211 k}{72}+\frac{101}{48}\right)\;,\\
\gamma &= -\frac{h}{48}+\left(-\frac{k^2}{1440}-\frac{k}{48}+\frac{5}{96}\right)\;.
}
We define $\mathfrak{h}= h+(k-3)$. The 4d class $\mathcal{S}$ theory has the punctures
\bes{
\left[(2\mathfrak{h})^4, 2\mathfrak{h}-1, 2\mathfrak{h}-(2k-5), 1^{2k}\right]~, \quad \left[(4\mathfrak{h}+1)^3, 1\right]~, \quad \left[(6\mathfrak{h}+2)^2\right]
}
and has the central charges
\bes{
&a = h^2 (6 k-12)+h \left(7 k^2-31 k+\frac{139}{4}\right)+\left(\frac{7 k^3}{3}-\frac{389 k^2}{24}+\frac{437 k}{12}-\frac{227}{8}\right)\;, \\
&c = h^2 (6 k-12)+h \left(7 k^2-31 k+36\right)+\left(\frac{7 k^3}{3}-\frac{97 k^2}{6}+\frac{113 k}{3}-\frac{63}{2}\right)\;,\\
&k_{\mathfrak{so}_{12}} = 12 h+(12 k-32)~, \qquad k_{\mathfrak{su}_{2}} = 12 h+(12 k-40)~, \qquad k_{\mathfrak{so}_{2k}} =4k+8\;,
}
in agreement with \eqref{anom4d}. The corresponding 3d mirror theory is
\bes{ \label{magquivso12su2odd}
\left[
\begin{array}{ccc @{} >{\color{cE8}}c >{\color{cE8}}c >{\color{cE8}}c >{\color{cE8}}c >{\color{cE8}}c >{\color{cE8}}c >{\color{cE8}}c >{\color{cE8}}c @{}}
	 & & & & & & & & 2 & & \\
	\ldots & 2k & & 4 & 4 & 4 & 4 & 4& 4 & 2 & 2 \\
	\end{array} \right]_\mathfrak{h}\;.
}
The quantity $24(c-a) = - 1440 \gamma= 30 h+\left(k^2+30 k-75\right)$ is equal to the Coulomb branch dimension of the corresponding 3d mirror theory \eqref{magquivso12su2odd}. The Kac-type labels are
\begin{equation}
    (s_1,s_2,s_3,s_4,s_5,s_6,s_{4'},s_{2'},s_{3'})=(k-2,0,0,0,0,0,1,-1,0)
\end{equation}
for odd $k$. Incidentally, for even $k$, we have the same Kac-type labels.

\section{Higgsings of the Class \texorpdfstring{$\mathcal{S}$}{S} Theories}\label{Higgsings}
As the Higgs branch should be preserved after reducing the 6d theory on a torus, we may compare the Higgsings from the 6d and the 4d perspectives. In this section, we shall discuss these Higgsings which may or may not be detected by the class $\mathcal{S}$ theory descriptions.

\subsection{Higgsing Orbits}\label{higgsingorbits}
Although we do not have the class $\mathcal{S}$ theory descriptions for all the $D_k$ orbi-instanton theories, those with the class $\mathcal{S}$ theory descriptions are connected by Higgsings. Under Higgsings, they form a closed orbit, which we shall refer to as the Higgsing orbit, or the class $\mathcal{S}$ orbit. Notice that the orbit is closed since the length of the 6d generalised quiver may be reduced after the Higgsing. In other words, if theory A can be Higgsed to theory B (possibly with multiple minimal steps), then theory B may also be Higgsed to theory A but with a smaller $n$.

It is not surprising that the theories in the class $\mathcal{S}$ subset form an orbit under Higgsings as the Higgsings can be translated into manipulations of the punctures. This also implies that the 6d theories living outside the orbit cannot be reached by moving boxes in the punctures, which is in line with the missing of their class $\mathcal{S}$ theory descriptions, as mentioned in \cref{6d4drelations}.

Moreover, we find that the minimal steps of Higgsings (aka atomic Higgsings) are always nilpotent and combo Higgsings in terms of the 6d language. We shall not expound the details of different types of Higgsings in 6d here. Readers can find the explanations in \cite{Bao:2024wls}. In short, the nilpotent Higgsing is performed by turning on the nilpotent VEV for the flavour symmetry. The combo Higgsing is the case when the individual step of the nilpotent or the plateau Higgsing is not allowed, but a combination of reducing the gauge algebras and blowing down the $-1$ curves is possible\footnote{The plateau Higgsing is triggered by the semisimple part of the flavour symmetry. Overall, there are four types of Higgsings, and the remaining one is called the endpoint-changing Higgsing which changes the endpoint of the generalised quiver.}.

\begin{figure}[ht]
    \centering
    \includegraphics[width=15cm]{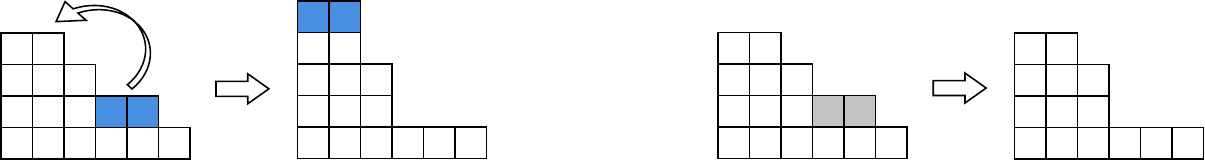}
    \caption{The moves of the boxes in the puncture. On the left, we have $\mathfrak{M}=2$. On the right, we have $\mathfrak{M}=-2$.}\label{movingboxes}
\end{figure}

Here, we shall exemplify the Higgsing/class $\mathcal{S}$ orbit with the $D_4$ case. As before, 6d SCFTs in this class are specified through the number given to them in \cref{examples}, see table \eqref{D4Kac} for the $s$-labels associated with each case. In the list of the Higgsings, we shall give the elementary transverse slices\footnote{The transverse slices here would include the closure of the minimal nilpotent orbit $\overline{\mathcal{O}}_\text{min}(\mathfrak{g})$ of a Lie algebra $\mathfrak{g}$ (which we shall simply denote as $g$) and the Klenian singularity $A_l$.}, the Higgsing types in 6d (nilpotent or combo), and the moves of the boxes in the punctures. For the moves of the boxes, we shall denote them as $(\mathfrak{M}_1,\mathfrak{M}_2,\mathfrak{M}_3)$. The notation is understood as follows:
\begin{itemize}
    \item The integer $\mathfrak{M}_i$ is positive if there are $\mathfrak{M}_i$ boxes moved from the right columns to the left columns in the $i^\text{th}$ puncture, corresponding to a partial closure of the puncture.
    \item The integer $\mathfrak{M}_i$ is zero if there is no change in the $i^\text{th}$ puncture.
    \item The integer $\mathfrak{M}_i$ is negative if there are $|\mathfrak{M}_i|$ boxes removed from the $i^\text{th}$ puncture\footnote{Since we only have untwisted regular punctures here, the $\mathfrak{M}_i$ would be the same if one of them is negative.}.
\end{itemize}
This is illustrated in \cref{movingboxes}.

\paragraph{Example} Let us now list the Higgsing orbit for the $D_4$ case. The atomic Higgsings are given as follows\footnote{Recall that there are two sets of punctures describing cases 16 and 17 respectively. We shall denote them as 16$^\text{I}$, 16$_\text{I}$, 17$^\text{II}$ and 17$_\text{II}$, corresponding to the same superscripts and subscripts as in the Kac-type labels $s$.}:
\begin{itemize}
    \item case 4 $\rightarrow$ case 6: $c_4$, nilpotent, $(-2,-2,-2)$;
    \item case 6 $\rightarrow$ case 8: $a_5$, nilpotent, $(-2,-2,-2)$;
    \item case 8 $\rightarrow$ case 9: $b_3$, nilpotent, $(-2,-2,-2)$;
    \item case 8 $\rightarrow$ case 10: $a_1$, combo, $(2,0,0)$;
    \item case 9 $\rightarrow$ case 11: $A_1$, combo, $(0,1,0)$;
    \item case 10 $\rightarrow$ case 13: $d_6$, nilpotent, $(-4,-4,-4)$;
    \item case 10 $\rightarrow$ case 12: $c_1$, nilpotent, $(1,0,0)$;
    \item case 11 $\rightarrow$ case 13: $d_4$, nilpotent, $(-2,-2,-2)$;
    \item case 12 $\rightarrow$ case 14: $d_6$, nilpotent, $(-4,-4,-4)$;
    \item case 13 $\rightarrow$ case 14: $c_1$, nilpotent, $(1,0,0)$;
    \item case 13 $\rightarrow$ case 15: $d_4$, nilpotent, $(-2,-2,-2)$;
    \item case 14 $\rightarrow$ case 16$_\text{I}$: $b_4$, nilpotent, $(-1,-1,-1)$;
    \item case 15 $\rightarrow$ case 16$^\text{I}$: $c_2$, nilpotent, $(-2,-2,-2)$;
    \item case 15 $\rightarrow$ case 16$_\text{I}$: $c_2$, nilpotent, $(1,0,0)$;
    \item case 16$^\text{I}$ $\rightarrow$ case 4: $c_4$, nilpotent, $(1,0,0)$;
    \item case 16$_\text{I}$ $\rightarrow$ case 4: $c_4$, nilpotent, $(-2,-2,-2)$;
    \item case 16$^\text{I}$ $\rightarrow$ case 17$^\text{II}$: $c_1$, nilpotent, $(0,1,0)$;
    \item case 16$_\text{I}$ $\rightarrow$ case 17$_\text{II}$: $c_1$, nilpotent, $(1,0,0)$;
    \item case 17$^\text{II}$ $\rightarrow$ case 6: $a_7$, nilpotent, $(-2,-2,-2)$;
    \item case 17$_\text{II}$ $\rightarrow$ case 6: $a_7$, nilpotent, $(-4,-4,-4)$;
    \item case 17$_\text{II}$ $\rightarrow$ case 18: unknown slice, combo, $(2,0,0)$;
    \item case 18 $\rightarrow$ case 12: $d_8$, nilpotent, $(-6,-6,-6)$.
\end{itemize}

In general, $(\mathfrak{M}_1,\mathfrak{M}_2,\mathfrak{M}_3)$ with $\mathfrak{M}_1=\mathfrak{M}_2=\mathfrak{M}_3<0$ corresponds to the nilpotent Higgsing in 6d while the other moves could be either nilpotent or combo Higgsings. The example above indeed agrees with this.

\subsection{Elementary Transverse Slices}\label{slices}
Although from the moves of the boxes in the punctures, it is hard to tell what the transverse slices could be, we can deduce the 3d mirror theories from the punctures. Therefore, we can see if we can extract any information by comparing the parent and the child 3d mirror theories. The transverse slice may then be represented by the difference of the two 3d mirror theories.

From the cases in this paper, we can list some transverse slices as follows:
\begin{itemize}
    \item $A_1=a_1=c_1$:
    \begin{equation}
        \mathfrak{c}_1\quad\text{or}\quad\mathfrak{d}_1\;;
    \end{equation}
    \item $d_{l\geq3}$:
    \begin{equation}
        \mathfrak{d}_1\,\,\mathfrak{c}_1\,\,\mathfrak{d}_1\,\,\dots\,\,\mathfrak{c}_1\,\,\mathfrak{d}_1
    \end{equation}
    of length $2l-3$;
    \item $b_{l\geq3}$:
    \begin{equation}
        \mathfrak{d}_1\,\,\mathfrak{c}_1\,\,\mathfrak{d}_1\,\,\dots\,\,\mathfrak{d}_1\,\,\mathfrak{c}_1
    \end{equation}
    of length $2l-2$;
    \item $a_{2l+1}$:
    \begin{equation}
        {\color{gray} \mathfrak{c}_1\,\,\mathfrak{d}_1\,\,\mathfrak{c}_1\,\,\dots\,\,\mathfrak{c}_1}\quad\text{or}\quad{\color{gray} \mathfrak{d}_1\,\,\mathfrak{c}_1\,\,\mathfrak{d}_1\,\,\dots\,\,\mathfrak{d}_1}
    \end{equation}
    of length $2l+1$;
    \item $a_7$:
    \begin{align}
        \begin{split}
            &\mathfrak{c}_1\\
            \mathfrak{d}_1\,\,\mathfrak{c}_1\,\,&\mathfrak{d}_2\,\,\mathfrak{c}_1\mathfrak{d}_1\;;
        \end{split}
    \end{align}
    \item $b_{l\geq3}$:
    \begin{align}
        \begin{split}
            &\mathfrak{c}_1\\
            \mathfrak{c}_1\,\,\mathfrak{d}_1\,\,\mathfrak{c}_1\,\,\dots\,\,&\mathfrak{d}_1\,\,\mathfrak{c}_1
        \end{split}
    \end{align}
    with $2l-2$ nodes in total;
    \item $c_2$:
    \begin{equation}
        \mathfrak{c}_1\,\,\mathfrak{d}_0\,\,\mathfrak{c}_1\quad\text{or}\quad\mathfrak{d}_1\,\,\mathfrak{c}_1\;;
    \end{equation}
    \item $d_4$:
    \begin{align}
        \begin{split}
            &\mathfrak{c}_1\\
            \mathfrak{d}_1\,\,\mathfrak{c}_1\,\,&\mathfrak{d}_1\,\,\mathfrak{c}_1\;;
        \end{split}
    \end{align}
    \item $d_6$:
    \begin{align}
        \begin{split}
            &\mathfrak{c}_1\\
            \mathfrak{c}_1\,\,\mathfrak{d}_1\,\,\mathfrak{c}_2\,\,&\mathfrak{d}_2\,\,\mathfrak{c}_1\,\,\mathfrak{d}_1\;;
        \end{split}
    \end{align}
    \item $d_8$:
    \begin{align}
        \begin{split}
            &\mathfrak{c}_1\\
            \mathfrak{d}_1\,\,\mathfrak{c}_1\,\,\mathfrak{d}_2\,\,\mathfrak{c}_2\,\,&\mathfrak{d}_3\,\,\mathfrak{c}_2\,\,\mathfrak{d}_1\;;
        \end{split}
    \end{align}
    \item unknown 1d slices:
    \begin{equation}
        \mathfrak{c}_1\quad\text{or}\quad\mathfrak{d}_1\;.
    \end{equation}
\end{itemize}
In this list, the first three cases agree with the results in \cite{Lawrie:2024wan}. The remaining cases appear to be new. The $a_{2l+1}$ case is in grey since we suspect that this may actually be understood as a chain of unitary nodes rather than orthosymplectic ones, which is the one obtained from decay and fission for 3d unitary mirror theories as in \cite{Bourget:2023dkj,Bourget:2024mgn} (thus is not really new). The reason is that if we think of the brane system, then this slice corresponds to the D-branes that are moved infinitely far away and do not see the orientifolds.

As we can see, unlike the unitary cases, the correspondence between the transverse slices and the differences of the parent and the child orthosymplectic mirror theories is neither injective nor surjective. We also remark that it is natural to expect an unknown 1-dimensional slice is represented by a single $\mathfrak{c}_1$ or $\mathfrak{d}_1$ since this is the only possibility if it can be expressed using the 3d mirror theory.

\subsection{Hidden Higgsings}\label{hidden}
As mentioned above, it is not yet clear how the class $\mathcal{S}$ theory descriptions would detect all the possible Higgsings due to the lack of the class $\mathcal{S}$ theory descriptions for some orbi-instanton theories. In this sense, such Higgsings are referred to as hidden Higgsings. We shall also mention the Higgsings whose parent and child theories both have known class $\mathcal{S}$ descriptions. However, they are theories of different types. For some cases, the Higgsings also seem to be ``hidden'' from the class $\mathcal{S}$ theory descriptions.

\paragraph{Theories outside the class \texorpdfstring{$\mathcal{S}$}{S} orbits} Since we only have the class $\mathcal{S}$ theory descriptions for a sub-family of theories for each $D_k$, the Higgsings for those outside the Higgsing orbit are not yet manifest from the 4d perspective. Starting from a class $\mathcal{S}$ theory description, the Higgsings may be performed via moving the boxes in the punctures. The resulting theory would stay in the class $\mathcal{S}$. Therefore, for fixed $k$ and $n$, it is possible to exhaust all the combinations of the punctures that would give rise to a good or an ugly theory. As the theories outside the Higgsing orbits cannot arise from this, it is unlikely that the torus reduction of these 6d theories are class $\mathcal{S}$ theories (at least for three untwisted regular punctures).

It is also not clear how to obtain the 3d mirror theories for these theories from the decay and fission algorithm. Nevertheless, this, together with the above discussions on changing the punctures, does not imply that the hidden Higgsings are not allowed. From 6d, these Higgsings are still ``visible''. If one considers the quiver subtractions \cite{Hanany:2018uhm, Cabrera:2018ann} on the 3d mirror theories, the expected Higgsings and slices would be consistently produced. From the perspective of the quiver decay and fission, the part that needs to be removed from the parent 3d mirror theory may vary depend on cases even if we know what the transverse slice is. The changes of the 3d mirror theories may be more intricate, and there might be some new rules for the algorithm.

There is another perspective to view this phenomenon. In general, Higgsing the 6d quiver requires giving a vev to a collection of hypers that are charged under a certain flavour symmetry, leading to its breaking. In the class $\mathcal{S}$ theory this is mapped to changes of the punctures that implement the required symmetry breaking. Moving the boxes in the punctures that implement different symmetry breaking patterns then maps to giving vevs to different hypers in the 6d SCFTs, according to the symmetry mapping. As such, we naively expect to be able to implement any transition in the 6d quiver associated to breaking the appropriate symmetries by shifting boxes in the punctures. However, in many cases the full symmetry of the 4d class $\mathcal{S}$ SCFT is not manifest in the punctures. Specifically, in D-type class $\mathcal{S}$ it sometimes happens for certain symmetries to ``double", where the symmetry visible in the punctures is only a diagonal mixing of the two; see \cite[Section 5.4.1]{Gaiotto:2008ak} from the perspective of the mirror dual. When this happens attempting to Higgs the symmetry in the 4d SCFT actually implements the Higgsing of both symmetries simultaneously in the 6d quiver, even if individual Higgsing is possible. This phenomenon seems to occur in theories that can be Higgsed to the cases that appear to lie outside of class $\mathcal{S}$, where the symmetry we need to Higgs does not explicitly appear in the punctures. Trying to implement the Higgsing then leads to vevs also to other hypers, charged under the diagonal combination that manifestly appears in the punctures, and the Higgsing just jumps over these cases.     

\paragraph{Induced Higgsings} As discussed in \cite{Bao:2024wls,Bao:2025pxe}, when an end of the 6d generalised quiver is associated with a nilpotent orbit, some of the Higgsings can be understood as the inductions of the nilpotent orbit. In other words, the corresponding end in the 6d quiver of the child theory is given by the orbit that induces the orbit in the parent theory. For the definition of induced orbits, see \cite{collingwood1993nilpotent}. Here, it suffices to know that given a Lie algebra $\mathfrak{g}$ and its Levi subalgebra $\mathfrak{g}'\oplus\mathfrak{sl}(L)$ (where $\mathfrak{g}'$ is of the same type as $\mathfrak{g}$), the orbit in $\mathfrak{g}'\oplus\mathfrak{sl}(L)$ labelled by the partitions $\bm{f}\oplus\bm{d}$ induces an orbit in $\mathfrak{g}$ as follows. Write
\begin{equation}
    \bm{p}=[p_1,p_2,\dots]=[2d_1+f_1,2d_2+f_2,\dots]\;.
\end{equation}
Then the induced orbit is $\bm{p}_\text{X}$, where the subscript denotes the X-collapse of $\bm{p}$ (with X being B, C or D). The X-collapse of $\bm{p}$ is the unique largest partition $\bm{p}_\text{X}\preceq\bm{p}$ such that it gives an orbit in the Lie algebra $\mathfrak{g}$ of type X.

For orbi-instanton theories, one end is associated with the nilpotent orbits while the other end is associated with the discrete homomorphisms. When a Higgsing corresponds to the induction of the nilpotent orbits, we also say that this is an induced Higgsing, yielding a physical definition of the ``inductions'' of the discrete homomorphisms.

As an example, we list the possible induced Higgsings for the $D_4$ case with the child theories being A-type orbi-instanton theories in \cref{D4_A3_orbi-instanton_table}. Notice that we have changed the end for the orbit $\left[1^8\right]_\text{D}$ to $\left[2^4\right]_\text{D}$ in order to have such inductions (where the subscripts are used to distinguish partitions of different types)\footnote{The corresponding part in the global symmetry is now $\USp(4)$ instead of $\SO(8)$. In the class $\mathcal{S}$ theory descriptions, we simply replace the part $1^8$ in the puncture $\lambda$ with $2^4$.}. The orbit\footnote{The subtlety of the very even orbits is not important here. Therefore, we shall ignore the colours of the partitions.} $\left[2^4\right]_\text{D}$ is induced by the orbit $\left[1^4\right]_\text{A}$ in $\mathfrak{sl}(3)$ (where we have omitted the trivial $\mathfrak{gl}(1)$ part). As we can see, the dimension of the Higgs branch is always reduced by 1 after Higgsing.
\begin{table}[ht]
\centering
  \begin{tabular}{c|c|c||c|c|c}
\hline
$D_4$ label & Global symmetry & $\dim_{\mathbb{H}}-30K$ & $[\bbZ_4, E_8]$ & Global symmetry & $\dim_{\mathbb{H}}-30K$ \\ \hline \hline
0 & $E_8$ & $190$ & $1^4$  &  $E_8$ & $189$ \\ \hline
1 & $E_7$ & $161$ & $1^2+2$  &  $E_7\times\text{U}(1)$ & $160$ \\ \hline
2 & $\SO(13)$ & $144$ & $1^2+2'$  &  $\SO(14)\times\text{U}(1)$ & $143$ \\ \hline
5 & $E_6$ & $133$ & $1+3$  &  $E_6\times\SU(2)\times\text{U}(1)$ & $132$ \\ \hline
{\color{orange} 6} & $\SU(6)\times\text{U}(1)^2$ & $122$ & $1+3'$  &  $\SU(8)\times\text{U}(1)$ & $121$  \\ \hline
7 & $E_7\times\SU(2)$ & $132$ & $2^2$  &  $E_7\times\SU(2)$ & $131$  \\ \hline
{\color{orange} 12} & $\SO(12)\times\SU(2)$ & $115$ & $2+2'$  &  $\SO(12)\times\SU(2)\times\text{U}(1)$ & $114$  \\ \hline
{\color{orange} 14} & $\SO(9)\times\USp(4)$ & $106$ & $4$  &  $\SO(10)\times\SU(4)$ & $105$ \\ \hline
{\color{orange} 17} & $\SU(8)\times\text{U}(1)$ & $99$ & $4'$  &  $\SU(8)\times\SU(2)$ & $98$ \\ \hline
{\color{orange} 18} & $\SO(16)$ & $98$ & $(2')^2$  &  $\SO(16)$ & $97$  \\ \hline
\end{tabular}
  \caption{Inductions from $\mathrm{Hom}(\mathbb{Z}_4, E_8)$ to $\Hom(\text{Dic}_2, E_8)$, realized by 6d SCFTs. We have omitted the parts associated with the nilpotent orbits in the global symmetries ($\USp(4)$ and $\SU(4)$ respectively). The dimension of the Higgs branch is given by the corresponding number plus $30K$, depending on the length of the 6d generalised quiver (or equivalently, $n$). For the same $K$, the $D_4$ theories with larger labels can be Higgsed from those with smaller labels, and $K$ is chosen such that case 18 starts to become the short quiver when $K=0$. The theories in orange are those with the D-type class $\mathcal{S}$ descriptions.}\label{D4_A3_orbi-instanton_table}
\end{table}

Among these D-type orbi-instanton theories, five of them have known class $\mathcal{S}$ descriptions. How they are changed under the listed induced Higgsings is given in \cref{D4_A3_Class_S_table}. They can be divided into two cases as follow:
\begin{itemize}
    \item For theories 6 and 17 (the second one in \cref{D4_A3_Class_S_table}), the induced Higgsings can also be seen from the punctures, and hence the 3d mirror theories. More concretely, for each partition $\bm{d}$ labelling an A-type puncture, its parent theory has the corresponding D-type puncture given by the partition $\bm{p}_\text{D}$, where
    \begin{equation}
        \bm{p}=[p_1,p_2,\dots]=[2d_1,2d_2,\dots]
    \end{equation}
    as $\bm{f}$ is trivial. In other words, the D-type partitions are induced by the A-type partitions. This justifies the induced Higgsings also in the class $\mathcal{S}$ descriptions.

    \item For theories 12, 14, 18 (and the first one for theory 17), the induced Higgsings are not manifest from the punctures and the 3d mirror theories. For some D-type partitions, one needs to take a partial closure after taking the induced partitions from the A-type ones. This renders the induced Higgsings subtle from the perspective of the punctures. In terms of the 3d mirror theories, some nodes in the A-type theories would have larger ranks compared to the corresponding nodes in the D-type theories. Therefore, we cannot see the proposed Higgsings from the decay and fission of the magnetic quivers obtained here. Likewise, there would be issues when one performs the quiver subtractions.
\end{itemize}
\begin{table}[ht]
\centering
\begin{tabular}{|c|c||c|c|}
\hline
        $D_4$ label          & D-type punctures &         $[\mathbb{Z}_4,E_8]$          & A-type punctures \\ \hline
\multirow{3}{*}{6} & $\left[(2K+6)^4,(2K+5)^2,2^4\right]$ & \multirow{3}{*}{$1+3'$} & $\left[(K+3)^5,K+2,1^4\right]$ \\
                  & $\left[(4K+14)^2,4K+13,1\right]$ &                   & $\left[(2K+7)^3\right]$ \\
                  & $\left[(6K+21)^2\right]$ &                   & $\left[3K+11,3K+10\right]$ \\ \hline
\multirow{3}{*}{12} & $\left[(2K+6)^4,2K+5,2K+3,2^4\right]$ & \multirow{3}{*}{$2+2'$} & $\left[(K+3)^4,(K+2)^2,1^4\right]$ \\
                  & $\left[(4K+13)^3,1\right]$ &                   & $\left[(2K+7)^2,2K+6\right]$ \\
                  & $\left[(6K+20)^2\right]$ &                   & $\left[(3K+10)^2\right]$ \\ \hline
\multirow{3}{*}{14} & $\left[(2K+5)^5,2K+3,2^4\right]$ & \multirow{3}{*}{$4$} & $\left[(K+3)^2,(K+2)^4,1^4\right]$ \\
                  & $\left[(4K+12)^2,4K+11,1\right]$ &                   & $\left[(2K+6)^3\right]$ \\
                  & $\left[(6K+18)^2\right]$ &                   & $\left[(3K+9)^2\right]$ \\ \hline
\multirow{3}{*}{17} & $\left[(2K+5)^4,(2K+3)^2,2^4\right]$ & \multirow{3}{*}{$4'$} & $\left[(K+2)^6,1^4\right]$ \\
                  & $\left[(4K+11)^3,1\right]$ &                   & $\left[2K+6,(2K+5)^2\right]$ \\
                  & $\left[(6K+17)^2\right]$ &                   & $\left[(3K+8)^2\right]$ \\ \hline
\multirow{3}{*}{17} & $\left[(2K+4)^6,2^4\right]$ & \multirow{3}{*}{$4'$} & $\left[(K+2)^6,1^4\right]$ \\
                  & $\left[(4K+11)^2,4K+9,1\right]$ &                   & $\left[2K+6,(2K+5)^2\right]$ \\
                  & $\left[(6K+16)^2\right]$ &                   & $\left[(3K+8)^2\right]$ \\ \hline
\multirow{3}{*}{18} & $\left[(2K+5)^5,2K+1,2^4\right]$ & \multirow{3}{*}{$(2')^2$} & $\left[(K+3)^4,1^4\right]$ \\
                  & $\left[(4K+11)^3,1\right]$ &                   & $\left[(2K+8)^2,2K+6\right]$ \\
                  & $\left[(6K+17)^2\right]$ &                   &  $\left[(3K+11)^2\right]$\\ \hline
\end{tabular}
\caption{The class $\mathcal{S}$ counterpart of the induced flows. The A-type class $\mathcal{S}$ theories was obtained following the algorithm in \cite{Mekareeya:2017jgc}. In the punctures, $N$ follows the one in \cref{D4_A3_orbi-instanton_table} (i.e., the dimension of the Higgs branch agrees with the corresponding number in \cref{D4_A3_orbi-instanton_table} when $K=0$).}\label{D4_A3_Class_S_table}
\end{table}

This raises the question whether such Higgsings would exist for the second case. After all, from 6d, this is obtained by blowing down a $-1$ curve and reducing the gauge algebras simultaneously, which is the same as the first case. It is tempting to assert that the answer is positive. This is supported by the two distinct class $\mathcal{S}$ descriptions of theory 17. From the second one in \cref{D4_A3_orbi-instanton_table}, this Higgsing is manifest from the punctures and the 3d mirror theories. Therefore, the Higgsing is also predicted from these perspectives. Since the other class $\mathcal{S}$ description is equivalent with the same Higgs branch, the Higgsing should exist even if it is hidden from the class $\mathcal{S}$ theory descriptions. One possibility is that there is a magnetic quiver equivalent to the quiver of the child theory here, and the Higgsing would be manifest from that potential magnetic quiver (although the equivalence to the magnetic quiver directly from the class $\mathcal{S}$ punctures here would be more difficult to see).

It is natural to ask what would happen if we simply take the induced partitions without any further partial closures. Then there would be no issue with the ranks in the 3d mirror theory. It turns out that the resulting theory would be bad. In particular, the dimension of the Higgs branch of the bad class $\mathcal{S}$ theory would be the same as the good theory listed here. Therefore, if it flows to some combination of good theories and ugly theories, the good part should not be the desired 4d reduction from the 6d theory. Nevertheless, from the 5d viewpoint, the brane webs of the two (the one with the induced partitions and the one with further partial closures) should be related by Hanany-Witten transitions. This also suggests that the Higgsing should exist as it is manifest from the one with the induced partitions. It is hidden from the class $\mathcal{S}$ description as the full global symmetry may not be directly reflected from the punctures.

\section*{Acknowledgement}
We would like to thank Jacques Distler, Deshuo Liu, and Zhenghao Zhong for enjoyable discussions. We are grateful to Hiroyuki Shimizu for collaboration in early stages of the project. We also gratefully acknowledge support from the Simons Center for Geometry and Physics, Stony Brook University, where various stages of this research were performed during the programme ``Supersymmetric Quantum Field Theories, Vertex Operator Algebras, and Geometry'' and the 22nd Simons Physics Summer Workshop (2025). N.M. also thanks Seoul National University for hospitality during the workshop ``Aspects of Supersymmetric Quantum Field Theory 2025''. Special thanks go to Carlotta Meneghini and Michele Sarzana for their warm hospitality during the completion of this project. J.B. is supported by JSPS Grant-in-Aid for Scientific Research (Grant No.~23K25865). N.M.’s research is partially supported by the MUR-PRIN grant No. 2022NY2MXY (Finanziato dall’Unione europea – Next Generation EU, Missione 4 Componente 1 CUP H53D23001080006, I53D23001330006). G.Z. is partially supported by the Israel Science Foundation under grant No.~759/23. H.Y.Z. is supported by WPI Initiative, MEXT, Japan at Kavli IPMU, the University of Tokyo.

\appendix

\section{Explicit Examples}\label{examples}
In this appendix, we discuss some explicit examples. In particular, we shall list all the specific cases we find for $\widehat{D}_3$, $\widehat{D}_4$ and $\widehat{D}_5$ that can be described by three untwisted regular D-type punctures on a sphere. In \cref{Higgsings}, we shall see that these cases are connected by Higgsings. A more systematic analysis of the punctures and the corresponding 6d theories will be given in \cref{6d4drelations}.

\subsection{The \texorpdfstring{$D_3$}{D3} Case}\label{D3}
We begin with the case of $\widehat{D}_3=\widehat{A}_3=\mathbb{Z}_4$. This case is special as one can also compare it with the A-type description given in \cite{Mekareeya:2017jgc}. Additionally, in this case, we have a complete classification in terms of the Kac labels. The classification gives 10 possible theories, which are summarised in \cref{D3table}. When listing the global symmetries, we shall always omit the common part that corresponds to the nilpotent orbit end in the 6d generalised quiver.
\begin{table}[ht]
  \centering
  \begin{tabular}{c|c||c|c}
      0 & $E_8$ & {\color{orange} 5} & $\SO(12)\times\SU(2)$ \\\hline
      1 & $E_7\times\SU(2)$ & 6 & $E_6\times\SU(2)$ \\\hline 
	   2 & $\SO(14)$ & {\color{orange} 7} & $\SU(8)$ \\\hline
		 3 & $E_7$ & {\color{orange} 8} & $\SO(10)\times\SU(4)$ \\\hline
	   {\color{orange} 4} & $\SO(16)$ & {\color{orange} 9} & $\SU(8)\times\SU(2)$
  \end{tabular}
  \caption{All the $\widehat{D}_3$ cases and their global symmetries. The numbering follows the order in \cite{Mekareeya:2017jgc}. The theories in orange are those with the D-type class $\mathcal{S}$ descriptions.}\label{D3table}
\end{table}

Out of these, we find the following cases:
\begin{itemize}
 \item $x = 1$, $m = 1$: case 7 for $n$ even, case 8 for $n$ odd;
 \item $n_1 = 2$, $x = 1$, $n$ odd: case 4;
 \item $n_2 = 1$, $x = 1$: case 5 for $n$ even, case 9 for $n$ odd;
 \item $n_2 = 1$, $x = 1$, $y = 1$, $n$ even: case 8 (ugly class $\mathcal{S}$ theory including 6 free hypers);
 \item $n_4 = 1$, $l = 1$: case 8 for $n$ even, case 7 for $n$ odd (ugly class $\mathcal{S}$ theory including 3 free hypers);
 \item $x = 1$, $m = 1$, $y = 1$, $n$ odd: case 9 (ugly class $\mathcal{S}$ theory including 6 free hypers);
 \item $n_3 = 1$, $n_1 = 1$, $l = 1$, $n$ odd: case 5 (ugly class $\mathcal{S}$ theory including 8 free hypers);
 \item $n_5 = 1$, $n_1 = 1$, $m = 1$, $x = -1$, $n$ odd: case 5 (ugly class $\mathcal{S}$ theory including 4 free hypers);
 \item $n_5 = 1$, $n_3 = 1$, $x = -1$, $n$ odd: case 8 (ugly class $\mathcal{S}$ theory including 7 free hypers).
\end{itemize}
Above we have written down only the non-vanishing variables. We also remind the reader that the three punctures defining the class $\mathcal{S}$ theory are determined from the numbers through \eqref{classSform}. All other viable cases that are not listed describe bad class $\mathcal{S}$ theories. Overall, we find 5 out of the 10 possible cases, where the missing cases are 0, 1, 2, 3 and 6. The Kac-type labels of the good theories are listed in \cref{D3Kac}.
\begin{table}[ht]
\centering
\begin{tabular}{c|c||cc}
    4 & $(2,-1,1,-1,1,-1,1,-1,1)$ & \multicolumn{1}{c|}{8} & $(1,-1,1,-1,1,0,0,0,0)$ \\ \hline
    5 & $(1,0,0,0,0,0,1,-1,0)$ & \multicolumn{1}{c|}{9} & $(0,0,1,-1,1,-1,1,-1,1)$ \\ \hline
    7 & $(0,0,0,0,0,0,0,0,1)$ &  & 
    \end{tabular}
    \caption{The Kac-type labels of the good theories for $D_3$.}\label{D3Kac}
\end{table}

\subsection{The \texorpdfstring{$D_4$}{D4} Case}
Next, we move to the case of $\widehat{D}_4$. There are 19 known cases \cite{Frey:2018vpw}, which are summarised in \cref{D4table}.
\begin{table}[ht]
  \centering
  \begin{tabular}{c|c||c|c||cc}
     0 & $E_8$ & 7 & $E_7\times\SU(2)$ & \multicolumn{1}{c|}{{\color{orange} 14}} & $\SO(9)\times\USp(4)$ \\\hline
     1 & $E_7\times\SU(2)$ & {\color{orange} 8} & $\SO(7)\times\SU(2)^3$ & \multicolumn{1}{c|}{{\color{orange} 15}} & $\USp(4)^3$ \\\hline
		 2 & $\SO(13)$ & {\color{orange} 9} & $\SO(8)\times\SU(2)$ & \multicolumn{1}{c|}{{\color{orange} 16}} & $\USp(8)\times\SU(2)$ \\\hline
		 3 & $F_4\times\SU(2)$ & {\color{orange} 10} & $\SO(12)\times\SU(2)^2$ & \multicolumn{1}{c|}{{\color{orange} 17}} & $\SU(8)\times\textup{U}(1)$ \\\hline
		 {\color{orange} 4} & $\USp(8)$ & {\color{orange} 11} & $\SO(8)^2$ & \multicolumn{1}{c|}{{\color{orange} 18}} & $\SO(16)$ \\\hline
		 5 & $E_6$ & {\color{orange} 12} & $\SO(12)\times\SU(2)$ & & \\\cline{1-4}
		 {\color{orange} 6} & $\SU(6)\times\textup{U}(1)^2$ & {\color{orange} 13} & $\SO(8)\times\SU(2)^3$ & & 
  \end{tabular}
  \caption{All the $\widehat{D}_4$ cases and their global symmetries. The numbering follows \cite{Bao:2025pxe}. The theories in orange are those with the D-type class $\mathcal{S}$ descriptions (case 4 only has ugly class $\mathcal{S}$ theory descriptions).}\label{D4table}
\end{table}

Out of these, we find the following cases\footnote{There could be different cases giving the same set of punctures. For instance, $n_6=1$, $l=1$ and $l=1$, $x=2$ have coincident partitions, and they both give theories 9 and 16.}:
\begin{itemize}
 \item $n_6 = 1$, $l = 1$: case 9 for $n$ even, case 16 for $n$ odd;
 \item $n_5 = 1$, $n_1 = 1$, $l = 1$, $n$ odd: case 4 (ugly class $\mathcal{S}$ theory including 1 free hyper);
 \item $n_4 = 1$, $x = 1$: case 8 for $n$ even, case 15 for $n$ odd;
 \item $n_4 = 1$, $n_2 = 1$, $l = 1$: case 13 for $n$ even, case 6 for $n$ odd (ugly class $\mathcal{S}$ theory including 3 free hyper);
 \item $n_4 = 1$, $n_1 = 2$, $l = 1$, $n$ even: case 14 (ugly class $\mathcal{S}$ theory including 3 free hypers);
 \item $n_3 = 2$, $l = 1$, $n$ odd: case 8 (ugly class $\mathcal{S}$ theory including 7 free hypers);
 \item $n_3 = 1$, $n_1 = 3$, $l = 1$, $n$ odd: case 12 (ugly class $\mathcal{S}$ theory including 8 free hypers);
 \item $n_1 = 4$, $x = 1$, $n$ odd: case 18;
 \item $n_1 = 2$, $l = 2$, $x = 1$, $n$ odd: case 16 (ugly class $\mathcal{S}$ theory including 5 free hypers);
 \item $n_3 = 1$, $n_1 = 1$, $x = 1$, $n$ odd: case 16;
 \item $l = 1$, $x = 2$: case 16 for $n$ even, case 9 for $n$ odd;
 \item $n_2 = 2$, $x = 1$: case 10 for $n$ even, case 17 for $n$ odd;
 \item $m = 2$, $x = 1$: case 17 for $n$ even, case 11 for $n$ odd;
 \item $n_2 = 1$, $m = 1$, $x = 1$: case 6 for $n$ even, case 13 for $n$ odd;
 \item $n_2 = 1$, $n_1 = 1$, $x = 1$, $n$ even: case 12;
 \item $n_2 = 1$, $x = 1$, $l = 2$: case 8 for $n$ even, case 13 for $n$ odd (ugly class $\mathcal{S}$ theory including 4 free hypers);
 \item $n_1 = 2$, $m = 1$, $x = 1$, $n$ odd: case 14;
 \item $n_6 = 1$, $l = 1$, $y = 1$, $n$ odd: case 4 (ugly class $\mathcal{S}$ theory including 3 free hypers);
 \item $n_4 = 1$, $x = 1$, $y = 1$, $n$ even: case 9 (ugly class $\mathcal{S}$ theory including 3 free hypers);
 \item $x = 2$, $l = 1$, $y = 1$, $n$ even: case 4 (ugly class $\mathcal{S}$ theory including 3 free hypers);
 \item $n_2 = 2$, $x = 1$, $y = 1$, $n$ even: case 13 (ugly class $\mathcal{S}$ theory including 8 free hypers);
 \item $m = 2$, $x = 1$, $y = 1$, $n$ even: case 6 (ugly class $\mathcal{S}$ theory including 6 free hypers);
 \item $n_2 = 1$, $m = 1$, $x = 1$, $y = 1$, $n$ odd: case 15 (ugly class $\mathcal{S}$ theory including 4 free hypers);
 \item $n_2 = 1$, $n_1 = 2$, $x = 1$, $y = 1$, $n$ even: case 14 (ugly class $\mathcal{S}$ theory including 8 free hypers);
 \item $n_1 = 2$, $m = 1$, $x = 1$, $y = 1$, $n$ odd: case 16 (ugly class $\mathcal{S}$ theory including 5 free hypers);
 \item $n_5 = 1$, $n_3 = 1$, $m = 1$, $x = -1$, $n$ odd: case 8 (ugly class $\mathcal{S}$ theory including 3 free hypers);
 \item $n_5 = 1$, $n_1 = 3$, $m = 1$, $x = -1$, $n$ odd: case 12 (ugly class $\mathcal{S}$ theory including 4 free hypers);
 \item $n_5 = 2$, $x = -1$, $n$ odd: case 9 (ugly class $\mathcal{S}$ theory including 3 free hypers);
 \item $n_5 = 1$, $n_3 = 1$, $n_2 = 1$, $x = -1$, $n$ odd: case 13 (ugly class $\mathcal{S}$ theory including 7 free hypers);
 \item $n_5 = 1$, $n_3 = 1$, $n_1 = 2$, $x = -1$, $n$ odd: case 14 (ugly class $\mathcal{S}$ theory including 7 free hypers).
\end{itemize}
Above we have written down only the non-vanishing variables. All other viable cases that are not listed describe bad class $\mathcal{S}$ theories. Overall, we find 13 out of the 19 possible cases, where the missing cases are 0, 1, 2, 3, 5 and 7. The Kac-type labels of the good theories are listed in \cref{D4Kac}.
\begin{table}[ht]
\centering
\begin{tabular}{c|c||c|c||cc}
    6 & $(1,0,0,0,0,0,0,0,1)$ & 12 & $(2,0,0,0,0,0,1,-1,0)$ & \multicolumn{1}{c|}{16} & $(1,0,1,-1,1,-1,1,-1,1)_\text{I}$ \\ \hline
    8 & $(1,-1,1,0,0,0,1,-1,0)$ & 13 & $(0,0,1,-1,1,0,0,0,0)$ & \multicolumn{1}{c|}{17} & $(0,0,0,0,0,0,1,0,0)^\text{II}$ \\ \hline
    9 & $(1,-1,1,-1,1,-1,1,0,1)^\text{I}$ & 14 & $(2,-1,1,-1,1,0,0,0,0)$ & \multicolumn{1}{c|}{17} & $(1,0,1,-1,1,-1,1,-1,1)_\text{II}$ \\ \hline
    10 & $(0,1,0,0,0,0,1,-1,0)$ & 15 & $(0,0,0,0,1,-1,1,-1,1)$ & \multicolumn{1}{c|}{18} & $(3,-1,1,-1,1,-1,1,-1,1)$ \\ \hline
    11 & $(1,-1,1,-1,1,-1,1,0,1)^\text{II}$ & 16 & $(0,0,0,0,0,0,1,0,0)^\text{I}$ &  & 
\end{tabular}
    \caption{The Kac-type labels of the good theories for $D_4$.}\label{D4Kac}
\end{table}

\subsection{The \texorpdfstring{$D_5$}{D5} Case}\label{D5}
Next, we move to the case of $\widehat{D}_5$. Here, there are 45 known cases \cite{Frey:2018vpw}, which are summarised in \cref{D5table}. We note that there are cases with the same global symmetry, but different 6d quivers. See \cite{Frey:2018vpw} (these correspond to different embeddings of $\widehat{D}_5$ into $E_8$ with the same centralizer).
\begin{table}[ht]
  \centering
  \begin{tabular}{c|c|c|c|c|c}
 0 & $E_8$ & 15 & $E_6\times\textup{U}(1)$ & {\color{orange} 30} & $\USp(8)\times\SU(2)$ \\\hline
 1 & $E_7\times\SU(2)$ & 16 & $\SO(10)\times\SU(2)\times\textup{U}(1)$ & {\color{orange} 31} & $\SU(6)\times\textup{U}(1)^2$ \\\hline
	 {\color{orange} 2} & $\SU(8)\times\SU(2)$ & {\color{orange} 17} & $\SO(7)^2$ & {\color{orange} 32} & $\SU(6)\times\textup{U}(1)$ \\\hline
	 3 & $E_7\times\textup{U}(1)$ & {\color{orange} 18} & $\SO(9)\times\USp(4)$ & {\color{orange} 33} & $\SU(4)\times\USp(4)\times\text{U}(1)$ \\\hline
	 4 & $E_7$ & 19 & $\SO(13)$ & {\color{orange} 34} & $\SU(4)\times\SU(2)^2$ \\\hline
	 5 & $E_6\times\SU(2)\times\textup{U}(1)$ & 20 & $\SO(12)\times\textup{U}(1)$ & {\color{orange} 35} & $\SU(4)\times\SU(2)^2$ \\\hline
	 {\color{orange} 6} & $\SO(16)$ & {\color{orange}21} & $\SO(8)\times\SU(2)^2\times\textup{U}(1)$ & {\color{orange} 36} & $\SU(4)\times\SU(2)^2\times\textup{U}(1)$ \\\hline
	 {\color{orange} 7} & $\SO(10)\times\SU(4)$ & 22 & $\SO(11)\times\SU(2)$ & {\color{orange} 37} & $\SO(7)\times\SU(2)^2\times\textup{U}(1)$ \\\hline
	 8 & $\SO(14)\times\textup{U}(1)$ & 23 & $F_4\times\SU(2)$ & {\color{orange} 38} & $\SO(7)\times\SU(2)^2$ \\\hline
	 {\color{orange} 9} & $\SU(8)\times\textup{U}(1)$ & 24 & $F_4\times\SU(2)$ & {\color{orange} 39} & $\SO(7)\times\SU(2)^2$ \\\hline
	 {\color{orange} 10} & $\SU(8)$ & 25 & $\SO(9)\times\SU(2)\times\textup{U}(1)$ & {\color{orange} 40} & $\SO(9)$ \\\hline
	 {\color{orange} 11} & $\SU(4)^2\times\SU(2)$ & {\color{orange} 26} & $\USp(4)^2\times\SU(2)\times\textup{U}(1)$ & {\color{orange} 41} & $\USp(8)$ \\\hline
	 {\color{orange} 12} & $\SO(12)\times\SU(2)\times\textup{U}(1)$ & {\color{orange} 27} & $\USp(4)^2$ & {\color{orange} 42} & $\USp(6)\times\SU(2)\times \textup{U}(1)$ \\\hline
	 {\color{orange} 13} & $\SO(12)\times\SU(2)$ & {\color{orange} 28} & $\USp(8)\times\SU(2)$ & {\color{orange} 43} & $\USp(6)\times\SU(2)$ \\\hline
	 {\color{orange} 14} & $\SU(6)\times\SU(2)$ & {\color{orange} 29} & $\USp(6)\times\SU(2)\times\textup{U}(1)$ & {\color{orange} 44} & $\SO(7)\times\SU(2)$
  \end{tabular}
  \caption{All the $\widehat{D}_5$ cases and their global symmetries. The numbering follows the inverse order in \cite{Frey:2018vpw}. The theories in orange are those with the D-type class $\mathcal{S}$ descriptions (cases 41 and 43 only have ugly class $\mathcal{S}$ theory descriptions).}\label{D5table}
\end{table}

Out of these, we find the following cases:
\begin{itemize}
 \item $n_6 = 1$, $x = 1$: case 35 for $n$ even, case 29 for $n$ odd;
 \item $n_6 = 1$, $n_2 = 1$, $l = 1$: case 34 for $n$ even, case 42 for $n$ odd;
 \item $n_6 = 1$, $l = 3$: case 34 for $n$ even, case 42 for $n$ odd (ugly class $\mathcal{S}$ theory including 6 free hypers);
 \item $n_5 = 1$, $n_1 = 1$, $x = 1$, $n$ odd: case 27;
 \item $n_5 = 1$, $n_3 = 1$, $l = 1$, $n$ odd: case 43 (ugly class $\mathcal{S}$ theory including 1 free hyper);
 \item $n_5 = 1$, $n_1 = 3$, $l = 1$, $n$ odd: case 41 (ugly class $\mathcal{S}$ theory including 1 free hyper);
 \item $n_4 = 2$, $l = 1$: case 33 for $n$ even, case 36 for $n$ odd (ugly class $\mathcal{S}$ theory including 3 free hypers);
 \item $n_4 = 1$, $n_3 = 1$, $n_1 = 1$, $l = 1$, $n$ even: case 38 (ugly class $\mathcal{S}$ theory including 3 free hypers);
 \item $n_4 = 1$, $n_2 = 2$, $l = 1$: case 21 for $n$ even, case 14 for $n$ odd (ugly class $\mathcal{S}$ theory including 3 free hypers);
 \item $n_4 = 1$, $n_2 = 1$, $n_1 = 2$, $l = 1$, $n$ odd: case 32 (ugly class $\mathcal{S}$ theory including 3 free hypers);
 \item $n_4 = 1$, $n_2 = 1$, $x = 1$: case 37 for $n$ even, case 26 for $n$ odd;
 \item $n_4 = 1$, $n_1 = 4$, $l = 1$, $n$ even: case 18 (ugly class $\mathcal{S}$ theory including 3 free hypers);
 \item $n_4 = 1$, $n_1 = 2$, $x = 1$, $n$ even: case 39;
 \item $n_4 = 1$, $m = 1$, $x = 1$: case 36 for $n$ even, case 33 for $n$ odd;
 \item $n_4 = 1$, $x = 1$, $l = 2$: case 35 for $n$ even, case 29 for $n$ odd (ugly class $\mathcal{S}$ theory including 2 free hypers);
 \item $n_3 = 2$, $n_2 = 1$, $l = 1$, $n$ odd: case 37 (ugly class $\mathcal{S}$ theory including 7 free hypers);
 \item $n_3 = 2$, $n_1 = 2$, $l = 1$, $n$ odd: case 39 (ugly class $\mathcal{S}$ theory including 7 free hypers);
 \item $n_3 = 2$, $x = 1$, $n$ odd: case 28;
 \item $n_3 = 1$, $n_1 = 5$, $l = 1$, $n$ odd: case 13 (ugly class $\mathcal{S}$ theory including 8 free hypers);
 \item $n_3 = 1$, $n_1 = 3$, $x = 1$, $n$ odd: case 30;
 \item $n_3 = 1$, $n_1 = 1$, $m = 1$, $x = 1$, $n$ odd: case 38;
 \item $n_3 = 1$, $n_1 = 1$, $x = 1$, $l = 2$, $n$ odd: case 27 (ugly class $\mathcal{S}$ theory including 3 free hypers;
 \item $n_2 = 3$, $x = 1$: case 12 for $n$ even, case 2 for $n$ odd;
 \item $n_2 = 2$, $n_1 = 2$, $x = 1$, $n$ odd: case 10;
 \item $n_2 = 2$, $m = 1$, $x = 1$: case 14 for $n$ even, case 21 for $n$ odd;
 \item $n_2 = 2$, $x = 1$, $l = 2$: case 37 for $n$ even, case 26 for $n$ odd (ugly class $\mathcal{S}$ theory including 4 free hypers);
 \item $n_2 = 1$, $n_1 = 4$, $x = 1$, $n$ even: case 13;
 \item $n_2 = 1$, $n_1 = 2$, $m = 1$, $x = 1$, $n$ even: case 32;
 \item $n_2 = 1$, $n_1 = 2$, $x = 1$, $l = 2$, $n$ even: case 39 (ugly class $\mathcal{S}$ theory including 4 free hypers);
 \item $n_2 = 1$, $x = 2$, $l = 1$: case 42 for $n$ even, case 34 for $n$ odd;
 \item $n_2 = 1$, $m = 2$, $x = 1$: case 31 for $n$ even, case 11 for $n$ odd;
 \item $n_1 = 6$, $x = 1$, $n$ odd: case 6;
 \item $n_1 = 4$, $m = 1$, $x = 1$, $n$ odd: case 18;
 \item $n_1 = 4$, $x = 1$, $l = 2$, $n$ odd: case 30 (Ugly class $\mathcal{S}$ theory including 5 free hypers);
 \item $n_1 = 2$, $x = 2$, $l = 1$, $n$ odd: case 44;
 \item $n_1 = 2$, $m = 2$, $x = 1$, $n$ odd: case 17;
 \item $x = 3$: case 29 for $n$ even, case 35 for $n$ odd;
 \item $x = 2$, $l = 3$: case 42 for $n$ even, case 34 for $n$ odd (ugly class $\mathcal{S}$ theory including 6 free hypers);
 \item $m = 3$, $x = 1$: case 9 for $n$ even, case 7 for $n$ odd;
 \item $m = 2$, $x = 1$, $l = 2$: case 31 for $n$ even, case 11 for $n$ odd (ugly class $\mathcal{S}$ theory including 6 free hypers);
 \item $n_6 = 1$, $x = 1$, $y = 1$, $n$ even: case 40;
 \item $n_6 = 1$, $x = 1$, $y = 2$, $n$ odd: case 43 (ugly class $\mathcal{S}$ theory including 9 free hypers);
 \item $n_6 = 1$, $n_2 = 1$, $l = 1$, $y = 1$, $n$ odd: case 43 (ugly class $\mathcal{S}$ theory including 2 free hypers);
 \item $n_4=1$, $n_2=1$, $x=1$, $y=1$, $n$ even: case 34;
 \item $n_4 = 1$, $n_1 = 2$, $x = 1$, $y = 1$, $n$ even: case 44 (ugly class $\mathcal{S}$ theory including 3 free hypers);
 \item $n_4 = 1$, $m = 1$, $x = 1$, $y = 1$, $n$ odd: case 29 (ugly class $\mathcal{S}$ theory including 2 free hypers);
 \item $n_3 = 1$, $n_1 = 1$, $m = 1$, $x = 1$, $y = 1$, $n$ odd: case 27 (ugly class $\mathcal{S}$ theory including 3 free hypers);
 \item $n_2 = 3$, $x = 1$, $y = 1$, $n$ even: case 21 (ugly class $\mathcal{S}$ theory including 8 free hypers);
 \item $n_2 = 2$, $m = 1$, $x = 1$, $y = 1$, $n$ odd: case 26 (ugly class $\mathcal{S}$ theory including 4 free hypers);
 \item $n_2 = 1$, $n_1 = 4$, $x = 1$, $y = 1$, $n$ even: case 18 (ugly class $\mathcal{S}$ theory including 8 free hypers);
 \item $n_2 = 1$, $x = 2$, $l = 1$, $y = 1$, $n$ even: case 43 (ugly class $\mathcal{S}$ theory including 2 free hypers);
 \item $n_2 = 1$, $m = 2$, $x = 1$, $y = 1$, $n$ even: case 36 (ugly class $\mathcal{S}$ theory including 4 free hypers);
 \item $n_1 = 4$, $m = 1$, $x = 1$, $y = 1$, $n$ odd: case 30 (ugly class $\mathcal{S}$ theory including 5 free hypers);
 \item $x = 3$, $y = 1$, $n$ odd: case 40;
 \item $x = 3$, $y = 2$, $n$ even: case 43 (ugly class $\mathcal{S}$ theory including 9 free hypers);
 \item $m = 3$, $x = 1$, $y = 1$, $n$ odd: case 11 (ugly class $\mathcal{S}$ theory including 6 free hypers);
 \item $n_5 = 1$, $n_3 = 2$, $n_1 = 1$, $x = -1$, $n$ odd: case 38 (ugly class $\mathcal{S}$ theory including 7 free hypers);
 \item $n_5 = 1$, $n_3 = 1$, $n_2 = 2$, $x = -1$, $n$ odd: case 21 (ugly class $\mathcal{S}$ theory including 7 free hypers);
 \item $n_5 = 1$, $n_3 = 1$, $n_1 = 4$, $x = -1$, $n$ odd: case 18 (ugly class $\mathcal{S}$ theory including 7 free hypers);
 \item $n_5 = 1$, $n_3 = 1$, $n_2 = 1$, $m = 1$, $x = -1$, $n$ odd: case 37 (ugly class $\mathcal{S}$ theory including 3 free hypers);
 \item $n_5 = 1$, $n_3 = 1$, $n_1 = 2$, $m = 1$, $x = -1$, $n$ odd: case 39 (ugly class $\mathcal{S}$ theory including 3 free hypers);
 \item $n_5 = 1$, $n_1 = 5$, $m = 1$, $x = -1$, $n$ odd: case 13 (ugly class $\mathcal{S}$ theory including 4 free hypers).
\end{itemize}
Above we have written down only the non-vanishing variables. All other viable cases that are not listed describe bad class $\mathcal{S}$ theories. Overall, we find 31 out of the 45 possible cases, where the missing cases are 0, 1, 3, 4, 5, 8, 15, 16, 19, 20, 22, 23, 24 and 25. The Kac-type labels of the good theories are listed in \cref{D5Kac}.
\begin{table}[ht]
\centering
\begin{tabular}{c|c||c|c||cc}
    2 & $(0,-1,1,-1,1,-1,1,-1,1)$ & 21 & $(1,0,1,-1,1,0,0,0,0)^\text{II}$ & \multicolumn{1}{c|}{36} & $(1,-1,1,0,0,0,0,0,1)$ \\ \hline
    6 & $(4,-1,1,-1,1,-1,1,-1,1)$ & 26 & $(1,0,0,0,1,-1,1,-1,-1)^\text{II}$ & \multicolumn{1}{c|}{37} & $(0,0,1,0,0,0,1,-1,0)$ \\ \hline
    7 & $(1,-1,1,-1,1,0,0,1,0)$ & 27 & $(1,0,0,0,1,-1,1,-1,1)^\text{I}$ & \multicolumn{1}{c|}{38} & $(1,0,1,-1,1,0,0,0,0)^\text{I}$ \\ \hline
    9 & $(0,0,0,0,0,0,0,0,1)$ & 28 & $(1,-1,2,-1,1,-1,1,-1,1)$ & \multicolumn{1}{c|}{39} & $(2,-1,1,0,0,0,1,-1,0)$ \\ \hline
    10 & $(2,0,1,-1,1,-1,1,-1,1)^\text{II}$ & 29 & $(0,0,0,0,0,0,1,-1,1)$ & \multicolumn{1}{c|}{40} & $(1,-1,1,-1,1,-1,1,-1,2)$ \\ \hline
    11 & $(0,0,1,-1,1,-1,1,0,1)^\text{II}$ & 30 & $(2,0,1,-1,1,-1,1,-1,1)^\text{I}$ & \multicolumn{1}{c|}{42} & $(1,0,0,0,0,0,1,0,0)^\text{I}$ \\ \hline
    12 & $(1,1,0,0,0,0,1,-1,0)$ & 31 & $(1,0,0,0,0,0,1,0,0)^\text{II}$ & \multicolumn{1}{c|}{44} & $(2,-1,1,-1,1,-1,1,0,1)^\text{I}$ \\ \hline
    13 & $(3,0,0,0,0,0,1,-1,0)$ & 32 & $(2,0,0,0,0,0,0,0,1)$ &  &  \\ \cline{1-4}
    14 & $(0,1,0,0,0,0,0,0,1)$ & 33 & $(0,0,0,0,1,0,0,0,0)$ &  &  \\ \cline{1-4}
    17 & $(2,-1,1,-1,1,-1,1,0,1)^\text{II}$ & 34 & $(0,0,1,-1,1,-1,1,0,1)^\text{I}$ &  &  \\ \cline{1-4}
    18 & $(3,-1,1,-1,1,0,0,0,0)$ & 35 & $(1,-1,1,-1,1,0,1,-1,0)$ &  & 
\end{tabular}
    \caption{The Kac-type labels of the good theories for $D_5$.}\label{D5Kac}
\end{table}

\addcontentsline{toc}{section}{References}
\bibliographystyle{ytamsalpha}
\bibliography{references}

\end{document}